%% file: ICF-SRSR.tex
\documentclass[10pt,twocolumn,letterpaper]{article}

\usepackage[pagenumbers]{wacv}      
\usepackage{times}
\usepackage{epsfig}
\usepackage{graphicx}
\usepackage{amsmath}
\usepackage{amssymb}
\usepackage{booktabs}

\usepackage[pagebackref,breaklinks,colorlinks,citecolor=blue]{hyperref}
\usepackage[capitalize]{cleveref}
\crefname{section}{Sec.}{Secs.}
\Crefname{section}{Section}{Sections}
\Crefname{table}{Table}{Tables}
\crefname{table}{Tab.}{Tabs.}

\newcommand{\Paragraph}[1]{\noindent \textbf{#1}}
\usepackage{appendix}
\usepackage{titling}
    \pretitle{\vspace*{-3pt}\begin{center}\Large \bf}
    \posttitle{\vspace*{10pt}\par\end{center}}
    \preauthor{\large\begin{center}\par}
    \postauthor{\vspace*{6pt}\par\end{center}}
    \predate{}
    \date{}
    \postdate{}

\usepackage[utf8]{inputenc} % allow utf-8 input
\usepackage[T1]{fontenc}    % use 8-bit T1 fonts
\usepackage{url}            % simple URL typesetting
\usepackage{booktabs}       % professional-quality tables
\usepackage{amsfonts}       % blackboard math symbols
\usepackage{nicefrac}       % compact symbols for 1/2, etc.
\usepackage{microtype}      % microtypography
\usepackage{graphicx}
\usepackage{amsmath}
\usepackage{amssymb}
\usepackage{cancel}
\usepackage{enumerate}
\usepackage{comment}
\usepackage{stackengine}
\usepackage[dvipsnames]{xcolor}
\usepackage{pifont}
\usepackage{tikz}
\usepackage[accsupp]{axessibility}
\usepackage{tabularx}
\usepackage{multirow}

\newcommand*\samethanks[1][\value{footnote}]{\footnotemark[#1]}
\usepackage[capitalize]{cleveref}
\usepackage{subcaption}
\newcommand\cmark[1][]{%
  \tikz[scale=0.4,#1]{\fill(0,.35) -- (.25,0) -- (1,.7) -- (.25,.15) -- cycle;}%
}
\newcommand\crossmark[1][]{%
  \tikz[scale=0.4,#1]{
    \fill(0,0)--(0.1,0) .. controls (0.5,0.4) .. (1,0.7)--(0.9,0.7) ..  controls (0.5,0.5) ..(0,0.1) --cycle;
    \fill(1,0.1)--(0.9,0.1) .. controls (0.5,0.3) .. (0,0.7)--(0.1,0.7) .. controls (0.5,0.4) ..(1,0.2) --cycle;
  }%
}

\def\wacvPaperID{1239} % *** Enter the WACV Paper ID here
\def\confName{WACV}
\def\confYear{2024}

\begin{document}

\title{ICF-SRSR: Invertible scale-Conditional Function \textit{for}\\ Self-Supervised Real-world Single Image Super-Resolution}

\author{Reyhaneh Neshatavar$^{1}$\thanks{equal contribution} \qquad Mohsen Yavartanoo$^{1}$\textcolor{red}{\samethanks} \qquad Sanghyun Son$^{1}$ \qquad Kyoung Mu Lee$^{1,2}$ \\$^{1}$Dept. of ECE \& ASRI, $^{2}$IPAI, Seoul National University, Seoul, Korea\\
{\tt\small \{reyhanehneshat,myavartanoo,thstkdgus35,kyoungmu\}@snu.ac.kr}}

\maketitle

%%%%%%%%% ABSTRACT
\input{sections/abstract}
%%%%%%%%% BODY TEXT
\input{sections/introduction}
\input{sections/related_work}

\input{sections/method}

\input{sections/expriment}

\input{sections/conclusion}

%%%%%%%%% REFERENCES
{\small
\bibliographystyle{ieee_fullname}
\bibliography{ICF-SRSR}
}

%\input{sections/appendix}
%%%%%%%%%%%%%%%%%%%%%%%%%%%%%%%%%%%%%%%%%%%%%%%%%%%%%%%%%%%%
\appendixpageoff
\appendixtitleoff

\begin{appendices}

%%%%%%%%% TITLE - PLEASE UPDATE
\title{Supplementary Material \textit{for} \\
ICF-SRSR: Invertible scale-Conditional Function \textit{for}\\ Self-Supervised Real-world Single Image Super-Resolution}

\author{Reyhaneh Neshatavar$^{1\textcolor{red}{\ast}}$ \qquad Mohsen Yavartanoo$^{1\textcolor{red}{\ast}}$ \qquad Sanghyun Son$^{1}$ \qquad Kyoung Mu Lee$^{1,2}$ 
\\$^{1}$Dept. of ECE \& ASRI, $^{2}$IPAI, Seoul National University, Seoul, Korea\\
{\tt\small \{reyhanehneshat,myavartanoo,thstkdgus35,kyoungmu\}@snu.ac.kr}}

\maketitle
\renewcommand{\thesection}{S\arabic{section}}
\input{sections/figures/net_arch}

\input{sections/figures/multi}

%%%%%%%%%%%%%%%%%%%%%%%%%%%%%%%%%%%%%%%%%%%%%%%%%%%%%%%%%
\section{Details of network architecture}
As described in \cref{sec:net_arch} of our main manuscript, our ICF-SRSR adopts EDSR~\cite{lim2017enhanced} as a baseline.
However, to handle both up-sampling and down-sampling operations with the same network, we slightly modify the tail part of the original EDSR architecture for each scaling factor, \eg, $\times2$ and $\times4$, and their inverses.
\cref{fig:supp_edsr} shows the original EDSR~(\cref{fig:supp_edsr_original}) and our modified EDSR~(\cref{fig:supp_edsr_developed}). 
We use the pixel-unshuffle operator to down-sample an input image and generate the corresponding LLR image.
For more stable optimization, we use the detach operator of PyTorch before passing the first outputs to the network again.

\section{Details of multi-scale augmentation strategy}
As we mention in \cref{sec:ablation} of our main manuscript, we can generate images with various scaling factors, \eg, $\times2$, $\times4$, and $\times8$ and their corresponding inverses from a single LR input.
\cref{fig:supp_multi1} shows our multi-tail architecture, which introduces a tail for each of the scale conditions.
Then, we pass the generated output images of different scales to the model $f_\theta$ with their inverse scaling factors.
By doing so, we reconstruct the input LR image as shown in \cref{fig:supp_multi2}.
Accordingly, to train our model $f_\theta$ under such a configuration, we minimize the loss functions $\mathcal{L}^{\text{Cons}}$ and $\mathcal{L}^{\text{Color}}$ defined in \cref{sec:loss} of our main manuscript between the generated images and the input LR image.

\input{tables/supp_benchmark.tex}

\input{tables/supp_baseline.tex}

%%%%%%%%%%%%%%%%%%%%%%%%%%%%%%%%%%%%%%%%%%%%%%%%%%%%%% 
\section{Evaluation by SSIM}
We quantitatively show the results of our ICF-SRSR and EDSR~(LLR,LR) methods compared to other supervised and unsupervised methods trained on DIV2K~\cite{agustsson2017ntire} dataset and tested on the five standard benchmarks~\cite{bevilacqua2012low, zeyde2010single, martin2001database, huang2015single, Manga109} by SSIM metric in \cref{tab:supp-benchmark}.
According to the results, our method outperforms unsupervised method~\cite{huang2015single} on both scaling factors $\times2$ and $\times4$ and supervised method~\cite{chen2021learning} on scaling factor $\times2$ and is comparable with other methods.
%%%%%%%%%%%%%%%%%%%%%%%%%%%%%%%%%%%%%%%%%%%%%%%%%%%%%%%%%
\section{Ablation on baseline model}
We employ different models LIIF~\cite{chen2021learning}, EDSR~\cite{lim2017enhanced}, RDN~\cite{zhang2018residual}, and RCAN~\cite{zhang2018image} as the baseline of our ICF-SRSR framework.
In the case of EDSR, RDN, and RCAN, we develop the original network architecture to generate multi-scale images by applying a tail for each scaling factor $s$ and its inverse $\nicefrac{1}{s}$, individually.
In the case of LIIF, we leverage its continuous attribute to generate any scale of images by sub-sampling from the reconstructed continuous image.
\cref{tab:supp-baseline} shows the results of our ICF-SRSR with different baselines.
We illustrate that our method is model-agnostic and can leverage different state-of-the-art~(SOTA) baseline models.
We note that our method can achieve better performance using advanced baselines except LIIF, which is not trained with continuous scales due to the limitation of the color loss $\mathcal{L}^{\text{Color}}$.
We select the model EDSR as our baseline due to its training time efficiency.

%%%%%%%%%%%%%%%%%%%%%%%%%%%%%%%%%%%%%
\section{Ablation on the hyperparameter $\lambda_{\text{Color}}$.}

We conduct an ablation study to investigate the importance of our color loss $\mathcal{L}^\text{Color}$ defined in \cref{sec:loss} by changing its weight $\lambda_{\text{Color}}$.
Specifically, We increase the weight from 0.1 to 10 and report the performance of our ICF-SRSR trained on the scale $\times2$ of test sets of both real-world dataset RealSR~\cite{cai2019toward} and synthetic datasets Set5~\cite{bevilacqua2012low} and DIV2K~\cite{agustsson2017ntire} validation in \cref{tab:ab-loss}.
The results indicate that $\lambda_{\text{Color}}=0.2$ achieves the best performance on different datasets.
\input{tables/ablation}

%
%%%%%%%%%%%%%%%%%%%%%%%%%%%%%%%%%%%%%
\section{Comparison with DASR}
We follow the official implementation of DASR~\cite{wang2021unsupervised} and train it using 1) HR images of DIV2K, 2) HR images of RealSR-V3, and 3) LR images of RealSR-V3~(self-supervised) and compare the results with our self-supervised method ICF-SRSR in \cref{tab:DASR}. The results demonstrate the superiority of our method to effectively learn from LR images compared to the DASR method.
\input{tables/DASR}

%%%%%%%%%%%%%%%%%%%%%%%%%%%%%%%%%%%%%%%%%%%%%%%%%%%%%
\section{Noise-free results}
\input{sections/figures/supp/noisy}

In \cref{sec:ex_syn} of our main manuscript, we note that the ground-truth images of Set5~\cite{bevilacqua2012low} and Set14~\cite{zeyde2010single} datasets are noisy while our SR images are noise-free.
We show the difference between our SR images and the noisy ground-truth images in \cref{fig:supp-noise}.
The results prove our claim and show that we can restore SR images without any noise.
%%%%%%%%%%%%%%%%%%%%%%%%%%%%%%%%%%%%%%%%%%%%%%%%%%%%%%%%

\section{Complicated down-sampling degradations}
As we show in \cref{sec:ex_real} of our main manuscript, the proposed method can learn from real-world datasets with unknown degradations~(real LR usually includes complicated degradations).
For example, we can train our model $f_\theta$ on images from RealSR-V3~\cite{cai2019toward} and DRealSR~\cite{wei2020component} datasets directly and achieve promising results. 
Furthermore, we train and test our method ICF-SRSR on a dataset with more complicated degradations generated by the Real-ESRGAN~\cite{wang2021realesrgan} down-sampling strategy.
We note that the generated LR images by the Real-ESRGAN~\cite{wang2021realesrgan} down-sampling model are synthesized by a sequence of classical degradations such as blur, resize, noise, JPEG compression, and artifacts to simulate more practical degradations.
\cref{fig:reb_degrade} demonstrates that our method ICF-SRSR can perform $\times 2$ SR faithfully even on images with mild noise and artifacts.
\input{sections/figures/rebuttal_degrade.tex}
\vspace{-2mm}
\section{Visualization of the generated images}
%\vspace{1mm}
In \cref{fig:supp_scales2} and \cref{fig:supp_scales4}, we visualize the generated down-sampled~(LLR) and up-sampled~(SR) images by our ICF-SRSR framework for different scaling factors $\times2$ and $\times4$, respectively on various benchmark datasets Set14~\cite{zeyde2010single}, BSD100~\cite{martin2001database}, and Urban100~\cite{huang2015single} and also real-world dataset RealSR-V3~\cite{cai2019toward}.
We further restore the down-sampled LR images given HR images for scaling factor $\times2$ of Canon and Nikon sets from the RealSR-V3~\cite{cai2019toward} dataset as shown in \cref{fig:reb_LRHR}.
The comparison demonstrates that the generated down-sampled LR images by our self-supervised method ICF-SRSR look similar to the real LR images, validating the ability of our method to synthesize realistic LR-HR image pairs.
Such generated paired images LR-HR are useful to train other off-the-shelf supervised methods, as evident in \cref{tab:reb-down} of our main manuscript.
\input{sections/figures/supp/scales2}
\input{sections/figures/supp/scales4}
\input{sections/figures/rebuttal_LRHR.tex}
%%%%%%%%%%%%%%%%%%%%%%%%%%%%%%%%%%%%%%%%%%%%%%%%%%%%%%

%%%%%%%%%%%%%%%%%%%%%%%%%%%%%%%%%%%%%%%%%%%%%%%%%%%%%%
\vspace{-2mm}
\section{Training on a single image}
\input{sections/figures/supp/single_image}

\input{sections/figures/supp/real_single}

In \ref{sec:ablation} of our main manuscript, we show that our method ICF-SRSR can learn to restore SR images by training on a small dataset and even a single image as shown in \cref{fig:zero-shot}. 
We show more samples to illustrate the ability of our method to learn from only a single image.
Therefore, we train and evaluate our ICF-SRSR model on a single LR image from the test set of the RealSR-V3~\cite{cai2019toward} dataset captured by the Nikon camera for scaling factor $\times2$.
Our results in \cref{fig:supp_single} demonstrate that our method can restore an SR image by training the model on only the same image.
Furthermore, our result for the single-image case is not only on par with the multi-image case but also shows better performance for some samples in terms of PSNR metric and visual appearance.
This attribute makes our method more practical in real-world scenarios where there are not many sample images for training.
Moreover, we train and evaluate our self-supervised method ICF-SRSR on a single real-world smartphone photo and show the results in \cref{fig:supp_real_single}.

\end{appendices}

\end{document}

%% file: sections/abstract.tex
\begin{abstract}
  Single image super-resolution~(SISR) is a challenging ill-posed problem that aims to up-sample a given low-resolution~(LR) image to a high-resolution~(HR) counterpart. 
 Due to the difficulty in obtaining real LR-HR training pairs, recent approaches are trained on simulated LR images degraded by simplified down-sampling operators, \eg, bicubic.
 Such an approach can be problematic in practice because of the large gap between the synthesized and real-world LR images.
 To alleviate the issue, we propose a novel Invertible scale-Conditional Function~(ICF), which can scale an input image and then restore the original input with different scale conditions.
 By leveraging the proposed ICF, we construct a novel self-supervised SISR framework~(ICF-SRSR) to handle the real-world SR task without using any paired/unpaired training data.
 Furthermore, our ICF-SRSR can generate realistic and feasible LR-HR pairs, which can make existing supervised SISR networks more robust.
 Extensive experiments demonstrate the effectiveness of the proposed method in handling SISR in a fully self-supervised manner.
 Our ICF-SRSR demonstrates superior performance compared to the existing methods trained on synthetic paired images in real-world scenarios and exhibits comparable performance compared to state-of-the-art supervised/unsupervised methods on public benchmark datasets.
\end{abstract}

%% file: sections/introduction.tex
\section{Introduction}\label{sec:introduction}
Single image super-resolution~(SISR) as a fundamental vision problem is a procedure to reconstruct a super-resolution~(SR) image from a single low-resolution~(LR) image. 
SISR is an active research topic and has attracted increasing attention in low-level computer vision. 
It has many applications in various fields such as medical imaging \cite{greenspan2001mri, peled2001superresolution}, face recognition \cite{gunturk2003eigenface, wheeler2007multi}, satellite image processing \cite{li2015sub, tatem2002super} and security video surveillance \cite{lin2005investigation, zhang2010super}. 
Recent state-of-the-art~(SOTA) SR methods have achieved remarkable progress due to the development of deep convolutional neural networks~(CNNs).
They are usually trained on synthetic inputs in a fully-supervised fashion where LR images are generated by bicubic down-sampling from their HR counterparts.
%
%%%%%%%%%%%%%%%%%%%%%%%%%%%%%%%%%%%%%%%%%%%%%
\begin{figure}[t]
\small
        %\centering
	\captionsetup[subfloat]{labelformat=empty,aboveskip=1pt}
	%\begin{center}
		\newcommand{\rowArg}{2.43cm}
		\newcommand{\fullSize}{5.78cm}
		\newcommand{\patchSize}{2.7cm}
		\setlength\tabcolsep{0.1cm}
    \scalebox{0.6}{
		\begin{tabular}[b]{c c c}
			\multicolumn{2}{c}{\multirow{2}{*}[\rowArg]{
					\subfloat[Input~(LR)]
					{\includegraphics[trim={20cm 0 2cm 5cm},clip, height=\fullSize]
						{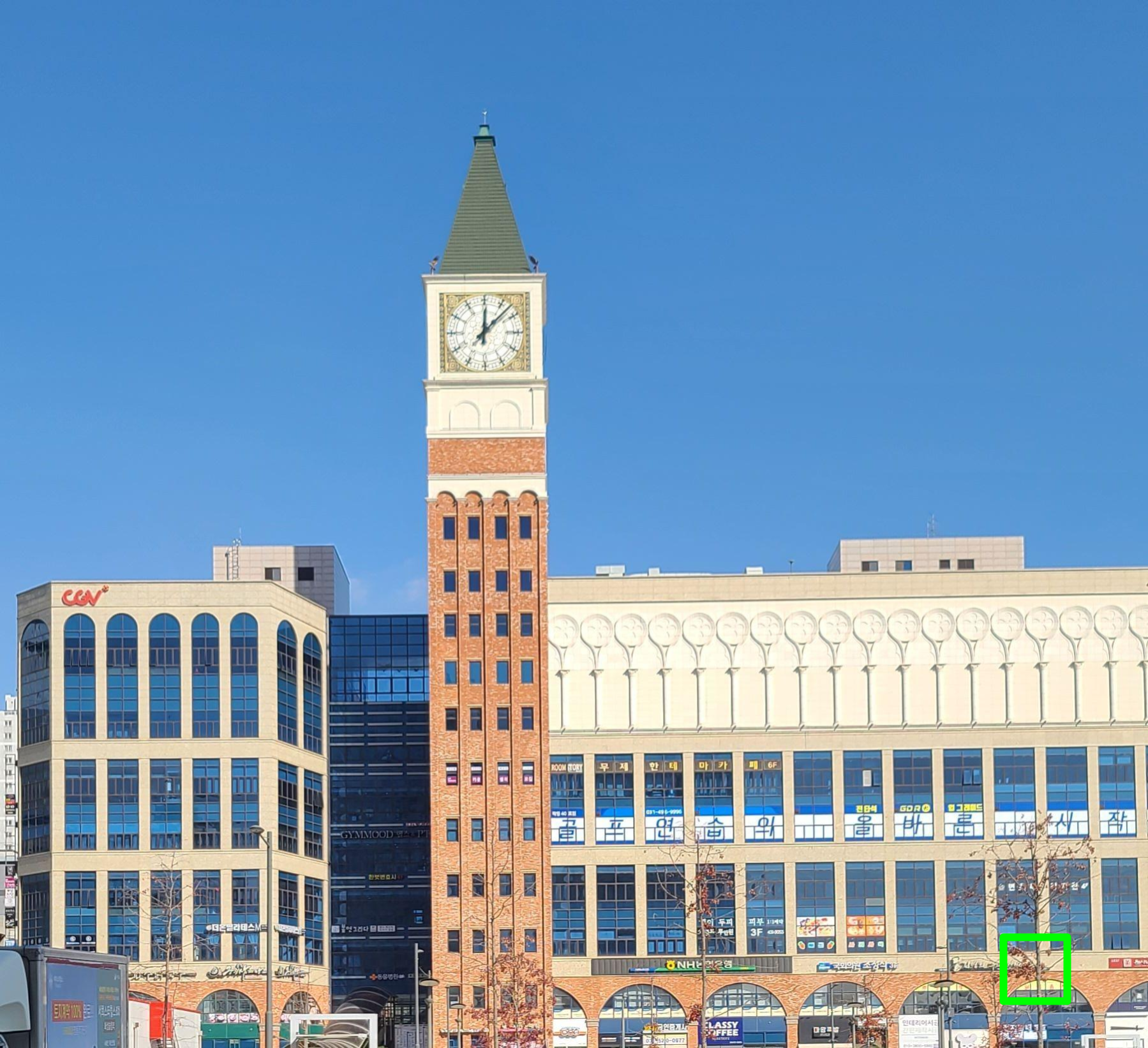}}}} & \hspace{2mm}
			\subfloat[\centering  LR  ]{
				\includegraphics[page=2, width = \patchSize]
				{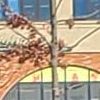}} 
            \hspace{0.5mm}
			\subfloat[\centering   Bicubic   ]{
				\includegraphics[page=3, width = \patchSize]
				{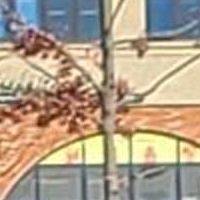}} 
            \hspace{0.5mm}
			\subfloat[\centering  ZSSR~\cite{shocher2018zero}]{
				\includegraphics[page=4, width = \patchSize]
				{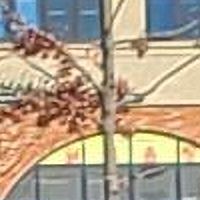}} \\  && \hspace{2mm}

			\subfloat[\centering   MZSR~\cite{soh2020meta}  ]{
				\includegraphics[page=5, width = \patchSize]
				{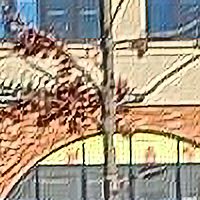}}
            \hspace{0.5mm}
			\subfloat[\centering DASR~\cite{wang2021unsupervised} ]{
				\includegraphics[page=6, width = \patchSize]
				{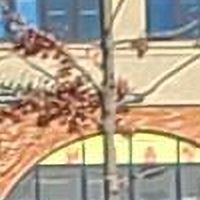}}
            \hspace{0.5mm}
			\subfloat[\centering  \textbf{ICF-SRSR}~(Ours) ]{
				\includegraphics[page=7, width = \patchSize]
				{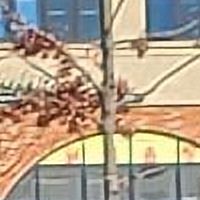}}
	\end{tabular}
 }	
	\vspace{-2mm}
	\caption{\textbf{Real-world image super-resolution}.
	We train our ICF-SRSR on a single real-world smartphone photo in a self-supervised manner to get the result for scale $\times 2$.
	The other listed methods are also zero-shot~\cite{shocher2018zero, soh2020meta} or unsupervised~\cite{wang2021unsupervised} methods.
	}
	\label{fig:zero-shot}
	\vspace{-4mm}
\end{figure}
%%%%%%%%%%%%%%%%%%%%%%%%%%%%%%%%%%%%%%%%%%%%%%%%%
%
Nevertheless, models trained on the synthetic datasets cannot generalize well when applied to real-world inputs~\cite{chen2019camera, cai2019toward}.
Another problem is that acquiring well-constructed LR-HR pairs from the real world is very challenging due to cost problems or hardware limitations~\cite{chen2019camera, cai2019toward, sr_zllz}.
Therefore, it is a common scenario that we have LR images only rather than having LR-HR training pairs.
Several approaches adopt unsupervised adversarial training~\cite{goodfellow2014generative} and leverage unpaired LR-HR images to alleviate the situation.
By jointly training down-sampling and up-sampling networks~\cite{yuan2018unsupervised, sr_deg, sr_gandeg, sr_unpaired_pesudo, sr_unsupervised}, those methods aim to generate synthetic LR images that have similar characteristics of given unpaired LR examples.
Then, the synthesized training pairs can be leveraged to optimize the up-sampling network.
However, such unsupervised strategies require appropriate HR images, even though those images are not paired with the given LR images.
Also, Son~\etal~\cite{son2021toward} have identified that those methods are biased toward some handcrafted functions \eg, nearest or bicubic interpolation, which limits the generalization.

In this paper, we present a novel self-supervised real-world SR framework, ICF-SRSR, to overcome the aforementioned challenges.
To this end, we first propose a concept of Invertible scale-Conditional Function~(ICF).
It is designed to perform up-sampling and down-sampling within a single model, conditioned by the scale arguments $s$ and $\nicefrac{1}{s}$, respectively.
Therefore, we can resize an input by a given scale $s$ and restore the initial input by taking the inverse scale $\nicefrac{1}{s}$.
Without utilizing paired/unpaired training images nor any specific down-sampling operator \eg, bicubic, ICF-SRSR containing a learnable ICF can be trained in a fully self-supervised manner.
Moreover, our method can generate realistic LR-HR image pairs from a set of given images useful for training the other off-the-shelf methods.
In the experiments, we demonstrate the ability of our ICF-SRSR to learn from real-world datasets, restore high-/lower-resolution images, and evaluate our method on other datasets in a self-supervised manner.
Our main contributions are threefold:
\begin{itemize}
    \item Our ICF-SRSR is a self-supervised framework for the SISR task that performs simultaneous SR and down-sampling based on the proposed ICF.
    \item Our ICF-SRSR can learn a feasible resizing function directly from real-world LR images.
    Our self-supervised approach performs better on real-world SR than existing methods trained on synthetic datasets, even with training on a single image, as evident in \cref{fig:zero-shot}.
    \item Our ICF-SRSR can also down-sample given natural images, which enables us to construct realistic training pairs.
    Therefore, we can train off-the-shelf SR methods using the generated pairs by our ICF-SRSR in the absence of real paired training samples.
\end{itemize}

%% file: sections/related_work.tex
\section{Related Works}
\label{gen_inst}
%
%%%%%%%%%%%%%%%%%%%%%%%%%%%%%%
\begin{figure*}
        \captionsetup[subfigure]{aboveskip=1pt}
     \centering
     \begin{subfigure}[b]{0.29\textwidth}
         \centering
         \includegraphics[width=\textwidth]{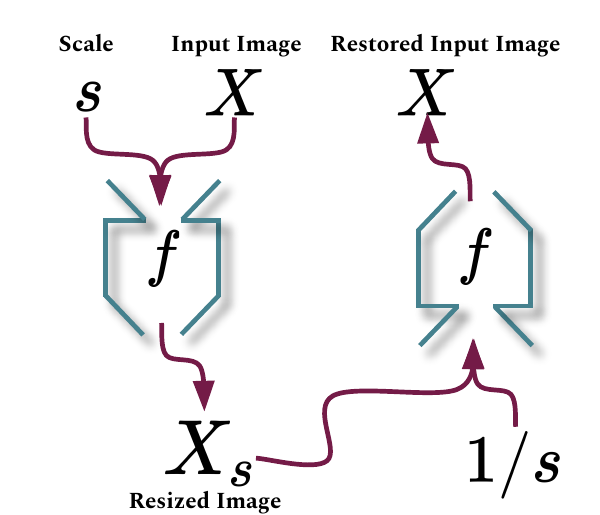}
         \caption{ICF}
         \label{fig:framework_a}
     \end{subfigure}
     \hfill
     \begin{subfigure}[b]{0.59\textwidth}
         \centering
         \includegraphics[width=\textwidth]{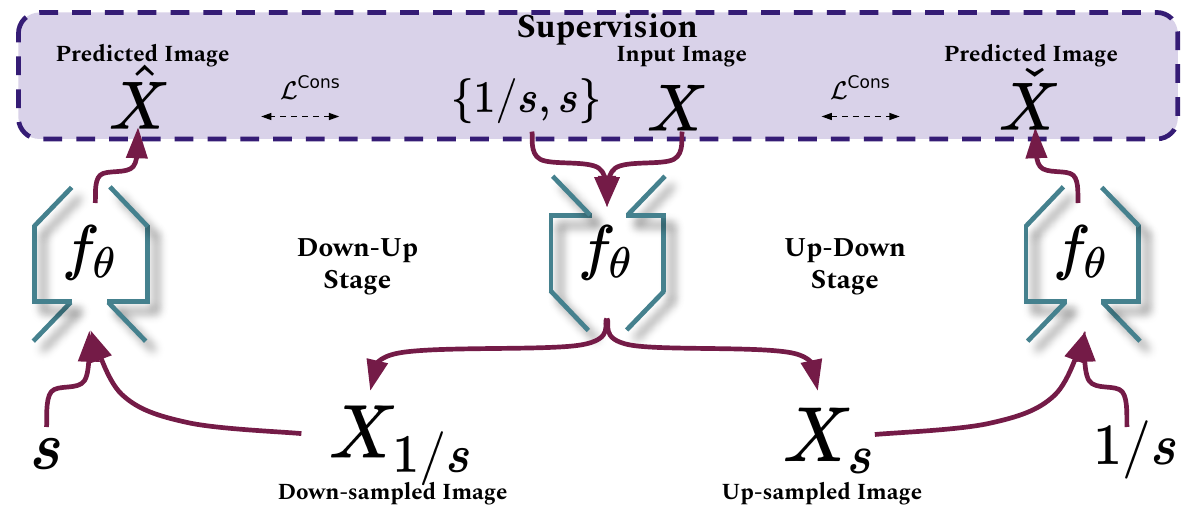}
         \caption{ICF-SRSR}
         \label{fig:framework_b}
     \end{subfigure}
     %\vspace{-2mm}
        \caption{\textbf{Overview of our proposed method.} (a) We introduce an invertible scale-conditional function~(ICF), which receives an input image and an arbitrary scale condition and generates a resized image. 
        It outputs the same input image for the resized image and the inverse scale condition. 
        (b) We propose a self-supervised SISR framework ICF-SRSR, in which a learnable ICF up-samples and down-samples a given image with different scale conditions and can reproduce the same input from the generated images by the inverse scales using the defined loss functions between the predicted images and the original input.}
        \label{fig:framework}
        \vspace{-4mm}
\end{figure*}
%%%%%%%%%%%%%%%%%%%%%%%%%%%%%%%%%%%%%%%%%%%%%%%%%%%%%%%%%
%
In this section, we review recent SR methods from the perspective of training supervision.
\subsection{Supervised image super-resolution}
Starting from Dong~\etal~\cite{dong2014learning}, CNNs~\cite{dong2016accelerating, shi2016realtime} have become a standard for SISR.
Following VDSR~\cite{kim2016accurate}, several methods such as LapSRN~\cite{lai2017deep}, EDSR~\cite{lim2017enhanced}, and SRGAN~\cite{ledig2017photo} have leveraged benefits of residual learning.
Advanced approaches utilize dense connections~\cite{wang2018esrgan, zhang2018residual}, channel attention~\cite{ zhang2018image, dai2019second, niu2020single}, and back-projection~\cite{haris2018deep, haris2019recurrent}, and even Transformers~\cite{dosovitskiy2020image, Mei_2021_CVPR, chen2021pre, wang2021uformer, zamir2021restormer, liang2021swinir} for high-performance SR architectures.
Furthermore, recent attempts extend the task toward continuous scaling factors~\cite{sr_meta, sr_arb, chen2021learning, inr_siren} and even arbitrary shapes~\cite{sr_warp}.

Nevertheless, supervised methods are still vulnerable when a given LR image is degraded by an unknown down-sampling function~\cite{son2021toward} that is not seen during training.
Therefore, several methods~\cite{gu2019blind, sr_blindsr, huang2020unfolding} jointly estimate latent kernel parameters and SR images to alleviate the issue.
Rather than up-sampling LR images directly, Correction filter~\cite{hussein2020correction} first converts a given input to resemble a bicubic down-sampled image and applies off-the-shelf SR methods.
Still, they require supervision from synthetic LR-HR pairs for training, which prevents their real-world applications.

\subsection{Unsupervised super-resolution}
To reduce biases from synthetic training data, zero-shot methods are trained on a given LR input only, without relying on supervision from large-scale data.
Ulyanov~\etal~\cite{ulyanov2018deep} has shown that the structure of CNNs can be prior for natural image representation which can be utilized for the SR task.
Based on internal patch recurrence~\cite{michaeli2013nonparametric}, ZSSR~\cite{shocher2018zero} is trained on numerous sub-patches of the given image to construct an input-specific SR model.
Later, there has been an attempt to integrate external and internal learning using model-agnostic meta-learning~\cite{finn2017model}.
MZSR~\cite{soh2020meta} is firstly trained on a large-scale paired dataset with multiple degradation parameters and then adopted to a given image during the inference time.

However, the zero-shot methods assume that the degradation pipeline for a given image is known, which is less practical.
To implement fully-blind SR methods, internal patch recurrence properties have played a critical role~\cite{michaeli2013nonparametric}.
Based on such a background, KernelGAN~\cite{bell2019blind} predicts a kernel that matches the distribution of the down-sampled image and the original input in an unsupervised manner.
The estimated kernel can also be utilized for several SR models~\cite{shocher2018zero, sr_srmd} for more accurate reconstruction.
Rather than explicitly utilize the concept of image distribution, we construct self-supervised chains to learn the SR model without assuming a specific degradation model.

\subsection{Cyclic architectures for super-resolution}
On the other hand, a class of methods interprets SR as a domain transfer problem between LR and HR distributions.
They introduce cyclic architectures~\cite{isola2017image} with adversarial loss~\cite{goodfellow2014generative, radford2015unsupervised, zhu2017unpaired} to train consecutive down-sampling and SR networks.
CinCGAN~\cite{yuan2018unsupervised} utilizes the concept of cycle consistency to train the model on unpaired LR-HR images.
Under the cyclic framework~\cite{sr_deg, sr_gandeg, sr_unpaired_pesudo, sr_unsupervised}, down-sampling models are trained to simulate the distribution of training LR images.
Then, the following SR network can learn to generalize on given LR images even if the corresponding HR pairs do not exist.
However, they are still biased toward handcrafted down-sampling functions~\cite{son2021toward} and lack generalization.
Without using adversarial loss, Guo~\etal~\cite{guo2020closed} combine paired and unpaired data to train a dual regression network with a loop.
In this paper, we further propose a self-supervised approach without requiring either paired/unpaired training data or a specific down-sampling operator.

\subsection{Real-world super-resolution}
To overcome the limitations of existing methods when handling real-world data, several approaches have captured paired LR-HR images in the wild.
While they are still limited due to scene diversity~\cite{chen2019camera}, accurate alignment~\cite{cai2019toward, wei2020component}, real-world datasets help generalization of existing SR models with more practical training data.
Zhang~\etal~\cite{sr_zllz} and Xu~\etal~\cite{sr_real_raw} leverage RAW and RGB images together to deliver better reconstruction quality.
Nevertheless, those pairs require careful alignment and complicated hardware setup, which are not scalable.
Recently, Real-ESRGAN~\cite{wang2021realesrgan} and BSRGAN~\cite{zhang2021designing} aim to synthesize more realistic and diverse LR images to improve the generalization ability of existing SR models. 
Still, they cannot leverage information from real-world images and heavily depend on such a synthesis process.
On the other hand, our fully self-supervised framework does not require synthetic or real-world pairs and can be trained on arbitrary LR images.

%% file: sections/method.tex
\section{Method}\label{sec:method}
We first introduce an Invertible scale-Conditional Function~(ICF) to design our self-supervised real-world single image super-resolution framework~(ICF-SRSR); then, we discuss our defined loss functions and the network architecture.
For convenience, we denote $X\in\mathbb{R}^{H\times W\times 3}$ as the input LR image with the arbitrary size of $H$ and $W$.
%%%%%%%%%%%%%%%%%%%%%%%%%%%%%%%%%%%%%%%%%%%%%%%%%%%%%
\subsection{Invertible scale-Conditional Function}
For a given input $X$, a conditional function $f(X|s)$ returns different outputs for different conditions $s$.
In this paper, we design an Invertible scale-Conditional Function~(ICF) as a specific conditional function, which can act as an operation and the inverse operation for different scale conditions. 
Without losing generality, we consider $f$ as an image-to-image mapping and $s$ as an arbitrary scaling factor, respectively.
Then, we can resize an arbitrary image $X$ as follows:
%%%%%%%%%%%%%%%%%%%%%%%%%%%%%%%%%%
\begin{equation}
    X_s = f \left( X| s \right),
\end{equation}
%%%%%%%%%%%%%%%%%%%%%%%%%%%%%%%%%%%
where $X_s \in \mathbb{R}^{sH \times sW \times 3}$ is a resized image.
Furthermore, for the same function $f$, we can get the original input $X$ again by the inverse scaling factor $\nicefrac{1}{s}$ as follows:
%%%%%%%%%%%%%%%%%%%%%%%%%%%%%%%%%%%
\begin{equation}
    X = f \left( X_s|\nicefrac{1}{s} \right).
\end{equation}
%%%%%%%%%%%%%%%%%%%%%%%%%%%%%%%%%%%
Therefore, $f$ as an ICF can project an image to its arbitrary-scale representation and back-project it to the original input for the scale conditions $s$ and $\nicefrac{1}{s}$, respectively.
\cref{fig:framework_a} illustrates the concept of our ICF.
We note that if $s=\nicefrac{1}{s}=1$ the function is identity which implies $f(X|1)=X$.

%%%%%%%%%%%%%%%%%%%%%%%%%%%%%%%%%%%%%%%%%%%%%%%%%%%%%

\subsection{Self-supervised SISR using ICF}
One of the challenges in real-world SR is that we cannot acquire the ground-truth HR image for an arbitrary LR image.
To overcome this limitation, we develop a novel self-supervised SR framework, ICF-SRSR, based on the concept of ICF.
As shown in \cref{fig:framework_b}, our method can simultaneously super-resolve and down-sample the given LR image $X$ with different scale conditions $s$ and $\nicefrac{1}{s}$, without requiring any paired/unpaired LR-HR training samples.
Specifically, we first parameterize an ICF $f_\theta$ with CNNs and utilize its property to optimize the model.
Then, we repeatedly apply $f_\theta$ to an LR image $X$ with different scale conditions to acquire two outputs $\check{X}, \hat{X} \in \mathbb{R}^{H \times W \times 3}$ as follows:
%
%%%%%%%%%%%%%%%%%%%%%%%%%%%%%%%%%%%%%%%%%%%%%%%%%%%%
%
\begin{equation}
    \begin{split}
        f_{\theta}(f_{\theta}(X|s)|\nicefrac{1}{s}) &= f_{\theta}(X_s|\nicefrac{1}{s})=\check{X}, \\
        f_{\theta}(f_{\theta}(X|\nicefrac{1}{s})|s) &= f_{\theta}(X_{\nicefrac{1}{s}}|s)=\hat{X},
    \end{split}
    \label{eq:icf}
\end{equation}
%%%%%%%%%%%%%%%%%%%%%%%%%%%%%%%%%%%%%%%%%%%%%%%%%%%%
%
where for $s>1$, $X_{s}\in \mathbb{R}^{sH \times sW \times 3}$ and $X_{\nicefrac{1}{s}} \in \mathbb{R}^{\nicefrac{H}{s} \times \nicefrac{W}{s} \times 3}$ are generated super-resolution~(SR) and low-low-resolution~(LLR) images, respectively.
For simplicity, we assume that both $\nicefrac{H}{s}$ and $\nicefrac{W}{s}$ are integers.

For an ideal ICF $f_\theta$, both $\check{X}$ and $\hat{X}$ in \cref{eq:icf} should be the same as the original LR image $X$. 
Therefore, we train $f_\theta$ in a self-supervised manner by reducing the distance between $X$ and the generated images $\check{X}$ and $\hat{X}$ in two stages simultaneously, as shown in \cref{fig:framework_b}.
In the up-down stage, we minimize the distance between $\check{X}$ and $X$.
By doing so, the network can learn to down-sample the generated SR image $X_s$ by restoring the output $\check{X}$ as the approximation of the original input $X$.
On the other hand, in the down-up stage, we aim to approximate the original input $X$ by reducing the distance between $\hat{X}$ and $X$.
Then, the network can learn to up-sample the generated LLR image $X_{\nicefrac{1}{s}}$.
Therefore, by leveraging the learned up-sampler and down-sampler applied on the generated images $X_{\nicefrac{1}{s}}$ and $X_s$, respectively, we can generate favorable SR and LLR images $X_s$ and $X_{\nicefrac{1}{s}}$ by employing the learned model $f_\theta$ on the input $X$ with the scale conditions $s$ and $\nicefrac{1}{s}$, respectively.

\input{tables/benchmark}

We also note that our method is different from CycleGAN~\cite{zhu2017unpaired}, which utilizes unpaired LR-HR images and performs two independent cycles, one on the LR and the other on the HR images.
Rather, our model is trained in a self-supervised manner by optimizing the $f_{\theta}$ jointly with two stages on LR images only, without requiring the adversarial loss.
In other words, $f_{\theta}$ can perform simultaneous up-sampling and down-sampling without requiring prior information or paired/unpaired data.
\input{sections/figures/b100}

%%%%%%%%%%%%%%%%%%%%%%%%%%%%%%%%%%%%%%%%%%%%%%%%%%%%%%%
\subsection{Training loss functions}
\label{sec:loss}
To train the proposed ICF $f_{\theta}$, we design a set of self-supervised loss functions.
First, we formulate the consistency loss $\mathcal{L}^{\text{Cons}}$, which preserves information during the simultaneous up-down and down-up stages.
The proposed consistency loss $\mathcal{L}^\text{Cons}$ on the approximated LR images $\hat{X}$ and $\check{X}$, and the original input $X$ is defined as follows:
%
%%%%%%%%%%%%%%%%%%%%%%%%%%%%%%%%%%%%%%%%%%%%%%%%%%%%
\begin{equation}\label{eq:loss_cons}
\begin{split}
\mathcal{L}^{\text{Cons}} &=\lVert \hat{X}-X \rVert + \lVert \check{X}-X \rVert.
\end{split}
\end{equation}
%%%%%%%%%%%%%%%%%%%%%%%%%%%%%%%%%%%%%%%%%%%%%%%%%%%%
%
For simplicity, we use $\lVert\cdot\rVert$ to represent the L1 norm.
The proposed consistency term $\mathcal{L}^{\text{Cons}}$ guarantees to generate reliable up-sampled and down-sampled images simultaneously.
%
%%%%%%%%%%%%%%%%%%%%%%%%%%%%%%%%%%%%%%%%%%%%%%%%%%%%
Furthermore, to stabilize the training and preserve colors between the input and intermediate images $X_s$ and $X_{\nicefrac{1}{s}}$, we utilize the low-frequency loss~\cite{son2021toward}.
We implement the low-pass filter with a spatial pooling operator $\mathbf{P} \left( \cdot, w, s \right)$, where $w$ and $s$ are window size and stride, respectively.
Our color-preserving loss $\mathcal{L}^\text{Color}$ is defined as follows:
%
%%%%%%%%%%%%%%%%%%%%%%%%%%%%%%%%%%%%%%%%%%%%%%%%%%%%
\begin{equation}\label{eq:loss_pool}
%\begin{aligned}
\begin{split}
\mathcal{L}^{\text{Color}} &= 
\lVert \mathbf{P} \left( X_s, 4s, 4s \right) - \mathbf{P} \left( X, 4, 4 \right) \rVert
\\ &+ \lVert \mathbf{P} \left( X_{\nicefrac{1}{s}}, 4, 4 \right) - \mathbf{P} \left( X, 4s, 4s \right) \rVert,
\end{split}
\end{equation}
%%%%%%%%%%%%%%%%%%%%%%%%%%%%%%%%%%%%%%%%%%%%%%%%%%%%
where the window size and stride are adjusted to match dimensions between each of $\left( X_s, X \right)$ and $\left( X_{\nicefrac{1}{s}}, X \right)$.
The total training objective $\mathcal{L}^{\text{Total}}$ is the combination of the aforementioned two loss terms, which is defined as follows: 
%%%%%%%%%%%%%%%%%%%%%%%%%%%%%%%%%%%%%%%%%%%%%%%%%%%%
\begin{equation}\label{eq:loss_total}
\mathcal{L}^{\text{Total}} = \mathcal{L}^{\text{Cons}}+\lambda_{\text{Color}}\mathcal{L}^{\text{Color}}.
\end{equation}
%%%%%%%%%%%%%%%%%%%%%%%%%%%%%%%%%%%%%%%%%%%%%%%%%%%%
 
\subsection{Network architecture}\label{sec:net_arch}
Our ICF-SRSR architecture leverages a single model to handle different scale conditions.
To implement the proposed method, we modify the existing SISR model, \eg, EDSR~\cite{lim2017enhanced} as our baseline backbone architecture.
Since the body part is invariant to the scale image (\ie, the input and output have the same resolution), we introduce multiple tail parts for different scale conditions.
Employing a single network with the shared body part is more efficient and can improve performance by observing more augmented data, \ie, images with different scales, during the training.
In the supplementary material, we provide the details of the network architecture and illustrate that our method is model-agnostic and can leverage different SOTA baseline models. 
We will also publish our ICF-SRSR implementation.

%% file: tables/benchmark.tex
\begin{table*}[t]
    \small
    \centering
    \begin{tabularx}{\linewidth}{c l >{\centering\arraybackslash}X >{\centering\arraybackslash}X >{\centering\arraybackslash}X >{\centering\arraybackslash}X
    >{\centering\arraybackslash}X >{\centering\arraybackslash}X
    }
    \toprule
    \multirow{2}{*}{\textbf{Supervision}} & 
    \multirow{2}{*}{\bf Method} & 
    \textbf{Set5} & 
    \textbf{Set14} &
    \textbf{BSD100} &
    \textbf{Urban100} &
    \textbf{Manga109} &
    \textbf{DIV2K}\\
    %\cline{4-7}
    & & $\times2$/$\times4$ & $\times2$/$\times4$ & $\times2$/$\times4$ & $\times2$/$\times4$ & $\times2$/$\times4$  & $\times2$/$\times4$ \\
    %\cline{4-7}
    
    \midrule
    %BM3D~\cite{sparse} & {25.65} & {0.685}& {34.51} & {0.850} & {25.65} & {0.685}& {34.51} & {0.850} \\
    %WNNM~\cite{6909762}& {25.78} & {0.809} & {34.67} & {0.864} & {25.65} & {0.685}& {34.51} & {0.850} \\
    %K-SVD~\cite{DBLP:journals/corr/abs-1909-13164} & {26.88} & {0.842}& \textbf{36.49} & \textbf{0.897}& {25.65} & {0.685}& {34.51} & {0.850} \\
    %EPLL~\cite{Hurault_2018} & \textbf{27.11} & \textbf{0.870}  & {33.51} & {0.824}& {25.65} & {0.685}& {34.51} & {0.850} \\
    %\hline
    & {\footnotesize Bicubic} & {33.66/28.42} & {30.24/26.00} & {29.56/25.96} & {26.88/23.14} & {30.80/24.89} & {31.01/26.66} \\
    \midrule
    \multirow{8}{*}{{\footnotesize Supervised}}
    & {\footnotesize VDSR~\cite{kim2016accurate}} & {37.53/31.35} & {33.03/28.01} & {31.90/27.29} & {30.76/25.18} & {37.22/28.83} & {33.66/28.17} \\
    & {\footnotesize EDSR~\cite{lim2017enhanced}} & {38.11/32.46} & {33.92/28.80} & {32.32/27.71} & {32.93/26.64} & {39.10/31.02} & \textbf{36.22}/{30.52} \\
    & {\footnotesize CARN~\cite{ahn2018fast}} & {37.76/32.13} & {33.52/28.60} & {32.09/27.58} & {31.92/26.07} & {38.36/30.47} & \hspace{9pt}{-\hspace{8pt}/30.10}  \\
    & {\footnotesize RCAN~\cite{zhang2018image}} & {38.27/32.63} & {34.12/28.87} & {32.41/27.77} & {33.34/26.82} & {39.44/31.19} & {36.13}/{30.52} \\
    & {\footnotesize RDN~\cite{zhang2018residual}} & {38.24/32.47} & {34.01/28.81} & {32.34/27.72} & {32.89/26.61} & {39.18/31.00} & {-\hspace{8pt}/\hspace{8pt}-} \\
    & {\footnotesize DRN-S~\cite{guo2020closed}} & {37.80/32.68} & {33.30/28.93} & {31.97/27.78} & {31.40/26.84} & {38.11/31.52} & {35.77}/\textbf{30.79} \\
    & {\footnotesize LIIF~\cite{chen2021learning}} & {38.17/32.50} & {33.97/28.80} & {32.32/27.74} & {32.87/26.68} & {-\hspace{8pt}/\hspace{8pt}-} & {34.99/29.27} \\
    & {\footnotesize ELAN~\cite{ELAN-light}} & \textbf{38.36}/\textbf{32.75} & \textbf{34.20}/\textbf{28.96} & \textbf{32.45}/\textbf{27.83} & \textbf{33.44}/\textbf{27.13} & \textbf{39.62}/\textbf{31.68} & {-\hspace{8pt}/\hspace{8pt}-} \\
    %&  \textbf{IMF-SRSR$^{\dagger}$} & {} & {-} & {-} & {-} & {-} & {-} \\
    %&  \textbf{IMF-SRSR$^{\ddagger}$} & {-} & {-} & {-} & {-} & {-} &  {-} \\
    \midrule
    \multirow{4}{*}{\footnotesize Unsupervised} 
    & {\footnotesize SelfExSR~\cite{huang2015single}}  & {36.49/30.31} & {32.22/27.40} & {31.18/26.84} & {29.54/24.82} & {35.78/27.82} & {-\hspace{8pt}/\hspace{8pt}-} \\
    & {\footnotesize ZSSR~\cite{shocher2018zero}}  & {37.37/31.13} & {33.00/28.01} & {31.65/27.12} & {29.34/24.12} & {35.57/27.04} & \textbf{34.45}/\textbf{29.08} \\
    &  {\footnotesize MZSR~\cite{soh2020meta}} &  {37.25/31.59} & {33.16/27.90} & \hspace{-9pt}{31.64/\hspace{8pt}-} & {30.41/25.52} & \textbf{36.70}/\textbf{29.58} & {-\hspace{8pt}/\hspace{8pt}-} \\
    &  {\footnotesize DASR~\cite{wang2021unsupervised}} &  \textbf{37.87}/\textbf{31.99} & \textbf{33.34}/\textbf{28.50} & {\textbf{32.03/27.52}} & \textbf{31.49}/\textbf{25.82} & {-\hspace{8pt}/\hspace{8pt}-} & {-\hspace{8pt}/\hspace{8pt}-} \\
    \midrule
    \multirow{2}{*}{\footnotesize Self-supervised}
    %&  \textbf{IMF-SRSR~(Test)} & {36.41/29.49} & {32.44/27.19} & {31.34/26.82} & {30.26/24.66} & {36.29/27.82} & {35.02/29.45} \\
    &  {\footnotesize \textbf{ICF-SRSR}~(Ours)} & {37.01/30.81} & {32.86/27.76} & {31.54/26.99} & {30.39/24.72} & {36.45/28.01} & {35.19/29.48} \\
    &  {\footnotesize \textbf{EDSR~(LLR,LR)}~(Ours)} & {\textbf{37.09/31.06}} & {\textbf{32.91/27.97}} & {\textbf{31.63/27.10}} & {\textbf{30.51/24.92}} & {\textbf{36.68/28.29}} & {\textbf{35.26/29.64}} \\ 
    %\multirow{3}{*}{(fully self-supervised)}
    \bottomrule
    \end{tabularx}
    \vspace{-2mm}
    \caption{
        \textbf{Quantitative comparisons on synthetic datasets.} 
        We compare ICF-SRSR with several supervised/unsupervised methods on the benchmarks~\cite{bevilacqua2012low, zeyde2010single, martin2001database, huang2015single, Manga109} and DIV2K~\cite{agustsson2017ntire} validation set for scales $\times 2$ and $\times 4$ with PSNR metric. 
        ICF-SRSR refers to our self-supervised method, while EDSR~(LLR,LR) is the model EDSR trained on our generated pairs (LLR,LR) of the DIV2K.
    }
    \label{tab:benchmark}
    \vspace{-2mm}
\end{table*}

%% file: sections/figures/b100.tex
\begin{figure*}[t]
    \centering
	\captionsetup[subfloat]{labelformat=empty,aboveskip=1pt}
	%\begin{center}		\newcommand{\rowArg}{1.5cm}
		\newcommand{\patchSize}{2.5cm}
		% 			\begin{adjustbox}{width=\linewidth, center=\linewidth}
		\setlength\tabcolsep{0.0cm}
  \scalebox{0.65}{
		\begin{tabular}[b]{c}
		%	\multicolumn{2}{c}{\multirow{2}{*}[\rowArg]{
		%			\subfloat[Input~(LR)]
		%			% 			{\includegraphics[width =    \fullSize, height = \fullSize]
		%			{\includegraphics[trim={9cm 0 0 0},clip, height=\fullSize]
		%				{figures/input_900.png}}}} & \hspace{4mm}
            \subfloat[\centering  LR ]{
				\includegraphics[page=1, width = \patchSize]
				{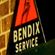}} 
				%\vspace{0mm}
            \hfill
			\subfloat[\centering  Bicubic ]{
				\includegraphics[page=1, width = \patchSize]
				{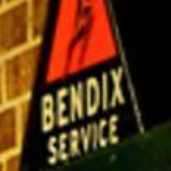}} 
				%\vspace{0mm}
            \hfill
			\subfloat[\centering  EDSR~\cite{lim2017enhanced}  ]{
				\includegraphics[page=2, width = \patchSize]
				{figures/img900.pdf}}
            \hfill
            \subfloat[\centering   DRN-S~\cite{guo2020closed}]{
				\includegraphics[page=1, width = \patchSize]
				{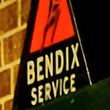}}
            \hfill
			\subfloat[\centering  LIIF~\cite{chen2021learning}  ]{
				\includegraphics[page=3, width = \patchSize]
				{figures/img900.pdf}} 
            \hspace{5mm}
            \subfloat[\centering  LR ]{
				\includegraphics[page=1, width = \patchSize]
				{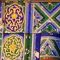}}
            \hfill
			\subfloat[\centering  Bicubic ]{
				\includegraphics[page=1, width = \patchSize]
				{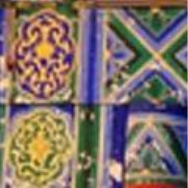}} 
				%\vspace{0mm}
            \hfill
			\subfloat[\centering  EDSR~\cite{lim2017enhanced}  ]{
				\includegraphics[page=2, width = \patchSize]
				{figures/img_826_3.pdf}} 
            \hfill
            \subfloat[\centering  DRN-S~\cite{guo2020closed} ]{
				\includegraphics[page=1, width = \patchSize]
				{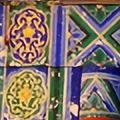}}
            \hfill
			\subfloat[\centering  LIIF~\cite{chen2021learning}  ]{
				\includegraphics[page=3, width = \patchSize]
				{figures/img_826_3.pdf}} \\   %\hspace{4mm}

            \subfloat[\centering  DASR~\cite{wang2021unsupervised} ]{
				\includegraphics[page=6, width = \patchSize]
				{figures/img900.pdf}} 	
            \hfill
			\subfloat[\centering   MZSR~\cite{soh2020meta}  ]{
				\includegraphics[page=5, width = \patchSize]
				{figures/img900.pdf}}
            \hfill
            \subfloat[\centering  ZSSR~\cite{shocher2018zero}  ]{
				\includegraphics[page=4, width = \patchSize]
				{figures/img900.pdf}}
				%\vspace{2mm}
			\hfill
			\subfloat[\centering  \textbf{ICF-SRSR}~(Ours) ]{
				\includegraphics[page=7, width = \patchSize]
				{figures/img900.pdf}}
            \hfill
			\subfloat[\centering  GT~(HR)]{
				\includegraphics[page=8, width = \patchSize]
				{figures/img900.pdf}}
            \hspace{5mm}
            \subfloat[\centering  DASR~\cite{wang2021unsupervised} ]{
				\includegraphics[page=6, width = \patchSize]
				{figures/img_826_3.pdf}}
            \hfill
			\subfloat[\centering   MZSR~\cite{soh2020meta}  ]{
				\includegraphics[page=5, width = \patchSize]
				{figures/img_826_3.pdf}}
            \hfill
            \subfloat[\centering  ZSSR~\cite{shocher2018zero}  ]{
				\includegraphics[page=4, width = \patchSize]
				{figures/img_826_3.pdf}}
				%\vspace{2mm}
			\hfill
			\subfloat[\centering \textbf{ICF-SRSR}~(Ours)]{
				\includegraphics[page=7, width = \patchSize]
				{figures/img_826_3.pdf}}
            \hfill
			\subfloat[\centering  GT~(HR) ]{
				\includegraphics[page=8, width = \patchSize]
				{figures/img_826_3.pdf}}            
	\end{tabular}}
    \vspace{-2mm}
	\caption{\textbf{Qualitative comparisons on a synthetic dataset.} 
	We compare our ICF-SRSR method with bicubic up-scaling, supervised methods EDSR~\cite{lim2017enhanced}, DRN-S~\cite{guo2020closed}, and LIIF~\cite{chen2021learning} and also unsupervised methods DASR~\cite{wang2021unsupervised}, MZSR~\cite{soh2020meta}, and ZSSR~\cite{shocher2018zero} trained on the DIV2K~\cite{agustsson2017ntire} training set and evaluated on the DIV2K validation set for scale $\times2$.
	}
	\label{fig:benchmark}
	\vspace{-4mm}
\end{figure*}

%% file: sections/expriment.tex
\section{Experiments}\label{sec:experiment}
We first introduce training and evaluation configurations of the proposed ICF-SRSR framework.
Then we conduct comprehensive experiments, extensive quantitative and qualitative comparisons with the other methods, and an in-depth analysis of our proposed method.

\subsection{Training details}
\Paragraph{Dataset.}
We train and evaluate our method on two scenarios. 1) Synthetic datasets, where the training and testing LR images are synthesized by a uniform degradation process~(\eg, bicubic down-sampling) from HR images. 2) Real-world datasets, which provide paired LR-HR images from the real-world captured by adjusting the focal length of a camera.

To train our ICF-SRSR, we use $800$ bicubic LR images from the DIV2K~\cite{agustsson2017ntire} dataset.
For evaluation, we adopt five standard benchmarks: Set5~\cite{bevilacqua2012low}, Set14~\cite{zeyde2010single}, BSD100~\cite{martin2001database}, Urban100~\cite{huang2015single}, and Manga109~\cite{Manga109}.
We also use the high-quality DIV2K validation set for evaluation.

To train and evaluate our ICF-SRSR under real-world scenarios, we utilize real-world datasets~\cite{cai2019toward, wei2020component} for the SISR task.
RealSR-V3~\cite{cai2019toward} includes paired LR-HR images captured by two different cameras, Canon and Nikon.
For each camera, about $200$ training images are captured from different scenes for each scaling factor $\times2$, $\times3$, and $\times4$.
We use only the LR images with scaling factors $\times2$ and $\times4$ for training and evaluate our method on the $50$ test pairs for each scale. 
DRealSR~\cite{wei2020component} also contains images captured by five DSLR cameras.
We conduct our experiments using images for $\times 2$ and $\times 4$ SR, containing 884 and 840 LR images, respectively.
For evaluation, we use $83$ and $93$ test pairs in DRealSR for $\times 2$ and $\times 4$, respectively.
\input{tables/real_data}

\Paragraph{Hyperparameters.}
During the training, we extract random patches of size $48\times48$ from LR images of both synthetic and real-world datasets.
For all our experiments, we set the batch size to $16$, and  $\lambda_{\text{Color}}=0.2$.
Random flip and rotation augmentations are applied to the input images to increase the number of effective training samples. 
We train our model using ADAM~\cite{kingma2017adam} optimizer with the initial learning rate $1 \times 10^{-4}$, which decays by a factor $0.5$ after every $200$ epochs.
For quantitative comparisons, we adopt structural similarity~(SSIM)~\cite{measure_ssim} and peak signal-to-noise ratio~(PSNR) on the luminance channel for the experiments on synthetic datasets and real-world dataset DRealSR~\cite{wei2020component} and also on RGB channels for dataset RealSR-V3~\cite{cai2019toward}.
All experiments are done using PyTorch 1.8.1 and Quadro RTX 8000 GPUs.
%%%%%%%%%%%%%%%%%%%%%%%%%%%%%%%%%%%%%%%%%%%%%%%%%%%%%%%%%%%%%%%%%%%%%%%%%%%%%%%%%%%%%%%%
\subsection{Evaluation on synthetic datasets}\label{sec:ex_syn}
We train our ICF-SRSR on the DIV2K~\cite{agustsson2017ntire} dataset with EDSR-baseline~\cite{lim2017enhanced} and test it on the public benchmark datasets~\cite{bevilacqua2012low, zeyde2010single, martin2001database, huang2015single, Manga109} and also the validation set of DIV2K.
We note that the proposed method is trained in a self-supervised manner by targeting a certain scale $s$.
Specifically, we train $\left( \times 2, \times \nicefrac{1}{2} \right)$ ICF and $\left( \times 4, \times \nicefrac{1}{4} \right)$ ICF independently.
\cref{tab:benchmark} shows extensive comparisons between the proposed self-supervised approach and the other representative supervised/unsupervised SR methods with PSNR metric.
We demonstrate that our ICF-SRSR approach achieves superior performance compared to the SelfExSR~\cite{huang2015single} model and comparable performance to the other unsupervised and supervised methods.
We note that ground-truth HR images in Set5 and Set14 are relatively noisier than the other datasets, preventing our self-supervised framework from learning accurate scaling functions.
We will discuss more details about the noisy cases in our supplementary material.
Notably, ICF-SRSR outperforms the unsupervised method ZSSR~\cite{shocher2018zero} by $1.05$dB on scale $\times2$ of Urban100 dataset and the supervised methods~\cite{kim2016accurate, chen2021learning} on both scales of DIV2K validation set.

Moreover, we apply the trained ICF-SRSR to LR images from the DIV2K training dataset and get LLR-LR paired images.
Then, we train off-the-shelf EDSR on the synthesized paired data from scratch and evaluate it on the test datasets as shown in \cref{tab:benchmark}.
The results demonstrate that EDSR~(LLR, LR) trained on our generated pairs~(LLR, LR) achieves superior performance than ICF-SRSR, which illustrates the merit of our method to generate useful training image pairs.

\cref{fig:benchmark} further visualizes the qualitative results of ICF-SRSR on two validation images from the DIV2K~\cite{agustsson2017ntire} dataset.
Our method achieves comparable results to the supervised methods~\cite{lim2017enhanced,chen2021learning} while restoring more details compared to the unsupervised methods~\cite{shocher2018zero,soh2020meta}. 
We note that the results on ZSSR~\cite{shocher2018zero} show lost information and scratched texts, and on MZSR~\cite{soh2020meta} include severe artifacts and color shifting.
For an in-depth comparison, we also provide quantitative results with SSIM metric in our supplementary material.

\subsection{Evaluation on real-world datasets}\label{sec:ex_real}
We train and evaluate ICF-SRSR for each scale $\times 2$ and $\times 4$ independently on the LR images of each Canon and Nikon camera from the real-world dataset RealSR-V3~\cite{cai2019toward} separately and also on the LR images of the real-world dataset DRealSR~\cite{wei2020component} in a self-supervised manner.
We further train the model EDSR~\cite{lim2017enhanced} on our generated~(LLR, LR) image pairs. 
We compare our method with the supervised methods~\cite{lim2017enhanced, wang2018esrgan, zhang2018image, cai2019toward, guo2020closed} trained on real paired images, which serve as the upper bounds for the SR problem.
\input{sections/figures/realsr2}

On the other hand, we employ the pre-trained supervised models EDSR~\cite{lim2017enhanced}, RRDB~\cite{wang2018esrgan}, IKC~\cite{gu2019blind}, BlindSR~\cite{sr_blindsr} and DRN-S~\cite{guo2020closed} on the synthetic DIV2K~\cite{agustsson2017ntire} dataset to super-resolve the LR images in the testing sets of RealSR-V3~\cite{cai2019toward} and DRealSR~\cite{wei2020component}.
Moreover, we utilize Kernel-GAN~\cite{bell2019blind} to approximate the down-sampling kernel from a single LR image and use ZSSR~\cite{shocher2018zero} as a zero-shot SR to apply to real LR images.
Our extensive comparisons with the various methods trained on real and synthetic datasets are summarized in \cref{tab:real_data}.
We illustrate that our self-supervised method can achieve superior performance compared to the methods pre-trained on the synthetic datasets and unsupervised method ZSSR~\cite{shocher2018zero}+Kernel-GAN~\cite{bell2019blind} in terms of both PSNR and SSIM metrics, which emphasizes the fact that the trained models on synthetic datasets with known degradations cannot perform well on real-world scenarios.
We qualitatively compare our method with the various existing methods on the RealSR-V3 dataset and visualize the SR results and their corresponding error maps with respect to the GT~(HR) in \cref{fig:realsr}. 
We demonstrate that our self-supervised method can achieve comparable and sometimes better performance to the supervised method LP-KPN~\cite{cai2019toward} trained on real paired images. 
We note that our method is generally more suitable for restoring the texture and preserving color compared to supervised method IKC~\cite{gu2019blind} and unsupervised method ZSSR~\cite{shocher2018zero}+Kernel-GAN~\cite{bell2019blind} as evident in appearance and PSNR, SSIM, and mean absolute error~(MAE) metrics.
We show more qualitative results in the supplementary material.
\subsection{Ablation study}\label{sec:ablation}
We conduct various ablation studies on the model design, down-sampling operators, few-shot learning, augmentation, and the effect of loss functions to better analyze our method.

\Paragraph{Model design.}
We conduct an experiment to show the superiority of a developed baseline as a single conditional model compared to two independent models and also the effect of training our two-stage framework compared to training each Up-Down and Down-Up stage separately.
Our results on synthetic dataset DIV2K~\cite{agustsson2017ntire} and Canon and Nikon images from real-world dataset RealSR-V3~\cite{cai2019toward} for scale $\times2$ show that training with two independent models or using only one stage~(half) results in unsatisfactory performance, demonstrating the uniqueness of our method in using a single invertible scale-conditional model as shown in \cref{tab:reb-two-models}.

\input{tables/reb_two-models.tex}

\Paragraph{Evaluation of down-sampling.} 
Due to the invertibility attribute of ICF, our method can be interpreted as a learnable down-sampler.
Therefore, we analyze our model $f_{\theta}$ as a down-sampling operator in three aspects.

\Paragraph{First.} We train ICF-SRSR on HR images from RealSR-V3~\cite{cai2019toward} and evaluate the model on HR images of the test dataset to gather the generated down-sampled images. 
Then, we compare ground-truth LR images with our generated LR images, as well as LR images obtained by down-sampling functions \eg, Nearest, Bicubic, Gaussian+Nearest, and Gaussian+bicubic~($\sigma=0.4$).
\cref{tab:downsampling} provides a comparison of LR images for different down-sampling models based on PSNR.
The values show the superiority of our learnable down-sampling method in generating more realistic LR images compared to ones with other down-sampling operators. 

\input{tables/downsampling.tex}

\Paragraph{Second.} 
We further analyze our learnable down-sampling operator $f_{\theta}$ compared to non-learnable down-sampling approaches.
We use our learnable down-sampling operator $f_{\theta}$, bicubic down-sampling, and Gaussian~($\sigma=0.4$) filtering followed by different nearest and bicubic down-sampling operators to generate the LLR images from given input LR images on the training sets.
Then, we train the model EDSR on the generated paired images~(LLR, LR) to learn generating SR images given LR counterparts.
We summarize the results for scale $\times2$ of the benchmarks Set5~\cite{bevilacqua2012low} and Set14~\cite{zeyde2010single}, and Canon and Nikon sets of RealSR-V3~\cite{cai2019toward} dataset for both non-learnable and our learnable down-sampling operators in \cref{tab:downsampling2}.
The results indicate the effect of our learnable down-sampling operator to generate appropriate image pairs for training, which results in a significant improvement compared to known down-sampling operators.

\input{tables/downsampling2.tex}

\Paragraph{Third.}
By using different down-sampling methods, we first generate LR samples from the real training HR images and then train a vanilla EDSR model using the generated pairs, \ie, (LR, HR).
As shown in \cref{tab:reb-down}, our synthesized pairs can provide more suitable training data compared to ones by previous learnable down-sampling methods ADL~\cite{son2021toward} and DRN-S~\cite{guo2020closed} as the EDSR performs much better for the $\times 2$ SR tasks on real dataset RealSR-V3~\cite{cai2019toward}.

\input{tables/reb_down.tex}

\paragraph{Few-shot learning.}
We train and evaluate our method on small datasets to show the advantage of our method to learning from only a few images without requiring a large-scale training dataset.
Therefore, we train the model ICF-SRSR~(Small) on the test sets of synthetic datasets Set14~\cite{zeyde2010single}, BSD100~\cite{martin2001database} and Urban100~\cite{huang2015single} and also real-world datasets RealSR-V3~\cite{cai2019toward} and DRealSR~\cite{wei2020component} and show their results on the corresponding test datasets in \cref{tab:ab-few-shot}. 
We demonstrate that our method can achieve slightly lower performance even when trained on very small datasets compared to our model ICF-SRSR~(Large) trained on large-scale training datasets.
\input{tables/few-shot.tex}

\Paragraph{Multi-scale augmentation.}
As we mention in \cref{sec:net_arch}, augmented data with different scales can lead to performance improvement.
Therefore, when we train ICF-SRSR directly on the test samples, we adopt diverse scaling factors as well as their reciprocals to compensate for the limited number of training data.
In \cref{tab:ab-scale}, we show that increasing the number of inputs induced by various scaling factors, \eg, $\times2$, $\times4$, and $\times8$, and their inverses can lead to obtaining superior performance on the RealSR-V3~\cite{cai2019toward} dataset.
More details about our multi-scale augmentation strategy are described in our supplementary material.

\input{tables/ab_scale}

\Paragraph{Effects of loss functions.}
\textcolor{blue}
We also analyze the effect of each loss function discussed in \cref{sec:loss}.
As shown in \cref{tab:ab-loss}, our novel self-supervised consistency loss $\mathcal{L}^{\text{Cons}}$ can drastically improve the model performance when it is added to color preserving loss $\mathcal{L}^{\text{Color}}$ on both synthetic and real-world datasets.
In our supplementary material, we further discuss the effect of the weight $\lambda_{\text{Color}}$.

\input{tables/ab_loss.tex}

%% file: tables/real_data.tex
\begin{table*}[t]
    \small
    \centering
    \begin{tabularx}{\linewidth}{c c l >{\centering\arraybackslash}X >{\centering\arraybackslash}X
    >{\centering\arraybackslash}X >{\centering\arraybackslash}X >{\centering\arraybackslash}X >{\centering\arraybackslash}X}
    \toprule
    \multirow{3}{*}{\textbf{Training Set}}& 
    \multirow{3}{*}{\textbf{Supervision}}&
    \multirow{3}{*}{\bf Method} & 
    %\multicolumn{3}{c}{\textbf{City100}}  & 
    %\multicolumn{3}{c}{\textbf{RealSR}} &
    \multicolumn{2}{c}{\textbf{RealSR~(Canon)}} & \multicolumn{2}{c}{\textbf{RealSR~(Nikon)}} &  \multicolumn{2}{c}{\textbf{DRealSR}} \\
    %\cline{4-7}
    & & &  \bf $\times2$  & \bf $\times4$ & \bf $\times2$ & \bf $\times4$  & \bf $\times2$ & \bf $\times4$\\ 
    & & &   {\footnotesize (PSNR/SSIM)} &  {\footnotesize (PSNR/SSIM)} & {\footnotesize (PSNR/SSIM)} & {\footnotesize (PSNR/SSIM)}  & {\footnotesize (PSNR/SSIM)} & {\footnotesize (PSNR/SSIM)}\\
    %\cline{4-7}
    \midrule
    & & {\footnotesize Bicubic} & {30.35}/{0.876}  &  {25.80}/{0.744} & {29.66}/{0.854} & {25.50}/{0.718} & {32.67}/{0.877} & {30.56}/{0.820}\\
    \midrule
    \multirow{5}{*}{\footnotesize Synthetic} & \multirow{5}{*}{\footnotesize Supervised} 
    %& Bicubic & {}  & {-}& {-} & {-} & {-}\\
    & {\footnotesize EDSR~\cite{lim2017enhanced}} & \textbf{30.58}/\textbf{0.880}  & {26.05}/{0.754} &  \textbf{30.00}/\textbf{0.861}  & {25.89}/{0.735}  & \textbf{32.82}/{0.869} & \textbf{30.64}/\textbf{0.821}\\
    & & {\footnotesize RRDB~\cite{wang2018esrgan}} & \hspace{6pt}{-\hspace{8pt}}/{\hspace{8pt}-} & \hspace{-2pt}{26.05}/{\hspace{8pt}-} &  \hspace{6pt}{-\hspace{8pt}}/{\hspace{8pt}-}  & \hspace{-2pt}{25.91}/{\hspace{8pt}-} & \hspace{6pt}{-\hspace{8pt}}/{\hspace{8pt}-}& \hspace{-2pt}{30.55}/{\hspace{8pt}-}\\
    &  & {\footnotesize IKC~\cite{gu2019blind}} & \hspace{6pt}{-\hspace{8pt}}/{\hspace{8pt}-} & {25.71}/{0.751} &  \hspace{6pt}{-\hspace{8pt}}/{\hspace{8pt}-}  & {25.27}/\textbf{0.740}& \hspace{6pt}{-\hspace{8pt}}/{\hspace{8pt}-}& \hspace{6pt}{-\hspace{8pt}}/{\hspace{8pt}-}\\
    & & {\footnotesize BilndSR~\cite{sr_blindsr}} & {27.99}/{0.822}   & \hspace{6pt}{-\hspace{8pt}}/{\hspace{8pt}-} &  {26.68}/{0.794} & \hspace{6pt}{-\hspace{8pt}}/{\hspace{8pt}-}& \hspace{6pt}{-\hspace{8pt}}/{\hspace{8pt}-}& \hspace{6pt}{-\hspace{8pt}}/{\hspace{8pt}-}\\
    & & {\footnotesize DRN-S~\cite{guo2020closed}} & {30.57}/{0.879} & \textbf{26.07}/\textbf{0.755} &  {29.99}/{0.860}  & \textbf{25.92}/{0.736} & {32.81}/\textbf{0.879}& {30.63}/\textbf{0.821}\\
    %&  & \textbf{IMF-SRSR$^{\dagger}$} & {30.62}  &  {-} & {30.00} & {-}& {32.84}& {-}\\
    %&  & \textbf{IMF-SRSR$^{\ddagger}$} & {}  &  {-} & {} & {-}& {-}& {-}\\
    %& \textbf{IMF-SRSR~(EDSR~(HR))} & {32.56}  &  {} & {} & {}\\    
    \midrule
    %BM3D~\cite{sparse} & {25.65} & {0.685}& {34.51} & {0.850} & {25.65} & {0.685}& {34.51} & {0.850} \\
    %WNNM~\cite{6909762}& {25.78} & {0.809} & {34.67} & {0.864} & {25.65} & {0.685}& {34.51} & {0.850} \\
    %K-SVD~\cite{DBLP:journals/corr/abs-1909-13164} & {26.88} & {0.842}& \textbf{36.49} & \textbf{0.897}& {25.65} & {0.685}& {34.51} & {0.850} \\
    %EPLL~\cite{Hurault_2018} & \textbf{27.11} & \textbf{0.870}  & {33.51} & {0.824}& {25.65} & {0.685}& {34.51} & {0.850} \\
    %\hline
    \multirow{8}{*}{\footnotesize Real-world} & \multirow{5}{*}{\footnotesize Supervised}
    & {\footnotesize EDSR~\cite{lim2017enhanced}} & {32.45}/{0.913}   & {27.59}/{0.792} &  {31.59}/\textbf{0.888} & {27.14}/{0.771} & {34.24}/\textbf{0.908} & \textbf{32.03}/{0.855}\\
    & & {\footnotesize RRDB~\cite{wang2018esrgan}} & \hspace{6pt}{-\hspace{8pt}}/{\hspace{8pt}-} & \hspace{-3pt}\textbf{27.90}/{\hspace{7pt}-} & \hspace{6pt}{-\hspace{8pt}}/{\hspace{8pt}-} & \hspace{-3pt}\textbf{27.39}/{\hspace{7pt}-} & {33.89}/{0.906}& {31.92}/{0.856}\\
    & & {\footnotesize RCAN~\cite{zhang2018image}} & \textbf{32.69}/\textbf{0.919} & {27.66}/{0.793} & \textbf{31.61}/\textbf{0.888} & {27.09}/{0.771} & \textbf{34.34}/\textbf{0.908} & {31.85}/\textbf{0.857}\\
    & & {\footnotesize LP-KPN~\cite{cai2019toward}} & \hspace{6pt}{-\hspace{8pt}}/{\hspace{8pt}-} & {27.76}/\textbf{0.807} & \hspace{6pt}{-\hspace{8pt}}/{\hspace{8pt}-} & {26.34}/\textbf{0.774} & \hspace{-3pt}{33.88}/{\hspace{8pt}-} & \hspace{-3pt}{31.58}/{\hspace{7pt}-}\\
    & & {\footnotesize DRN-S~\cite{guo2020closed}} & {32.50}/{0.912} & {27.79}/{0.805} & {31.43}/{0.884} & \hspace{6pt}{-\hspace{8pt}}/{\hspace{8pt}-} & {33.91}/{0.898} & \hspace{6pt}{-\hspace{8pt}}/{\hspace{8pt}-} \\
    %& & \textbf{IMF-SRSR}$^{\dagger}$ & {\textbf{32.58}} & {-} & {-} & {-} & {-} & {-} \\
    %& & IMF-SRSR$^{\ddagger}$ & {\textbf{32.68}} & {-} & {-} & {-} & {-} & {-} \\
    \cmidrule{2-9}
    & {\footnotesize Unsupervised}
    & {\footnotesize ZSSR~\cite{shocher2018zero}+~\cite{bell2019blind}} & {28.79}/{0.826} & {23.68}/{0.673} &  {27.54}/{0.799}  & {22.46}/{0.645}& \hspace{6pt}{-\hspace{8pt}}/{\hspace{8pt}-} & \hspace{6pt}{-\hspace{8pt}}/{\hspace{8pt}-}\\
    \cmidrule{2-9}
    & \multirow{2}{*}{\footnotesize Self-supervised} 
    %\textbf{IMF-SRSR~(Test)} & {30.67}  &  {26.08} & {29.99} & {25.76}& {32.83}& {30.62}\\
    & {\footnotesize \textbf{ICF-SRSR}~(Ours)} & {30.98}/{0.885}  &  {26.27}/{0.763} & {30.31}/{0.864} & {25.89}/\textbf{0.742}& {32.87}/{0.880}& {30.65}/{0.821}\\
    & & {\footnotesize\textbf{EDSR~(LLR,LR)}~(Ours)} & \textbf{31.13}/\textbf{0.888}  &  \textbf{26.32}/\textbf{0.764} & \textbf{30.33}/\textbf{0.865} & \textbf{25.92}/\textbf{0.742} & \textbf{32.91}/\textbf{0.881}& \textbf{30.68}/\textbf{0.823}\\
    \bottomrule
    \end{tabularx}
    \vspace{-2mm}
    \caption{
        \textbf{Quantitative comparison on real-world datasets.} We compare our self-supervised ICF-SRSR and EDSR~(LLR,LR), \ie, the model EDSR~\cite{lim2017enhanced} trained on our generated paired dataset (LLR,LR), to several supervised/unsupervised methods trained on synthetic DIV2K~\cite{agustsson2017ntire}, real-world RealSR-V3~\cite{cai2019toward} and DRealSR~\cite{wei2020component} datasets for scales $\times2$ and $\times4$ with PSNR and SSIM metrics. 
    }
    \label{tab:real_data}
    \vspace{-4mm}
\end{table*}

%% file: sections/figures/realsr2.tex
\begin{figure*}
    \centering
	\captionsetup[subfloat]{labelformat=empty,aboveskip=1pt,justification=centering}
	%\begin{center}
		\newcommand{\rowArg}{1.2cm}
		\newcommand{\fullSize}{6cm}
		\newcommand{\patchSize}{2 cm}
		% 			\begin{adjustbox}{width=\linewidth, center=\linewidth}
		\setlength\tabcolsep{0.1cm}
        \hspace{115pt}
        \textbf{\scriptsize Supervised~Real.} 
        \hspace{5pt}
        \textbf{\scriptsize Supervised~Syn.} 
        \hspace{7.5pt}
        \textbf{\scriptsize Unsupervised} 
        \hspace{12.5pt}
        \textbf{\scriptsize Self-Supervised}
        \begin{tabular}[b]{c c c}
			\multicolumn{2}{c}{\multirow{3}{*}[\rowArg]{
					\subfloat[Input~(LR)]
					% 			{\includegraphics[width = \fullSize, height = \fullSize]
					{\includegraphics[page=1, width=\fullSize, trim={0px 0px 0px 0px}, clip]
						{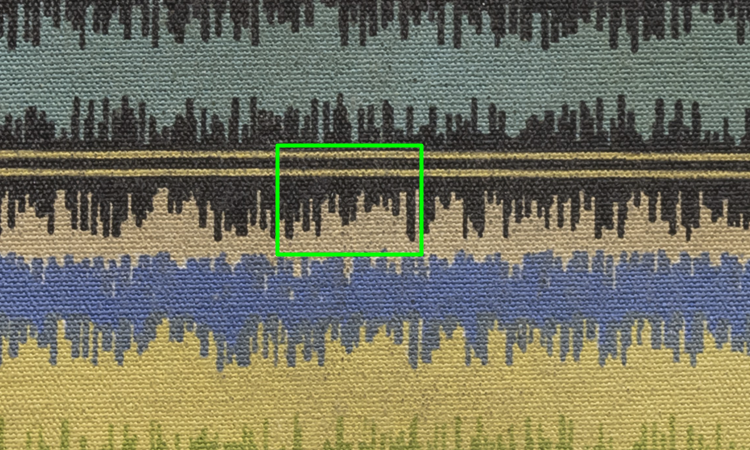}}}} 
            &
			\subfloat[\scriptsize LP-KPN~\cite{cai2019toward} \\ (\textcolor{Orchid}{22.01}/\textcolor{ForestGreen}{0.672})]{
				\includegraphics[width = \patchSize]
				{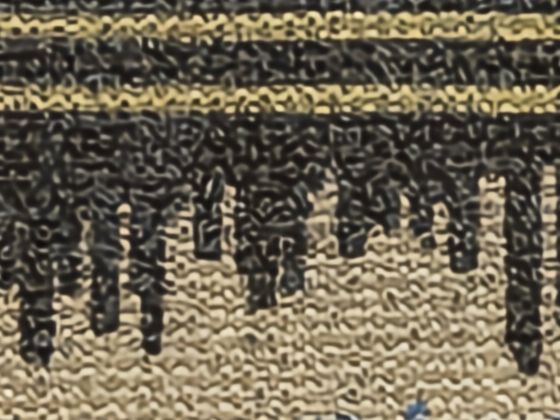}} 
			\subfloat[\scriptsize IKC~\cite{gu2019blind} \\ (\textcolor{Orchid}{24.95}/\textbf{\textcolor{ForestGreen}{0.745}})]{
				\includegraphics[width = \patchSize]
				{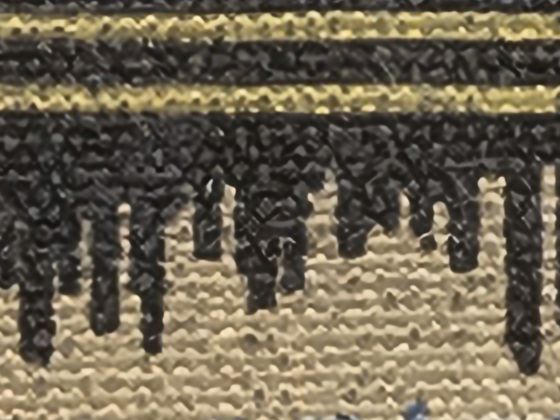}}
            \subfloat[\scriptsize ZSSR~\cite{shocher2018zero}+~\cite{bell2019blind} \\ (\textcolor{Orchid}{16.66}/\textcolor{ForestGreen}{0.507})]{
				\includegraphics[width = \patchSize]
				{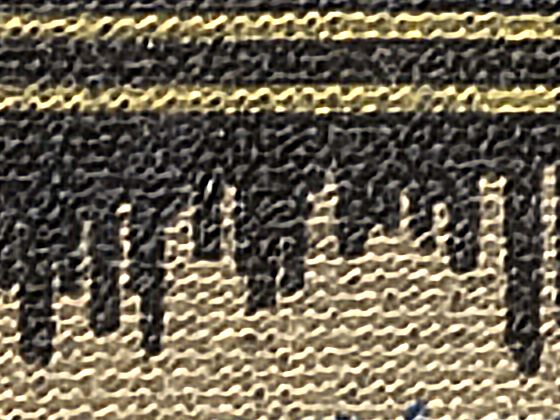}}
            \subfloat[\scriptsize ICF-SRSR~(Ours) \\ (\textbf{\textcolor{Orchid}{25.82}}/\textcolor{ForestGreen}{0.731})]{
				\includegraphics[width = \patchSize]
				{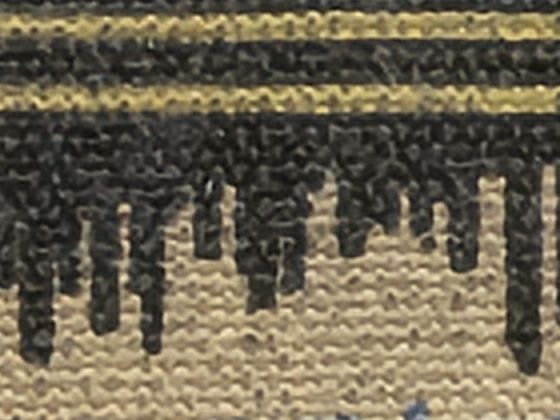}}
            \subfloat[\scriptsize GT~(HR)\\ \scriptsize  ({\textcolor{Orchid}{PSNR}}/{\textcolor{ForestGreen}{SSIM}})]{
				\includegraphics[width = \patchSize]
				{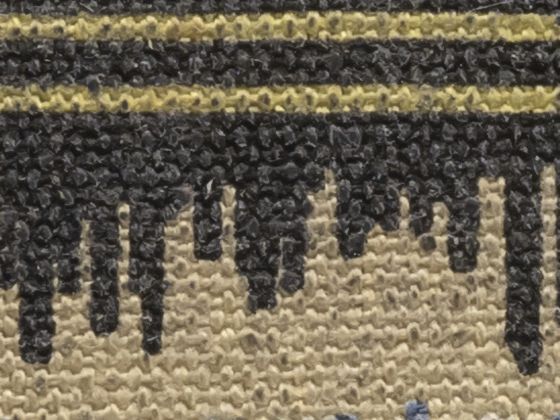}}   \vspace{-.040cm}
\\
            && 
			\subfloat[\scriptsize Error$_\text{LP-KPN~\cite{cai2019toward}}$\\(={\textcolor{BrickRed}{15.78}})]{
				\includegraphics[width = \patchSize]
				{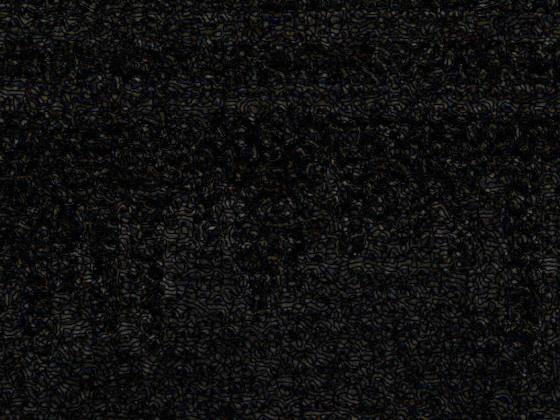}} 
			\subfloat[\scriptsize Error$_\text{ IKC~\cite{gu2019blind}}$\\(={\textcolor{BrickRed}{11.13}})]{
				\includegraphics[width = \patchSize]
				{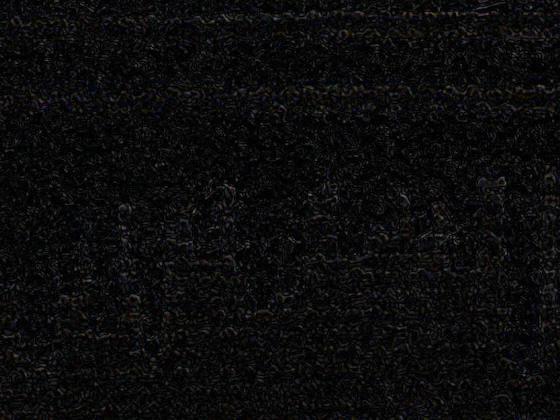}}
            \subfloat[\scriptsize Error$_\text{ ZSSR~\cite{shocher2018zero}+~\cite{bell2019blind}}$\\(={\textcolor{BrickRed}{30.29}})]{
				\includegraphics[width = \patchSize]
				{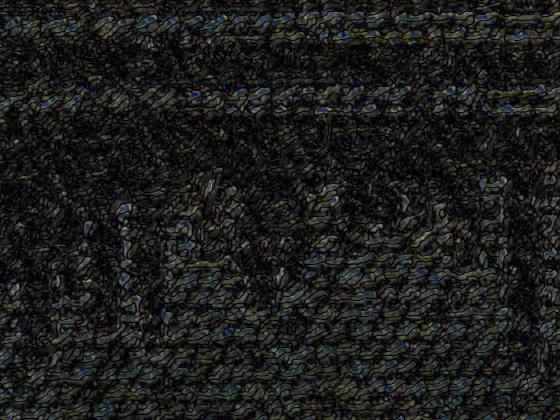}}
            \subfloat[\scriptsize Error$_\text{ICF-SRSR}$\\(=\textbf{\textcolor{BrickRed}{10.13}})]{
				\includegraphics[width = \patchSize]
				{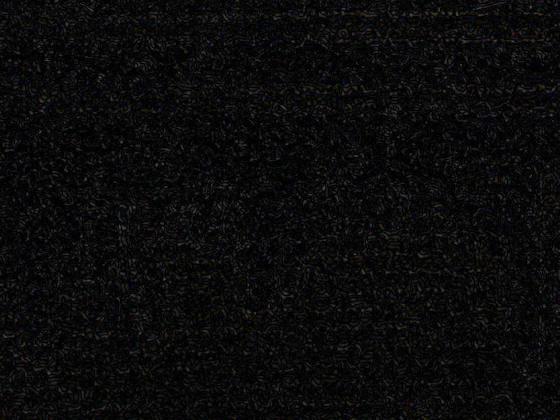}}
            \subfloat[\scriptsize Zero\\(            {\textcolor{BrickRed}{MAE}})]{
				\includegraphics[width = \patchSize]
				{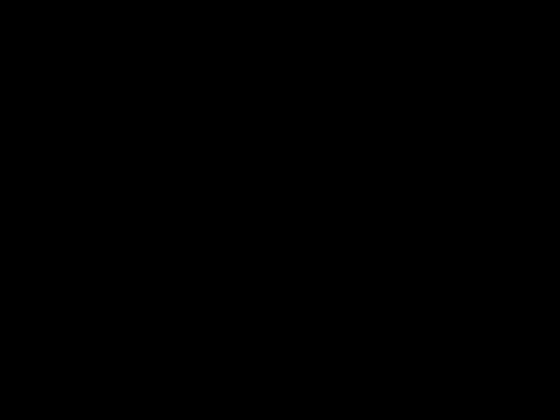}}               
	\end{tabular}\textbf{} \vspace{0.25cm}

        \begin{tabular}[b]{c c c}
			\multicolumn{2}{c}{\multirow{2}{*}[\rowArg]{
					\subfloat[Input~(LR)]
					% 			{\includegraphics[width = \fullSize, height = \fullSize]
					{\includegraphics[page=1, width=\fullSize, trim={0px 0px 0px 20px}, clip]
						{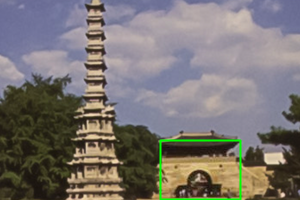}}}} 
            &
			\subfloat[\scriptsize LP-KPN~\cite{cai2019toward} \\ \scriptsize (\textbf{\textcolor{Orchid}{30.24}}/\textbf{\textcolor{ForestGreen}{0.871}})]{
				\includegraphics[width = \patchSize]
				{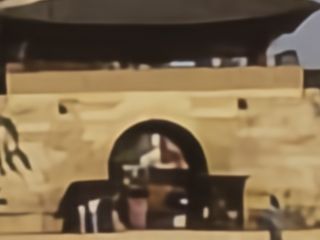}} 
			\subfloat[\scriptsize IKC~\cite{gu2019blind} \\ \scriptsize  ({\textcolor{Orchid}{27.96}}/{\textcolor{ForestGreen}{0.832}})]{
				\includegraphics[width = \patchSize]
				{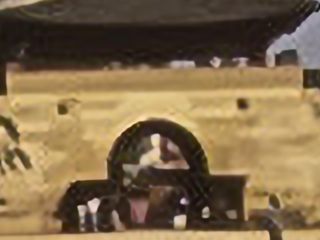}}
            \subfloat[\scriptsize ZSSR~\cite{shocher2018zero}+~\cite{bell2019blind} \\ \scriptsize  ({\textcolor{Orchid}{23.82}}/{\textcolor{ForestGreen}{0.741}})]{
				\includegraphics[width = \patchSize]
				{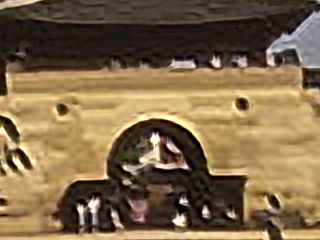}}
            \subfloat[\scriptsize ICF-SRSR~(Ours) \\ \scriptsize  ({\textcolor{Orchid}{29.49}}/{\textcolor{ForestGreen}{0.852}})]{
				\includegraphics[width = \patchSize]
				{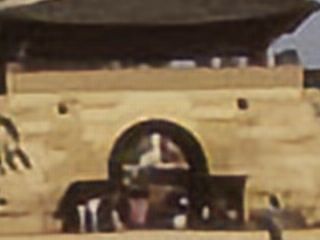}}
            \subfloat[\scriptsize GT~(HR) \\ \scriptsize  ({\textcolor{Orchid}{PSNR}}/{\textcolor{ForestGreen}{SSIM}})]{
				\includegraphics[width = \patchSize]
				{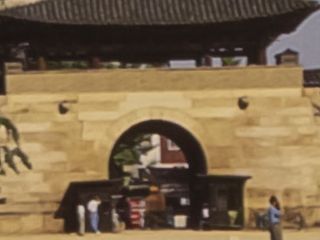}}   \vspace{-.040cm}
\\
            &&
			\subfloat[\scriptsize Error$_\text{ LP-KPN~\cite{cai2019toward}}$\\(=\textbf{\textcolor{BrickRed}{4.66}})]{
				\includegraphics[width = \patchSize]
				{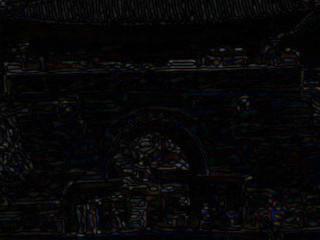}} 
			\subfloat[\scriptsize Error$_\text{ IKC~\cite{gu2019blind}}$\\(={\textcolor{BrickRed}{6.15}})]{
				\includegraphics[width = \patchSize]
				{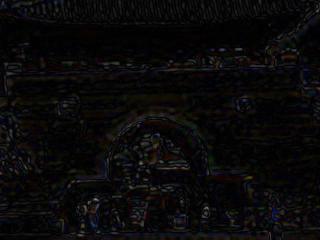}}
            \subfloat[\scriptsize Error$_\text{ ZSSR~\cite{shocher2018zero}+~\cite{bell2019blind}}$\\(={\textcolor{BrickRed}{9.52}})]{
				\includegraphics[width = \patchSize]
				{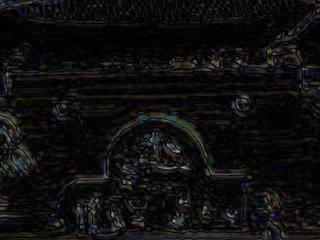}}
            \subfloat[\scriptsize Error$_\text{ ICF-SRSR}$\\(={\textcolor{BrickRed}{5.24}})]{
				\includegraphics[width = \patchSize]
				{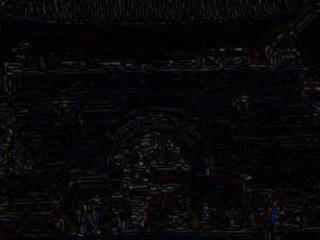}}
            \subfloat[\scriptsize Zero\\(            {\textcolor{BrickRed}{MAE}})]{
				\includegraphics[width = \patchSize]
				{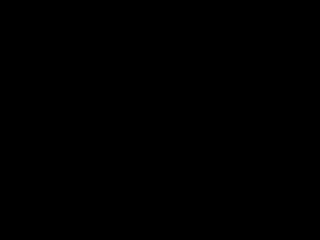}}                    
	\end{tabular}\textbf{}
 %\vspace{-2mm}
	%\captionsetup{justification=raggedright,singlelinecheck=false}
	\vspace{-2mm}
	\caption{\textbf{Qualitative comparisons on a real-world dataset.} 
        We visualize the super-resolution results~(first row) and their corresponding error maps with respect to the GT~(second row) for an image captured by each Nikon and Canon camera. 
        We compare our self-supervised method ICF-SRSR with the supervised method LP-KPN~\cite{cai2019toward} and the unsupervised method ZSSR~\cite{shocher2018zero}+~\cite{bell2019blind} trained on the RealSR-V3~\cite{cai2019toward} dataset and the supervised method IKC~\cite{gu2019blind} trained on synthetic dataset DIV2K~\cite{agustsson2017ntire} for scale $\times4$ with \textcolor{Orchid}{PSNR}, \textcolor{ForestGreen}{SSIM}, and \textcolor{BrickRed}{MAE} metrics.}
        \label{fig:realsr}
	\vspace{-4mm}
\end{figure*}

%% file: tables/reb_two-models.tex
\begin{table}[h]
    %\vspace{-3mm}
    \small
    \centering
    %\setlength\tabcolsep{0.5pt}
    %\setlength\tabcolsep{0.0001pt}
    %\resizebox{0.65\columnwidth}{!}{
    \begin{tabularx}{\linewidth}{l 
    >{\centering\arraybackslash}X 
    >{\centering\arraybackslash}X 
    >{\centering\arraybackslash}X  
    }
    \toprule
    \textbf{Method} 
    & \textbf{DIV2K~($\times 2$)}&
    \textbf{Canon~($\times 2$)} & \textbf{Nikon~($\times 2$)}  \\
    %\cline{4-7}
    \midrule
     Two Models & {34.81} & {30.61} & {30.01} \\
     Up-Down & {29.92}& {28.56} & {27.52}  \\
     Down-Up & {34.59}& {30.58} & {30.00} \\
     \midrule
     \textbf{ICF-SRSR} & \textbf{35.19} & \textbf{30.98} & \textbf{30.31} \\
    \bottomrule
    \end{tabularx}
    \vspace{-2mm}
    \caption{
    \textbf{Ablation on model design.}
    }
    \label{tab:reb-two-models}
    \vspace{-4mm}
\end{table}

%% file: tables/downsampling.tex
\begin{table}[h]
    \small
    \centering
    \begin{tabularx}{\linewidth}{l >{\centering\arraybackslash}X >{\centering\arraybackslash}X >{\centering\arraybackslash}X  >{\centering\arraybackslash}X
    }
    \toprule
    \multirow{2}{*}{\textbf{Down-sampling}} & 
    \multicolumn{2}{c}{\textbf{Canon}} & 
    \multicolumn{2}{c}{\textbf{Nikon}} \\
    %\cline{4-7}
    & $\times2$ & $\times4$ & $\times2$ & $\times4$\\
    %\cline{4-7}
    
    \midrule
     Nearest & {29.35} & {24.51} & {28.54} & {23.91} \\
     Bicubic & {30.27} & {25.76} & {29.71} & {25.56} \\
     Gaussian+Nearest & {29.62} & {24.65} & {28.87} & {24.09} \\
     Gaussian+Bicubic & {30.61} & {25.95} & {30.12} & {25.81} \\
     
     \midrule
     \textbf{ICF-SRSR} &  \textbf{32.46} & \textbf{28.93} & \textbf{32.12} & \textbf{29.15}\\
    \bottomrule
    \end{tabularx}
    
    \vspace{-2mm}
    \caption{
        \textbf{Ablation on down-sampling performance.} %Comparisons of the generated LR images from the given HR images using various known down-sampling operators with our learnable down-sampling method ICF-SRSR.
    }
    \label{tab:downsampling}
    \vspace{-4mm}
\end{table}

%% file: tables/downsampling2.tex
\begin{table}[h]
    \small
    \centering
    \begin{tabularx}{\linewidth}{l >{\centering\arraybackslash}X >{\centering\arraybackslash}X  >{\centering\arraybackslash}X 
    >{\centering\arraybackslash}X
    }
    \toprule
    \textbf{Down-sampling}  & 
    \textbf{Set5} & \textbf{Set14} & 
    \textbf{Canon} & 
    \textbf{Nikon}  \\
    \midrule
     Bicubic  & {35.30} & {31.53} & {30.41} & {29.80}\\
     Gaussian+Nearest  & {30.79} & {28.39} & {29.41} & {28.60}  \\
     Gaussian+Bicubic  & {35.43} & {31.84} & {30.47} & {29.86} \\
     \midrule
     \textbf{ICF-SRSR} & \textbf{37.09} & \textbf{32.91} &  \textbf{31.13} & \textbf{30.33}\\
    \bottomrule
    \end{tabularx}
    
    \vspace{-2mm}
    \caption{\textbf{Comparison with non-learnable down-sampling operators to generate paired training data for SR task.}
        %\textbf{SISR results using different down-sampling operators.}
        %Effects of different down-sampling operators to generate appropriate paired images for training.
        %
    }
    \label{tab:downsampling2}
    \vspace{-4mm}
\end{table}

%% file: tables/reb_down.tex
\begin{table}[h]
    %\vspace{-3mm}
    \small
    \centering
    %\setlength\tabcolsep{0.5pt}
    %\setlength\tabcolsep{0.0001pt}
    %\resizebox{0.65\columnwidth}{!}{
    \begin{tabularx}{\linewidth}{l 
    >{\centering\arraybackslash}X 
    >{\centering\arraybackslash}X 
    }
    \toprule
    \textbf{Downsampling} 
    & \textbf{Canon~$(\times2)$} & \textbf{Nikon~$(\times2)$} \\
    %\cline{4-7}
    \midrule
     ADL~[\textcolor{blue}{49}] & {30.76} & {30.44}  \\
     DRN-S~[\textcolor{blue}{20}] & {30.82} & {30.24}  \\
     \midrule
    \textbf{ICF-SRSR} & \textbf{31.94} & \textbf{31.24}\\
    \bottomrule
    \end{tabularx}
    \vspace{-2mm}
    \caption{\textbf{Comparison with learnable down-sampling operators to generate paired training data for SR task.}
    %    \textbf{}
        %
    }
    \label{tab:reb-down}
    \vspace{-5mm}
\end{table}

%% file: tables/few-shot.tex
\begin{table}[h]
    %\vspace{-3mm}
    \small
    \centering
    \begin{tabularx}{\linewidth}{l >{\centering\arraybackslash}X
    >{\centering\arraybackslash}X
    >{\centering\arraybackslash}X >{\centering\arraybackslash}X  >{\centering\arraybackslash}X >{\centering\arraybackslash}X     }
    \toprule
    \multirow{2}{*}{\textbf{Training set}} &\multicolumn{2}{c}{\textbf{Set14}} & 
    \multicolumn{2}{c}{\textbf{BSD100}} & \multicolumn{2}{c}{\textbf{Urban100}} \\
    %\cline{4-7}
    & $\times2$ & $\times4$ & $\times2$ & $\times4$ & $\times2$ & $\times4$\\
    \midrule
     Large & {32.86} & {27.76} & {31.54} & {26.99} & {30.39} & {24.72} \\
     Small & {32.44} & {27.19} & {31.34} & {26.82} & {30.26} & {24.66}  \\
     \midrule
    \multirow{2}{*}{\textbf{Training set}} &
    \multicolumn{2}{c}{\textbf{Canon}} & 
    \multicolumn{2}{c}{\textbf{Nikon}} & \multicolumn{2}{c}{\textbf{DRealSR}} \\
    %\cline{4-7}
    & $\times2$ & $\times4$ & $\times2$ & $\times4$ & $\times2$ & $\times4$\\
    %\cline{4-7}
    \midrule
     Large & {30.98} & {26.26} & {30.31} & {25.89} & {32.87} & {30.65} \\
     Small & {30.67} & {26.08} & {29.99} & {25.76} & {32.83} & {30.62}  \\
    \bottomrule
    \end{tabularx}
    \vspace{-2mm}
    \caption{
        \textbf{Few-shot learning.}
        %
        %For Set14, BSD100, and Urban100, Large denotes the DIV2K training set, while Small corresponds to each test dataset.
        %Similarly, for Canon, Nikon, and DRealSR, Large denotes training splits of the corresponding dataset and Small is their test splits.
    }
    \label{tab:ab-few-shot}
    \vspace{-4mm}
\end{table}

%% file: tables/ab_scale.tex
\begin{table}[t]
    %\vspace{-3mm}
    \small
    \centering
    %\setlength\tabcolsep{0.5pt}
    %\setlength\tabcolsep{0.0001pt}
    %\resizebox{0.65\columnwidth}{!}{
    \begin{tabularx}{\linewidth}{l 
    >{\centering\arraybackslash}X 
    >{\centering\arraybackslash}X 
    >{\centering\arraybackslash}X 
    >{\centering\arraybackslash}X 
    }
    \toprule
    \textbf{Scale} 
    & \textbf{Canon~$(\times2)$} & \textbf{Nikon~$(\times2)$} \\
    %\cline{4-7}
    \midrule
     \text{2} & {30.67} & {29.99}  \\
     \text{2,4} & {30.75} & {30.09} \\
      \text{2,4,8} & \textbf{30.78} & \textbf{30.11} \\
    \bottomrule
    \end{tabularx}
    \vspace{-2mm}
    \caption{
        \textbf{Multi-scale augmentation.}
    }
    \label{tab:ab-scale}
    \vspace{-4mm}
\end{table}

%% file: tables/ab_loss.tex
\begin{table}[h]
    %\vspace{-3mm}
    \small
    \centering
    %\setlength\tabcolsep{0.5pt}
    %\setlength\tabcolsep{0.0001pt}
    %\resizebox{0.65\columnwidth}{!}{
    \begin{tabularx}{\linewidth}{l 
    >{\centering\arraybackslash}X 
    >{\centering\arraybackslash}X 
    >{\centering\arraybackslash}X  
    }
    \toprule
    \textbf{Loss} 
    & \textbf{DIV2K~$(\times2)$} & \textbf{Canon~$(\times2)$} & \textbf{Nikon~$(\times2)$}\\
    %\cline{4-7}
    \midrule
     $\mathcal{L}^{\text{Color}}$ only & {30.31} & {29.12} & {28.38} \\
     $\mathcal{L}^{\text{Color}}$,$\mathcal{L}^{\text{Cons}}$ & \textbf{35.19} & \textbf{30.98} & \textbf{30.31}\\
    \bottomrule
    \end{tabularx}
    \vspace{-2mm}
    \caption{
        \textbf{Effect of loss functions.}
    }
    \label{tab:ab-loss}
    \vspace{-5mm}
\end{table}

%% file: sections/conclusion.tex
\section{Conclusion}\label{sec:conclusion}
We propose ICF, a novel invertible scale-conditional function that receives an image and an arbitrary scaling factor and generates the resized image, and can reconstruct the same input image by the given resized image and the inverse scaling factor. 
Then, we utilize ICF to design a self-supervised real-world single-image super-resolution framework ICF-SRSR. 
Accordingly, our framework is able to generate up-sampled and down-sampled images simultaneously, where the generated down-sampled images can be used to construct paired images appropriate for training existing models.
Extensive experiments demonstrate the strengths of our self-supervised method on both synthetic and real-world datasets and superior performance on the real-world dataset compared to supervised models trained on the synthetic datasets.

\Paragraph{Limitations and future works.}
One remaining limitation is that we only apply our method to a few real-world datasets due to the lack of aligned LR-HR image pairs for evaluation in other real-world datasets.
Therefore, we aim to provide a large-scale real-world dataset from various scenes for better evaluation in our future work.
Moreover, we will investigate the applications of our defined ICF to self-supervised image warping and other image restoration tasks.

%% file: sections/figures/net_arch.tex
\renewcommand{\thefigure}{S1}
\begin{figure*}[h]
     \centering
     \begin{subfigure}[b]{0.45\textwidth}
         \centering
         \includegraphics[page=1, width=\textwidth]{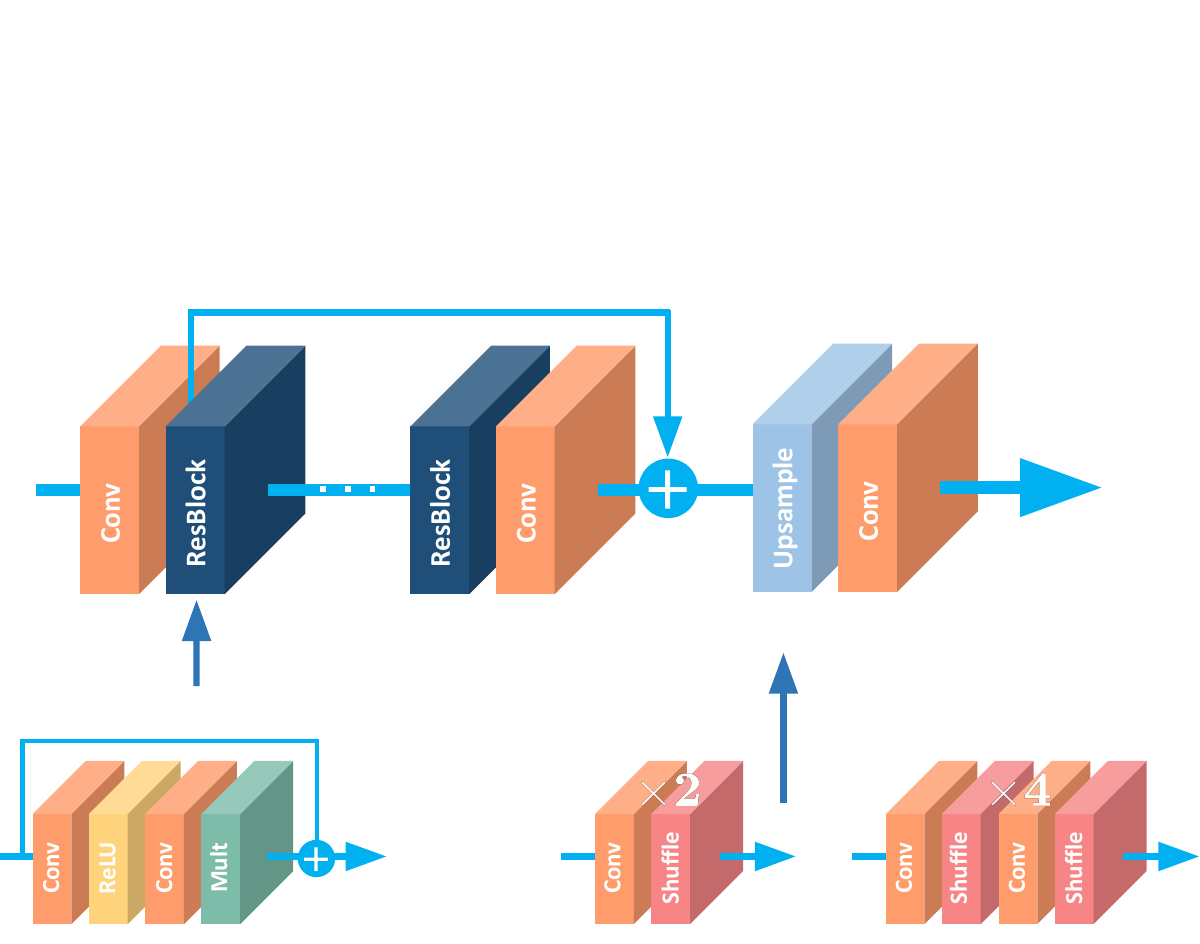}
         \caption{Original EDSR.}
         \label{fig:supp_edsr_original}
     \end{subfigure}
     \hfill
     \begin{subfigure}[b]{0.45\textwidth}
         \centering
         \includegraphics[page=2, width=\textwidth]{figures/supp/EDSR.pdf}
         \caption{Our modified EDSR for our ICF-SRSR.}
         \label{fig:supp_edsr_developed}
     \end{subfigure}
        \vspace{-2mm}
        \caption{\textbf{The network architecture of our modified EDSR.}   
        }
        \label{fig:supp_edsr}
        \vspace{-4mm}
\end{figure*}

%% file: sections/figures/multi.tex
\renewcommand{\thefigure}{S2}
\begin{figure*}[t]
     \centering
     \begin{subfigure}[b]{0.25\textwidth}
         \centering
         \includegraphics[width=\textwidth]{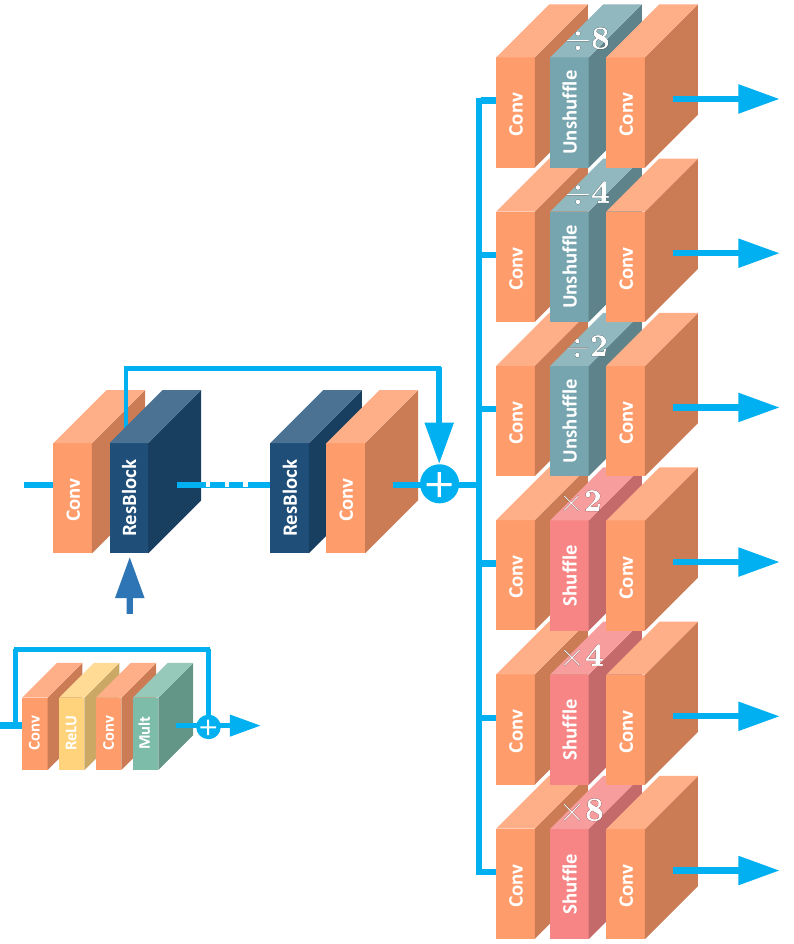}
         \caption{Multi-tail modified EDSR.}
         \label{fig:supp_multi1}
     \end{subfigure}
     \hfill
     \begin{subfigure}[b]{0.7\textwidth}
         \centering
         \includegraphics[width=\textwidth]{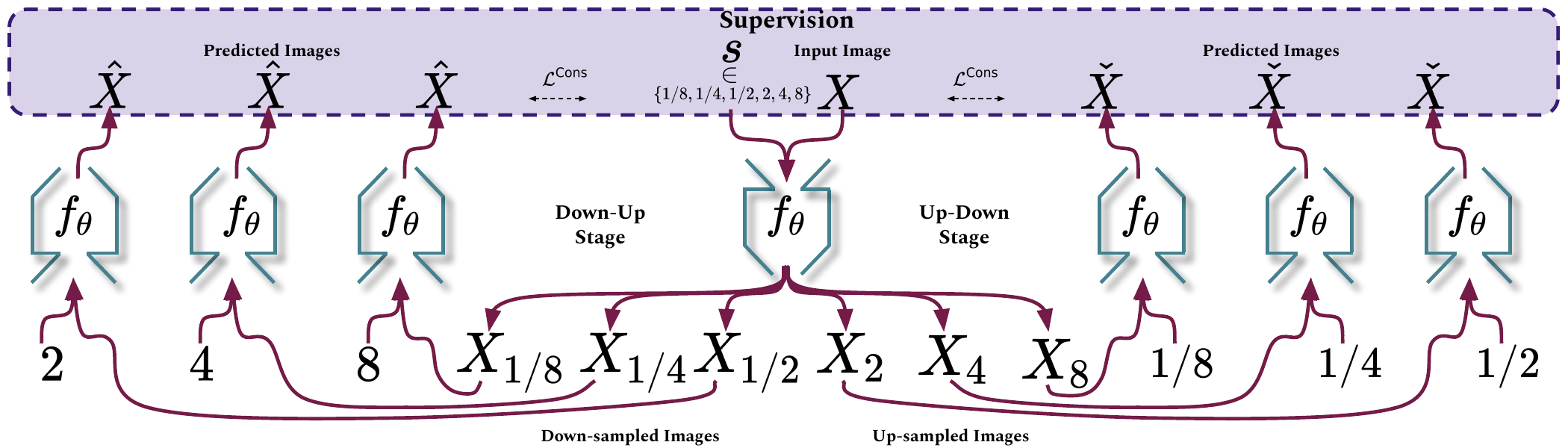}
         
         \caption{Multi-scale augmentation strategy.}
         \label{fig:supp_multi2}
     \end{subfigure}
        \vspace{-2mm}
        \caption{\textbf{The overview of our multi-scale augmentation strategy.} (a) Our multi-tail EDSR for $\times2$, $\times4$, $\times8$ and their inverse scaling factors. (b) An overview of the proposed multi-scale augmentation.   
        }
        \label{fig:supp_multi}
        \vspace{-5mm}
\end{figure*}

%% file: tables/supp_benchmark.tex
\renewcommand{\thetable}{S1}
\begin{table*}[t]
    \small
    \centering
    \begin{tabularx}{\linewidth}{c l >{\centering\arraybackslash}X >{\centering\arraybackslash}X >{\centering\arraybackslash}X >{\centering\arraybackslash}X
    >{\centering\arraybackslash}X 
    }
    \toprule
    \multirow{2}{*}{\textbf{Supervision}} & 
    \multirow{2}{*}{\bf Method} & 
    \textbf{Set5} & 
    \textbf{Set14} &
    \textbf{BSD100} &
    \textbf{Urban100} &
    \textbf{Manga109}\\
    & & $\times2$/$\times4$ & $\times2$/$\times4$ & $\times2$/$\times4$ & $\times2$/$\times4$ & $\times2$/$\times4$   \\

    \midrule
    & {\footnotesize Bicubic} & {0.929/0.810} & {0.868/0.702} & {0.843/0.667} & {0.840/0.657} & {0.933/0.789}  \\
    \midrule
    \multirow{8}{*}{Supervised}
    & {\footnotesize VDSR~\cite{kim2016accurate}} & {0.959/0.884} & {0.912/0.768} & {0.896/0.725} & {0.914/0.752} & {0.975/0.887}  \\
    & {\footnotesize EDSR~\cite{lim2017enhanced}} & {0.960/0.898} & {0.919/0.787} & {0.901/0.742} & {0.935/0.803} & {0.977/0.915}  \\
    & {\footnotesize CARN~\cite{ahn2018fast}} & {0.959/0.894} & {0.916/0.781} & {0.897/0.735} & {0.925/0.784} & {0.976/0.908}  \\
    & {\footnotesize RCAN~\cite{zhang2018image}} & {0.961/0.900} & {0.921/0.788} & {0.902/0.743} & {0.938/0.806} & {0.978/0.917} \\
    & {\footnotesize RDN~\cite{zhang2018residual}} & {0.961/0.899} & {0.921/0.787} & {0.901/0.741} & {0.935/0.802} & {0.978/0.915}  \\
    & {\footnotesize DRN-S~\cite{guo2020closed}} & {0.960/0.901} & {0.910/0.790} & {0.900/0.744} & {0.920/0.807} & \textbf{0.980}/{0.919}  \\
    & {\footnotesize LIIF~\cite{chen2021learning}} & {0.933/0.898} & {0.882/0.788} & {0.871/0.742} & {0.905/0.805} & {~~~~-~~~~/~~~~-~~~~}  \\
    &{\footnotesize ELAN~\cite{ELAN-light}} & \textbf{0.962}/\textbf{0.902} & \textbf{0.922}/\textbf{0.791} & \textbf{0.903}/\textbf{0.745} & \textbf{0.939}/\textbf{0.816} & {0.979}/\textbf{0.922} \\
    \midrule
    \multirow{3}{*}{Unsupervised} 
    & {\footnotesize SelfExSR~\cite{huang2015single}} & {0.953/0.861} & {0.903/0.751} & {0.885/0.710} & {0.897}/{0.740} & \textbf{0.968}/{0.718} \\
    & {\footnotesize ZSSR~\cite{shocher2018zero}}  & \textbf{0.957}/\textbf{0.879} & \textbf{0.910}/\textbf{0.765} & \textbf{0.892}/\textbf{0.721} & {0.894}/{0.682} & {0.957}/\textbf{0.813} \\
    &  {\footnotesize MZSR~\cite{soh2020meta}} &  {0.956/~~~~-~~~~} & {~~~~-~~~~/~~~~-~~~~} & {\textbf{0.892}/~~~~-~~~~} & {\textbf{0.909}/~~~~-~~~~} & {~~~~-~~~~/~~~~-~~~~}  \\
    \midrule
    \multirow{2}{*}{Self-supervised}
    & {\footnotesize \textbf{ICF-SRSR}~(Ours)} & {0.956/0.874} & {0.908/0.760} & {0.888/0.715} & {0.910/0.740} & {0.970/0.872}  \\
    & {\footnotesize \textbf{EDSR~(LLR,LR)}~(Ours)} & \textbf{0.957}/\textbf{0.876} & \textbf{0.909}/\textbf{0.763} & \textbf{0.889}/\textbf{0.717} & \textbf{0.911}/\textbf{0.745} & \textbf{0.971}/\textbf{0.876} \\ 
    \bottomrule
    \end{tabularx}
    
    \vspace{-2mm}
    \caption{
        \textbf{Quantitative comparisons of different methods on synthetic datasets by SSIM.} 
        We compare our ICF-SRSR with several supervised and unsupervised methods on the five standard benchmark datasets~\cite{bevilacqua2012low, zeyde2010single, martin2001database, huang2015single, Manga109} on scales $\times 2$ and $\times 4$. 
        ICF-SRSR refers to our self-supervised method, while EDSR~(LLR,LR) is the model EDSR trained on our generated pairs (LLR,LR) of the DIV2K dataset.
        We also note that MZSR does not report SSIM for $\times 4$ SR in the original paper.
    }
    \label{tab:supp-benchmark}
    \vspace{-4mm}
\end{table*}

%% file: tables/supp_baseline.tex
\renewcommand{\thetable}{S2}
\begin{table*}%[t]
    \small
    \centering
    \begin{tabularx}{\linewidth}{l >{\centering\arraybackslash}X >{\centering\arraybackslash}X >{\centering\arraybackslash}X  >{\centering\arraybackslash}X
    >{\centering\arraybackslash}X
    >{\centering\arraybackslash}X
    }
    \toprule
    {\textbf{Baseline}} & 
    {\textbf{Set5}} &
    {\textbf{Set14}} & 
    {\textbf{BSD100}} & 
    {\textbf{Urban100}} &
    {\textbf{DIV2K}}
    \\
    %& PSNR  & PSNR  & PSNR & SSIM & PSNR & PSNR
    \midrule
    \textbf{ICF-SRSR~(LIIF)} & {36.46} & {32.39} & {31.18} & {29.74} & {34.52}  \\
    \textbf{ICF-SRSR~(EDSR)} & {37.01} & {32.86} & {31.54} & {30.39} & {35.19} \\
    \textbf{ICF-SRSR~(RDN)} & {37.03} & {32.87} & {31.56} & {30.42} & {35.18} \\
     \textbf{ICF-SRSR~(RCAN)} & \textbf{37.12} &  \textbf{32.92} &  \textbf{31.59} &  \textbf{30.50} &  \textbf{35.21}  \\

    \bottomrule
    \end{tabularx}
    
    \vspace{-2mm}
    \caption{
        \textbf{Evaluation of our ICF-SRSR with different baselines by PSNR metric on scale $\times 2$.} 
    }
    \label{tab:supp-baseline}
    \vspace{-4mm}
\end{table*}

%% file: tables/ablation.tex
\renewcommand{\thetable}{S3}
\begin{table*}[h!]
    %\vspace{-3mm}
    \small
    \centering
    \begin{tabularx}{\linewidth}{l 
    >{\centering\arraybackslash}X 
    >{\centering\arraybackslash}X 
    >{\centering\arraybackslash}X  
    >{\centering\arraybackslash}X
    }
    \toprule
    \textbf{$\lambda_{\text{Color}}$} 
    & \textbf{Canon} & \textbf{Nikon} & \textbf{Set5} & \textbf{DIV2K}\\
    %\cline{4-7}
    
    \midrule
     0.1 & {30.62} & {29.97} & {36.24} & \textbf{35.03}\\
     0.2 & \textbf{30.67} & {29.99} & \textbf{36.41} & {35.02}\\
     1 & {30.63} & \textbf{30.02} & {36.38} & {34.93}\\
     10 & {30.61} & {29.98} & {36.35} & {34.82}\\

    \bottomrule
    \end{tabularx}
    
    \vspace{-2mm}
    \caption{
        \textbf{Ablation on the hyperparameter $\lambda_{\text{Color}}$.}
    }
    \label{tab:ab-loss}
    \vspace{-4mm}
\end{table*}

%% file: tables/DASR.tex
\renewcommand{\thetable}{S4}
\begin{table*}[h!]
    %\vspace{-7mm}
    \small
    \centering
    %\setlength\tabcolsep{3.5pt}
    %\resizebox{0.65\columnwidth}{!}{
    \begin{tabularx}{\linewidth}{l 
    >{\centering\arraybackslash}X 
    >{\centering\arraybackslash}X 
    >{\centering\arraybackslash}X  
    >{\centering\arraybackslash}X
    >{\centering\arraybackslash}X
    >{\centering\arraybackslash}X
    }
    \toprule
    \textbf{Method} & \textbf{Self-Supervised} & \textbf{Set}  
    %\cline{4-7}
    & \textbf{Canon($\times2$)} & \textbf{Canon($\times4$)} & \textbf{Nikon($\times2$)} & \textbf{Nikon($\times4$)} \\
    \midrule
     DASR~\cite{wang2021unsupervised} & $\crossmark[red, scale=0.5]$ & DIV2K~(HR) &  {30.66} & {25.98} & {29.74} & 25.25  \\
     DASR~\cite{wang2021unsupervised} & $\crossmark[red, scale=0.5]$ & RealSR-V3~(HR) & {30.76} & {26.09} & {30.15} & \textbf{25.94}\\
     \midrule
     DASR~\cite{wang2021unsupervised} & $\cmark[blue, scale=0.5]$ & RealSR-V3~(LR) & {30.68} & {25.38} & {30.08} & {25.13}  \\
     \textbf{ICF-SRSR}& $\cmark[blue, scale=0.5]$ & RealSR-V3~(LR) & \textbf{30.98} & \textbf{26.26} & \textbf{30.31} & {25.89}  \\
    \bottomrule
    \end{tabularx}
     \vspace{-2mm}
    \caption{\textbf{Quantitative comparison with DASR~\cite{wang2021unsupervised} method trained on different training datasets.}}
    \label{tab:DASR}
    \vspace{-4mm}
\end{table*}

%% file: sections/figures/supp/noisy.tex
\renewcommand{\thefigure}{S3}
\begin{figure*}[h]
    \centering
	\captionsetup[subfloat]{labelformat=empty,aboveskip=1pt}
	%\begin{center}
		\newcommand{\rowArg}{1.74cm}
		\newcommand{\fullSize}{4.50cm}
		\newcommand{\patchSize}{2.72cm}
		% 			\begin{adjustbox}{width=\linewidth, center=\linewidth}
		\setlength\tabcolsep{0.15cm}
		\begin{tabular}[b]{c c c}
			\multicolumn{2}{c}{\multirow{2}{*}[\rowArg]{
					\subfloat[Input~(LR)]
					% 			{\includegraphics[width = \fullSize, height = \fullSize]
					{\includegraphics[height=\fullSize]
						{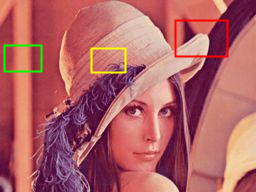}}}} &
			\subfloat[\centering  Ours ]{
				\includegraphics[width = \patchSize]
				{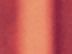}} 
                \hspace{1.25mm}
			\subfloat[\centering  Ours   ]{
				\includegraphics[width = \patchSize]
				{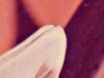}}
                \hspace{1.25mm}
			\subfloat[\centering  Ours   ]{
				\includegraphics[width = \patchSize]
				{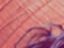}} \\

			 & & \subfloat[\centering   GT ]{
				\includegraphics[width = \patchSize]
				{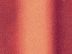}}
                \hspace{1.25mm}
			\subfloat[\centering   GT ]{
				\includegraphics[width = \patchSize]
				{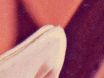}}
                \hspace{1.25mm}
			\subfloat[\centering  GT]{
				\includegraphics[width = \patchSize]
				{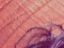}}
	\end{tabular}
  \vspace{-8mm}
\end{figure*}

\begin{figure*}[h]
    \centering
	\captionsetup[subfloat]{labelformat=empty,aboveskip=1pt}
	%\begin{center}
		\newcommand{\rowArg}{1.74cm}
		\newcommand{\fullSize}{4.50cm}
		\newcommand{\patchSize}{2.72cm}
		% 			\begin{adjustbox}{width=\linewidth, center=\linewidth}
		\setlength\tabcolsep{0.15cm}
	%%%%%%%%%%%%%%%%%%%%%%%%%%%%%%%%%
		\begin{tabular}[b]{c c c}
			\multicolumn{2}{c}{\multirow{2}{*}[\rowArg]{
					\subfloat[Input~(LR)]
					% 			{\includegraphics[width = \fullSize, height = \fullSize]
					{\includegraphics[height=\fullSize]
						{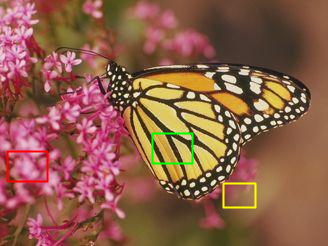}}}} &
			\subfloat[\centering  Ours ]{
				\includegraphics[width = \patchSize]
				{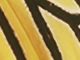}} 
                \hspace{1.25mm}
			\subfloat[\centering  Ours   ]{
				\includegraphics[width = \patchSize]
				{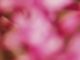}} 
                \hspace{1.25mm}
			\subfloat[\centering  Ours   ]{
				\includegraphics[width = \patchSize]
				{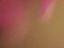}} \\ 												

			& & \subfloat[\centering   GT ]{
				\includegraphics[width = \patchSize]
				{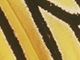}}
                \hspace{1.25mm}
			\subfloat[\centering   GT ]{
				\includegraphics[width = \patchSize]
    			{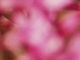}}
                \hspace{1.25mm}
			\subfloat[\centering  GT]{
				\includegraphics[width = \patchSize]
				{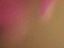}}
	\end{tabular}
  \vspace{-8mm}
\end{figure*}
 
\begin{figure*}[h]
    \centering
	\captionsetup[subfloat]{labelformat=empty,aboveskip=1pt}
	%\begin{center}
		\newcommand{\rowArg}{1.74cm}
		\newcommand{\fullSize}{4.50cm}
		\newcommand{\patchSize}{2.72cm}
		% 			\begin{adjustbox}{width=\linewidth, center=\linewidth}
		\setlength\tabcolsep{0.15cm}
		\begin{tabular}[b]{c c c}
			\multicolumn{2}{c}{\multirow{2}{*}[\rowArg]{
					\subfloat[Input~(LR)]
					{\includegraphics[height=\fullSize]
						{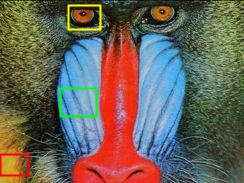}}}} &
			\subfloat[\centering  Ours  ]{
				\includegraphics[width = \patchSize]
				{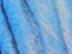}} 
                \hspace{1.25mm}
			\subfloat[\centering  Ours   ]{
				\includegraphics[width = \patchSize]
				{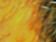}} 
                \hspace{1.25mm}
			\subfloat[\centering  Ours   ]{
				\includegraphics[width = \patchSize]
				{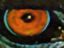}} \\

			& & \subfloat[\centering  GT ]{
				\includegraphics[width = \patchSize]
				{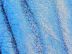}}
                \hspace{1.25mm}
			\subfloat[\centering   GT ]{
				\includegraphics[width = \patchSize]
				{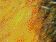}}
				\hspace{1.25mm}
			\subfloat[\centering  GT]{
				\includegraphics[width = \patchSize]
				{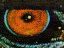}}
	\end{tabular}
 \vspace{-8mm}
\end{figure*}

\setcounter{figure}{2}
\begin{figure*}[h]
    \centering
	\captionsetup[subfloat]{labelformat=empty,aboveskip=1pt}
	%\begin{center}
		\newcommand{\rowArg}{1.74cm}
		\newcommand{\fullSize}{4.50cm}
		\newcommand{\patchSize}{2.72cm}
		% 			\begin{adjustbox}{width=\linewidth, center=\linewidth}
		\setlength\tabcolsep{0.15cm}
	%%%%%%%%%%%%%%%%%%%%%%%%%%%%%%%%%
		\begin{tabular}[b]{c c c}
			\multicolumn{2}{c}{\multirow{2}{*}[\rowArg]{
					\subfloat[Input~(LR)]
					{\includegraphics[height=\fullSize]
						{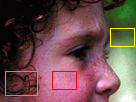}}}} &
			\subfloat[\centering  Ours ]{
				\includegraphics[width = \patchSize]
				{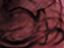}} 
                  \hspace{1.25mm}
			\subfloat[\centering  Ours   ]{
				\includegraphics[width = \patchSize]
				{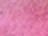}} 
                 \hspace{1.25mm}
			\subfloat[\centering  Ours   ]{
				\includegraphics[width = \patchSize]
				{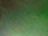}}  \\

			& & \subfloat[\centering   GT ]{
				\includegraphics[width = \patchSize]
				{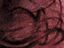}}
                 \hspace{1.25mm}
			\subfloat[\centering   GT ]{
				\includegraphics[width = \patchSize]
				{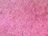}}
				\hspace{1.25mm}
			\subfloat[\centering  GT]{
				\includegraphics[width = \patchSize]
				{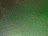}}
	\end{tabular}
  %\vspace{2mm}
    %\fi
	%\setlength{\abovecaptionskip}{0cm}
	%\vspace{-2mm}
%\captionsetup{justification=raggedright,singlelinecheck=false}
	\caption{\textbf{Visualization of noise-free super-resolved images on scale $\times 2$.} 
	}
	\label{fig:supp-noise}
	%\vspace{2mm}
\end{figure*}

%% file: sections/figures/rebuttal_degrade.tex
\renewcommand{\thefigure}{S4}
\begin{figure*}[h]
     %\vspace{-3mm}
     \centering
     %%%%%%%%%%%%%%%%%%%%%%%%%%% Airplane1 %%%%%%%%%%%%%%%%%%%%%%%%%%
     \begin{subfigure}[b]{0.22\textwidth}
         \centering
         \includegraphics[page=1, width=\textwidth]{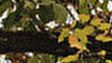}
     \end{subfigure}
     \hspace{0.1cm}
     \begin{subfigure}[b]{0.22\textwidth}
         \centering
         \includegraphics[page=1, width=\textwidth]{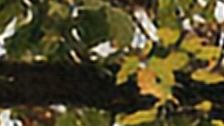}
     \end{subfigure}
     \hspace{0.5cm}
     \begin{subfigure}[b]{0.22\textwidth}
         \centering
         \includegraphics[page=1, width=\textwidth]{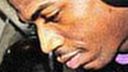}
     \end{subfigure} 
     \hspace{0.1cm}
     \begin{subfigure}[b]{0.22\textwidth}
         \centering
         \includegraphics[page=1,width=\textwidth]{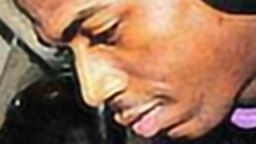}
     \end{subfigure}
     \vspace{0.5mm}
     
    %%%%%%%%%%%%%%%%%%%%%%%%%%% Cabinet %%%%%%%%%%%%%%%%%%%%%%%%%%
     \begin{subfigure}[b]{0.22\textwidth}
         \centering
         \includegraphics[page=1, width=\textwidth]{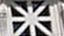}
         \caption*{LR}
     \end{subfigure}  
     \hspace{0.1cm}
     \begin{subfigure}[b]{0.22\textwidth}
         \centering
         \includegraphics[page=1, width=\textwidth]{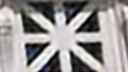}
         \caption*{SR}
     \end{subfigure}
     \hspace{0.5cm}   
     \begin{subfigure}[b]{0.22\textwidth}
         \centering
         \includegraphics[page=1, width=\textwidth]{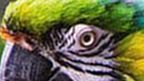}
         \caption*{LR}
     \end{subfigure}
     \hspace{0.1cm}
     \begin{subfigure}[b]{0.22\textwidth}
         \centering
         \includegraphics[page=1, width=\textwidth]{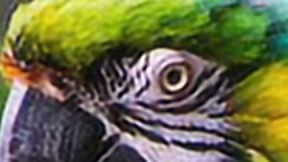}
         \caption*{SR}
     \end{subfigure}
     %\vspace{-5mm}
     \caption{\textbf{Visualization of SR performance on images with more complicated down-sampling degradations.} 
	}
        \label{fig:reb_degrade}
\end{figure*}     

%% file: sections/figures/supp/scales2.tex
\renewcommand{\thefigure}{S5}
\begin{figure*}
     \centering
     %%%%%%%%%%%%%%%%%%%%%%%%%%% Set14 %%%%%%%%%%%%%%%%%%%%%%%%%%
     \begin{subfigure}[b]{0.24\textwidth}
         \centering
         \includegraphics[width=\textwidth]{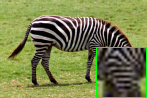}
     \end{subfigure}
     \hfill
     \begin{subfigure}[b]{0.24\textwidth}
         \centering
         \includegraphics[width=\textwidth]{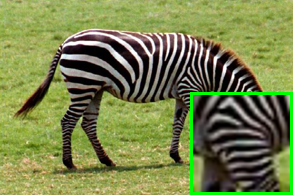}
     \end{subfigure}
     \hfill
     \begin{subfigure}[b]{0.24\textwidth}
         \centering
         \includegraphics[width=\textwidth]{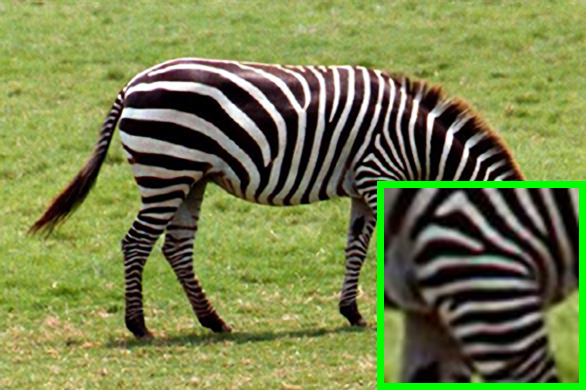}
     \end{subfigure}
     \hfill
     \begin{subfigure}[b]{0.24\textwidth}
         \centering
         \includegraphics[width=\textwidth]{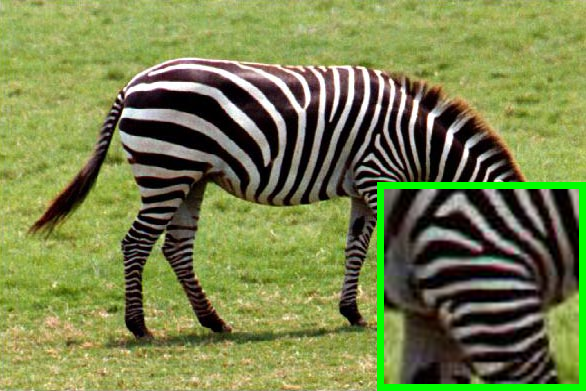}
     \end{subfigure}
     \\
    %%%%%%%%%%%%%%%%%%%%%%%%%%% B100 %%%%%%%%%%%%%%%%%%%%%%%%%%
     \begin{subfigure}[b]{0.24\textwidth}
         \centering
         \includegraphics[width=\textwidth]{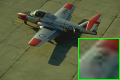}
     \end{subfigure}
     \hfill
     \begin{subfigure}[b]{0.24\textwidth}
         \centering
         \includegraphics[width=\textwidth]{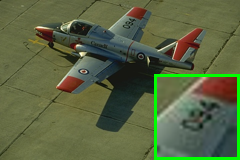}
     \end{subfigure}
     \hfill
     \begin{subfigure}[b]{0.24\textwidth}
         \centering
         \includegraphics[width=\textwidth]{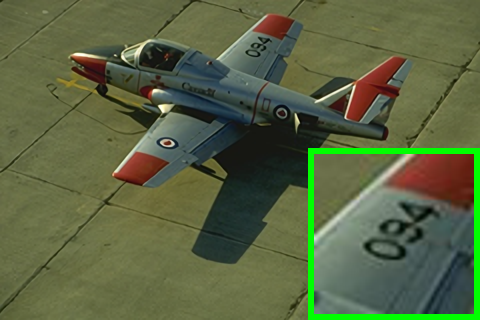}
     \end{subfigure}
     \hfill
     \begin{subfigure}[b]{0.24\textwidth}
         \centering
         \includegraphics[width=\textwidth]{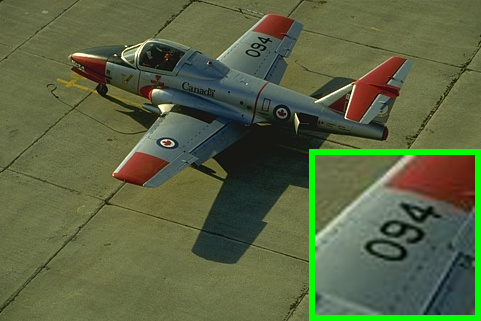}
     \end{subfigure}
     \\
     %%%%%%%%%%%%%%%%%%%%%%%%%%% Urban100 %%%%%%%%%%%%%%%%%%%%%%%%%%
     \begin{subfigure}[b]{0.24\textwidth}
         \centering
         \includegraphics[width=\textwidth]{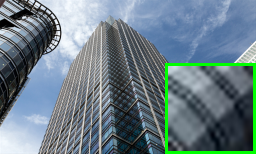}
     \end{subfigure}
     \hfill
     \begin{subfigure}[b]{0.24\textwidth}
         \centering
         \includegraphics[width=\textwidth]{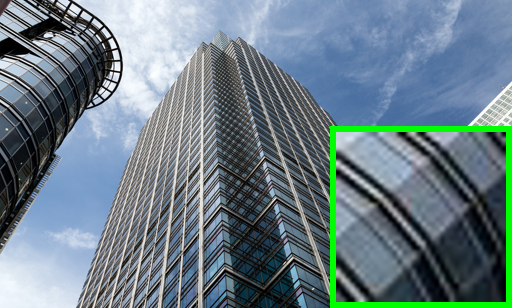}
     \end{subfigure}
     \hfill
     \begin{subfigure}[b]{0.24\textwidth}
         \centering
         \includegraphics[width=\textwidth]{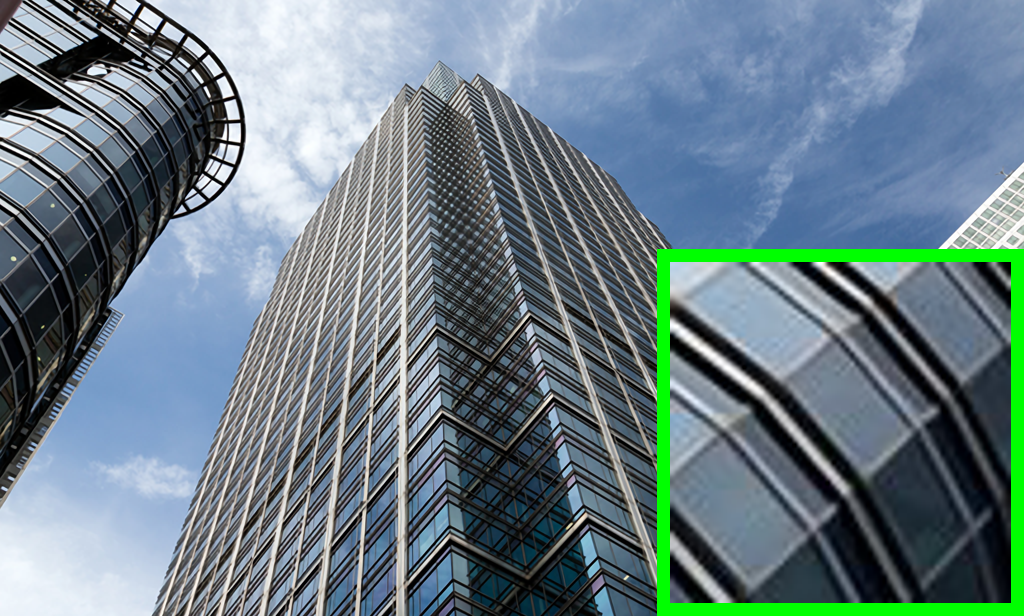}
     \end{subfigure}
     \hfill
     \begin{subfigure}[b]{0.24\textwidth}
         \centering
         \includegraphics[width=\textwidth]{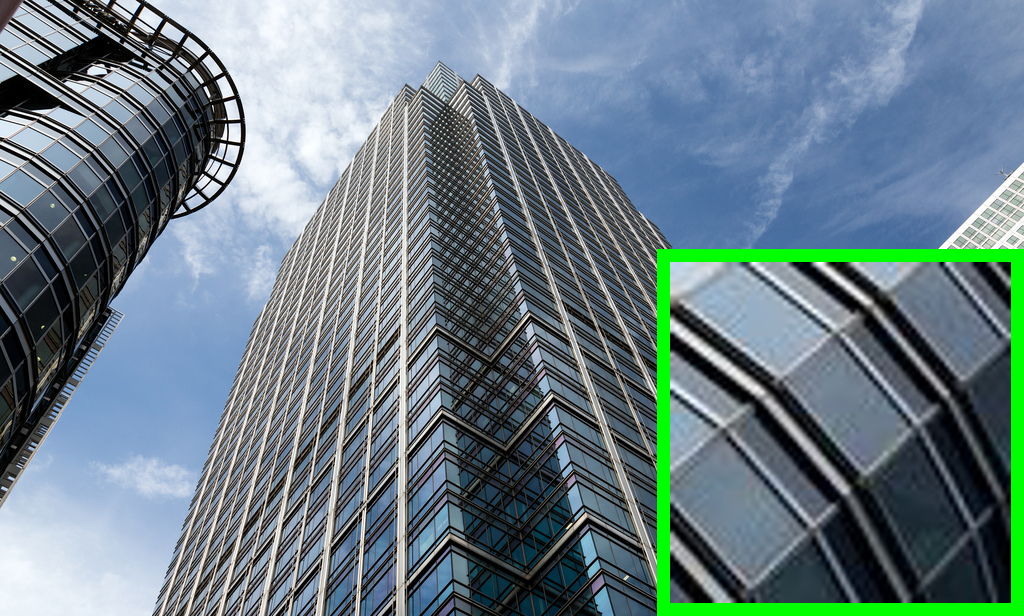}
     \end{subfigure}
     \\
     %%%%%%%%%%%%%%%%%%%%%%%%%%% RealSR %%%%%%%%%%%%%%%%%%%%%%%%%%
     \begin{subfigure}[b]{0.24\textwidth}
         \centering
         \includegraphics[width=\textwidth]{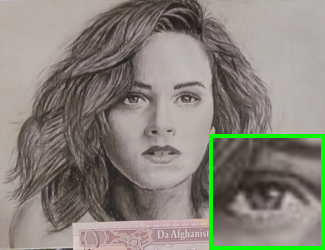}
         \caption*{LLR~(Ours)}
     \end{subfigure}
     \hfill
     \begin{subfigure}[b]{0.24\textwidth}
         \centering
         \includegraphics[width=\textwidth]{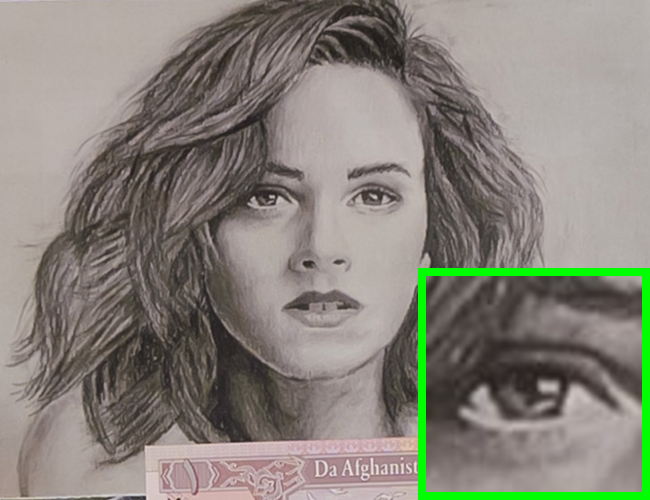}
         \caption*{LR~(Input)}
     \end{subfigure}
     \hfill
     \begin{subfigure}[b]{0.24\textwidth}
         \centering
         \includegraphics[width=\textwidth]{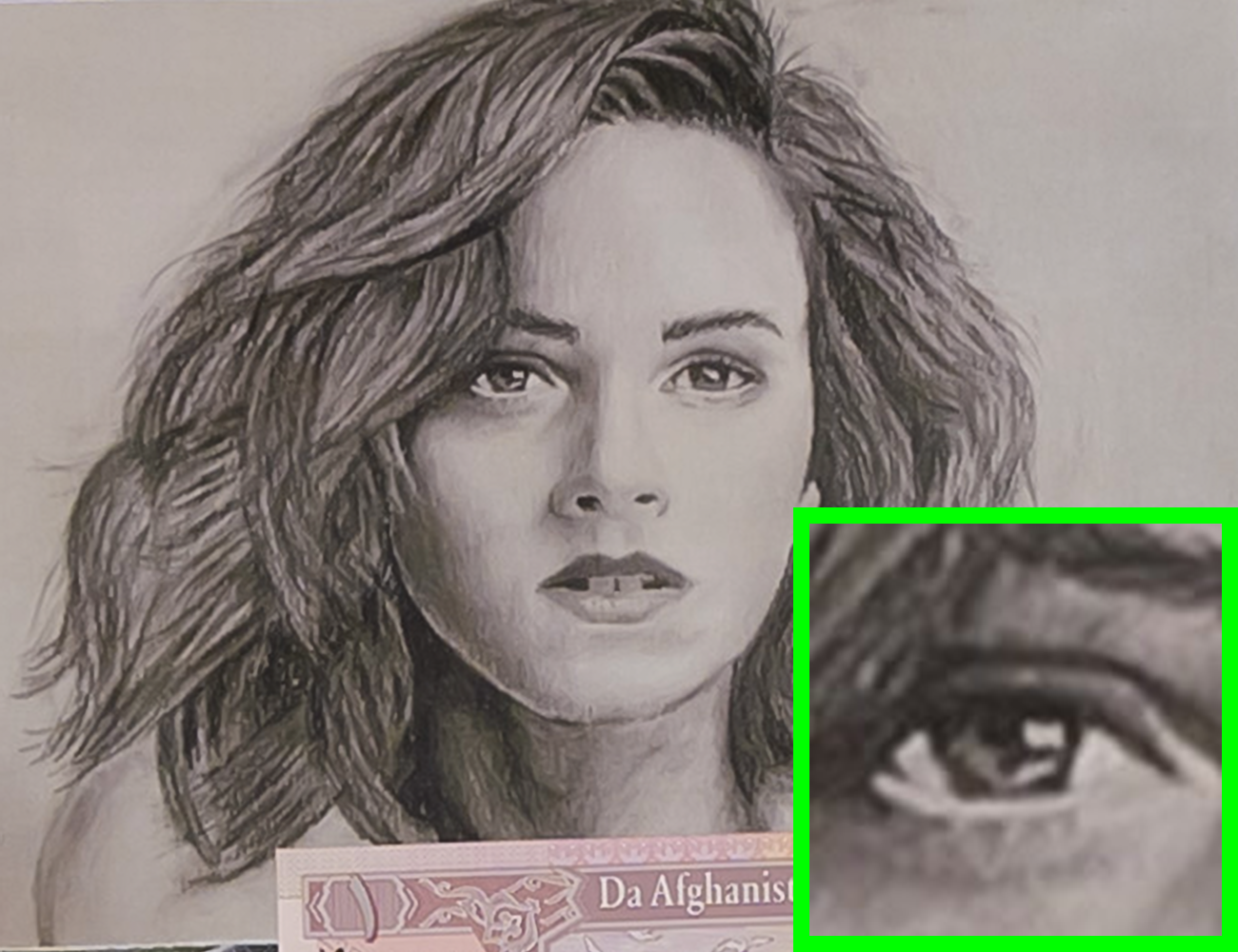}
         \caption*{SR~(Ours)}
     \end{subfigure}
     \hfill
     \begin{subfigure}[b]{0.24\textwidth}
         \centering
         \includegraphics[width=\textwidth]{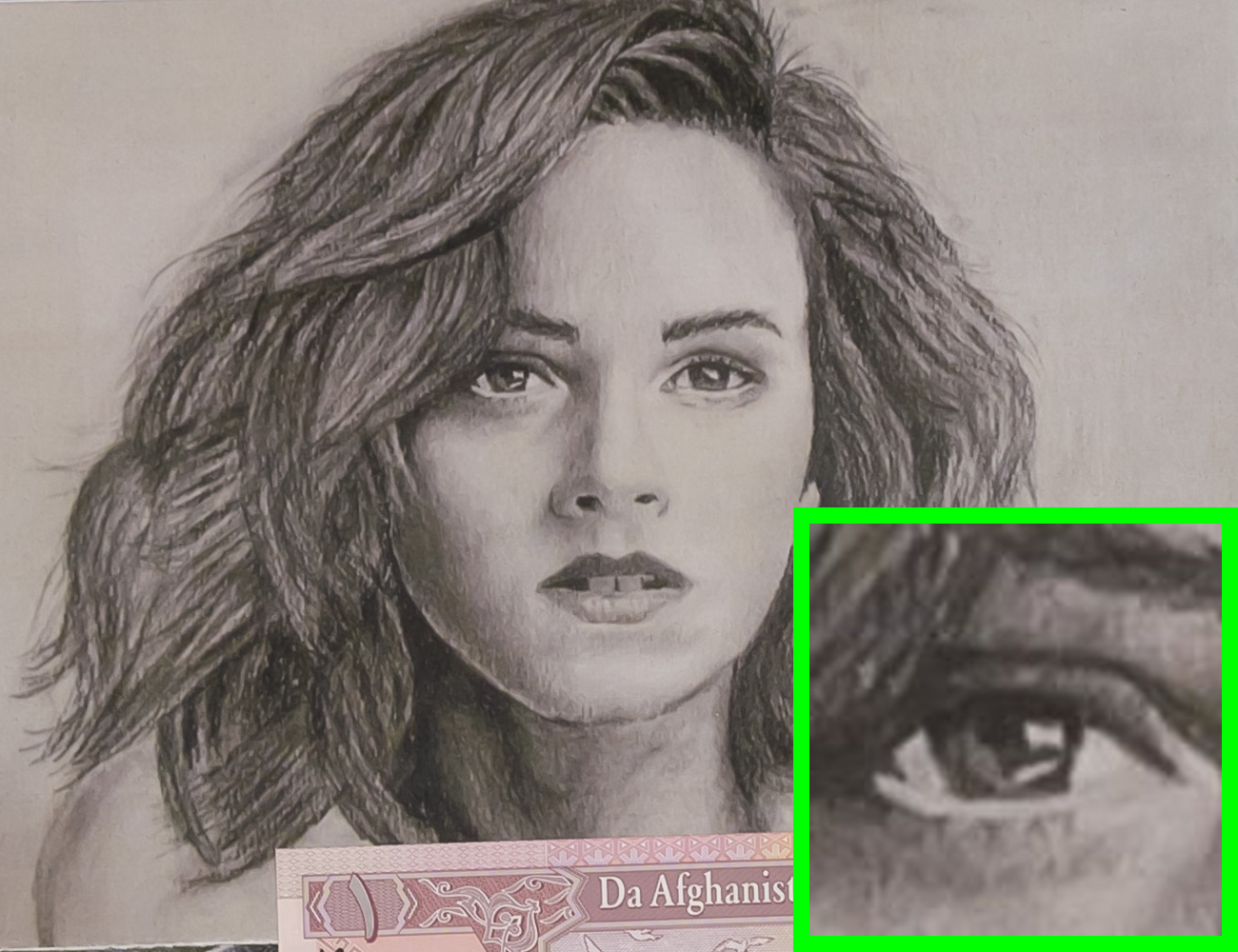}
         \caption*{GT~(HR)}
     \end{subfigure}
     \\
        \caption{
        \textbf{Qualitative comparisons of the generated images (LLR and SR) by ICF-SRSR for scale $\times2$.} 
        }
        \label{fig:supp_scales2}
\end{figure*}     

%% file: sections/figures/supp/scales4.tex
\renewcommand{\thefigure}{S6}
\begin{figure*}
     \centering
     %%%%%%%%%%%%%%%%%%%%%%%%%%% Set14 %%%%%%%%%%%%%%%%%%%%%%%%%%
     \begin{subfigure}[b]{0.24\textwidth}
         \centering
         \includegraphics[width=\textwidth]{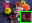}
     \end{subfigure}
     \hfill
     \begin{subfigure}[b]{0.24\textwidth}
         \centering
         \includegraphics[width=\textwidth]{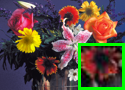}
     \end{subfigure}
     \hfill
     \begin{subfigure}[b]{0.24\textwidth}
         \centering
         \includegraphics[width=\textwidth]{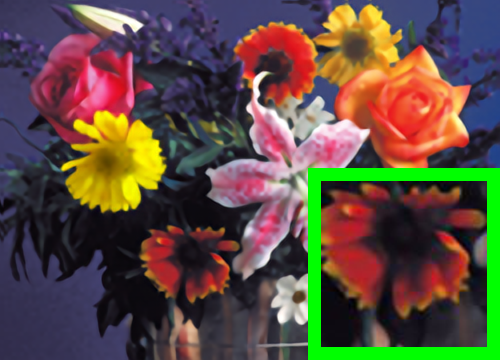}
     \end{subfigure}
     \hfill
     \begin{subfigure}[b]{0.24\textwidth}
         \centering
         \includegraphics[width=\textwidth]{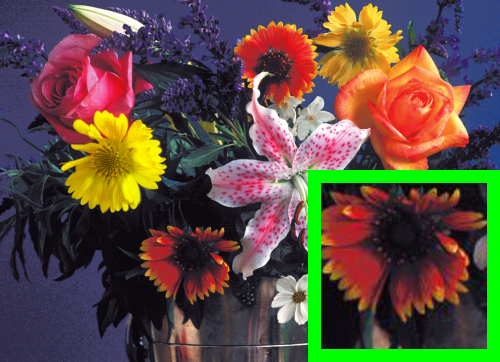}
     \end{subfigure}
     \\
    %%%%%%%%%%%%%%%%%%%%%%%%%%% B100 %%%%%%%%%%%%%%%%%%%%%%%%%%
     \begin{subfigure}[b]{0.24\textwidth}
         \centering
         \includegraphics[width=\textwidth]{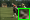}
     \end{subfigure}
     \hfill
     \begin{subfigure}[b]{0.24\textwidth}
         \centering
         \includegraphics[width=\textwidth]{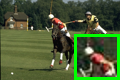}
     \end{subfigure}
     \hfill
     \begin{subfigure}[b]{0.24\textwidth}
         \centering
         \includegraphics[width=\textwidth]{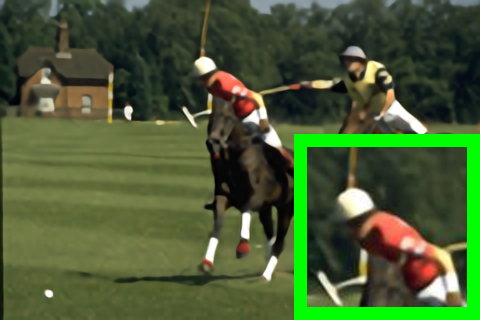}
     \end{subfigure}
     \hfill
     \begin{subfigure}[b]{0.24\textwidth}
         \centering
         \includegraphics[width=\textwidth]{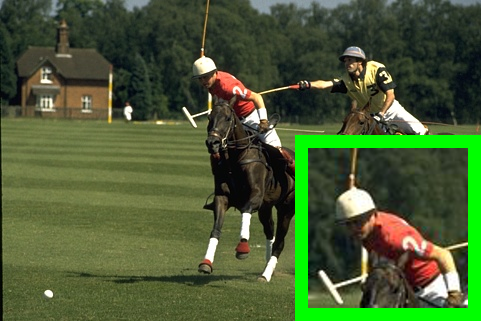}
     \end{subfigure}
     \\
     %%%%%%%%%%%%%%%%%%%%%%%%%%% Urban100 %%%%%%%%%%%%%%%%%%%%%%%%%%
     \begin{subfigure}[b]{0.24\textwidth}
         \centering
         \includegraphics[width=\textwidth]{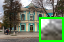}
     \end{subfigure}
     \hfill
     \begin{subfigure}[b]{0.24\textwidth}
         \centering
         \includegraphics[width=\textwidth]{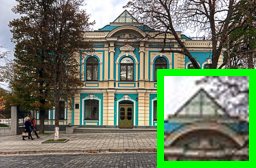}
     \end{subfigure}
     \hfill
     \begin{subfigure}[b]{0.24\textwidth}
         \centering
         \includegraphics[width=\textwidth]{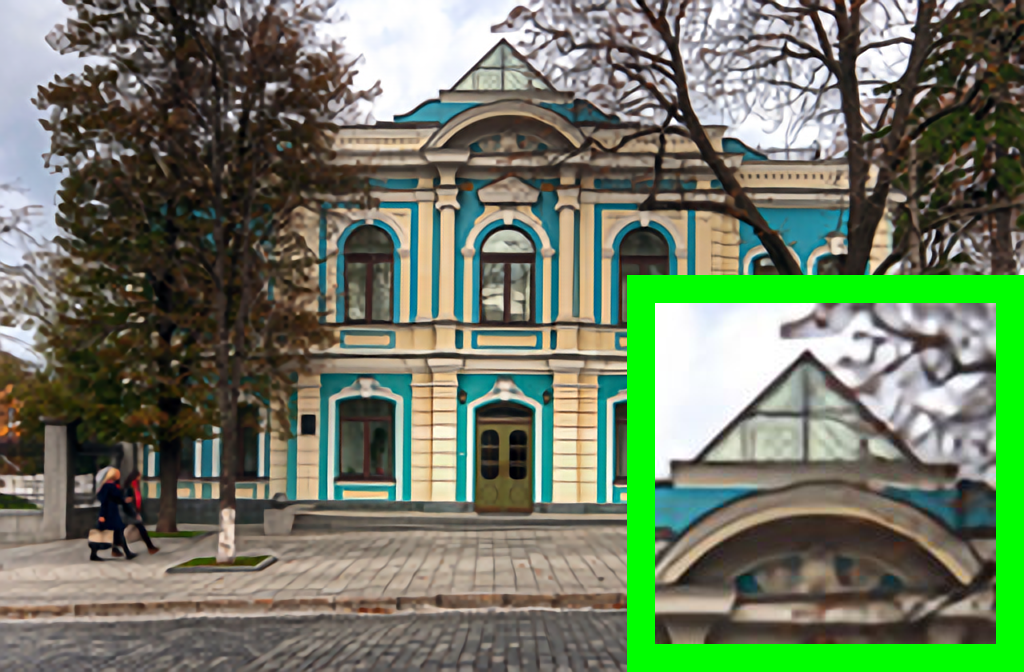}
     \end{subfigure}
     \hfill
     \begin{subfigure}[b]{0.24\textwidth}
         \centering
         \includegraphics[width=\textwidth]{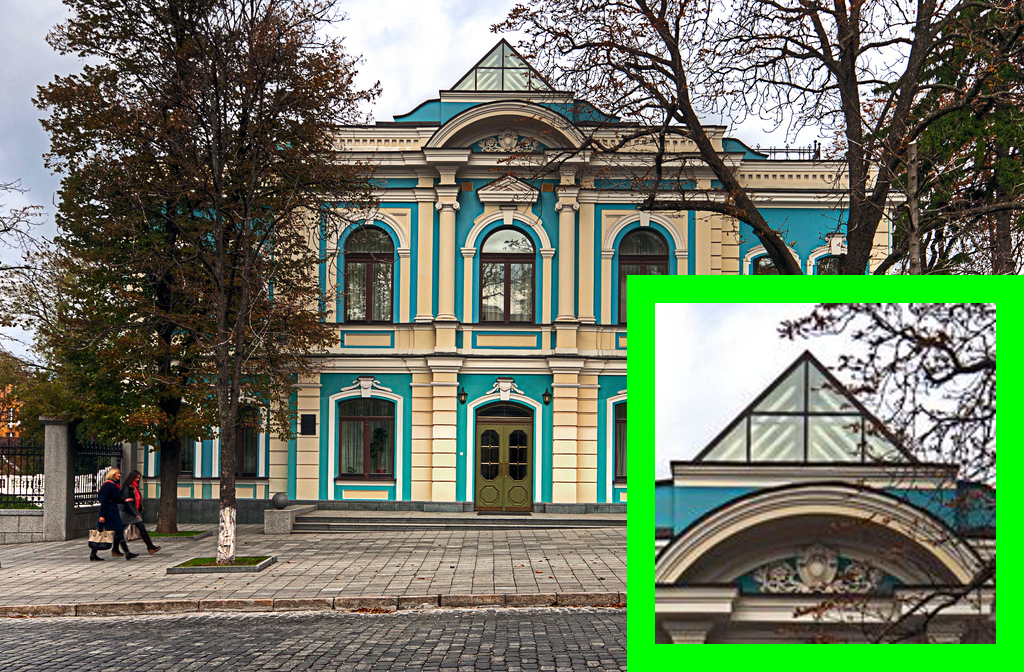}
     \end{subfigure}
     \\
     %%%%%%%%%%%%%%%%%%%%%%%%%%% RealSR %%%%%%%%%%%%%%%%%%%%%%%%%%
     \begin{subfigure}[b]{0.24\textwidth}
         \centering
         \includegraphics[trim={0 0cm 0 0},clip, width=\textwidth]{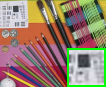}
         \caption*{LLR~(Ours)}
     \end{subfigure}
     \hfill
     \begin{subfigure}[b]{0.24\textwidth}
         \centering
         \includegraphics[trim={0 0 0 0},clip, width=\textwidth]{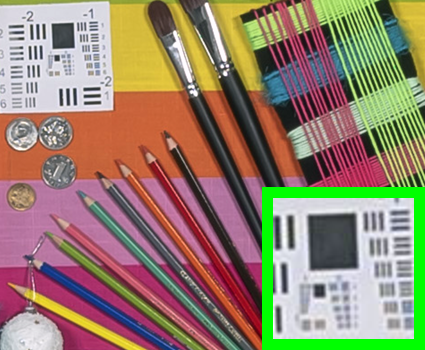}
         \caption*{LR~(Input)}
     \end{subfigure}
     \hfill
     \begin{subfigure}[b]{0.24\textwidth}
         \centering
         \includegraphics[trim={0 0 0 0},clip, width=\textwidth]{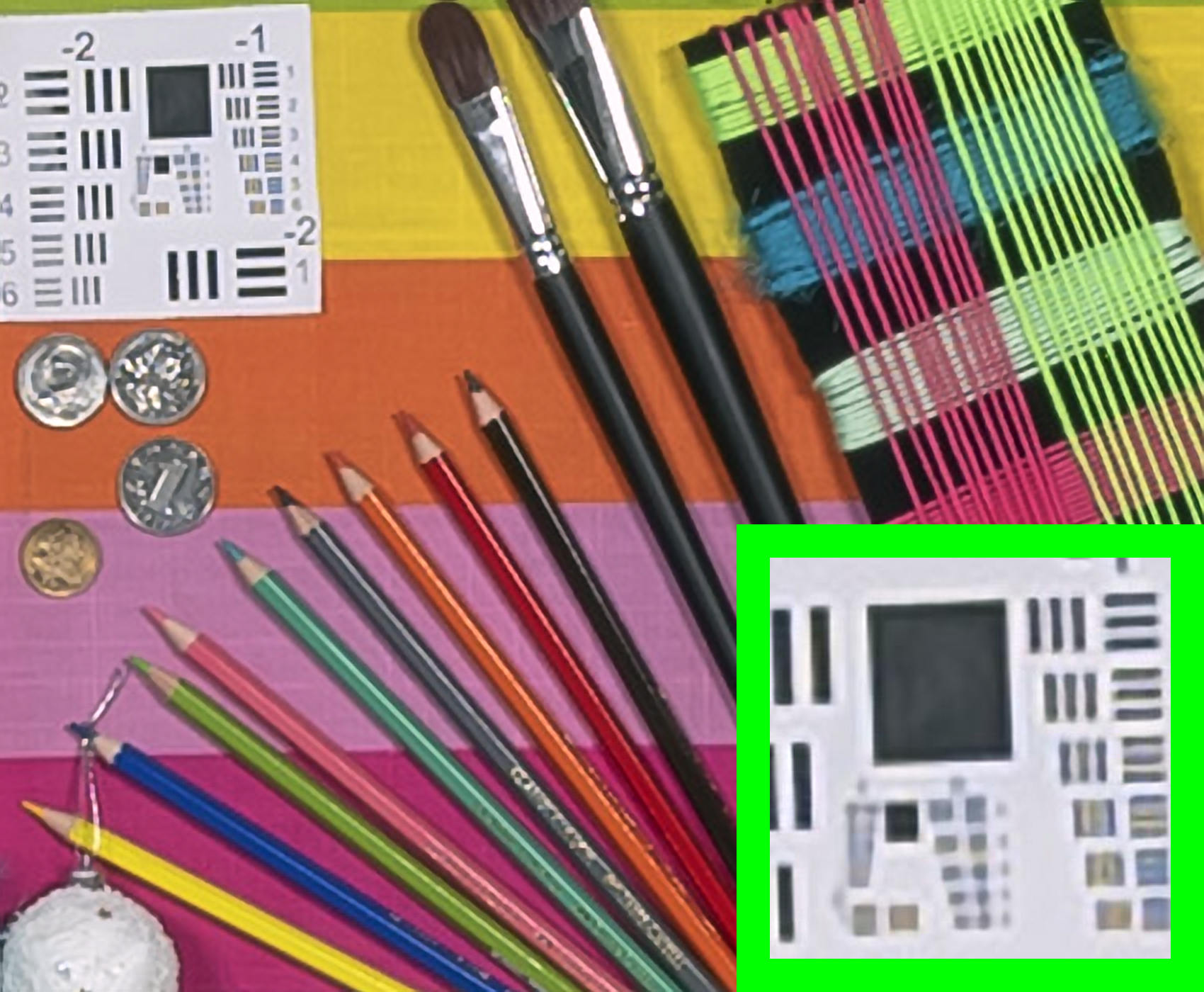}
         \caption*{SR~(Ours)}
     \end{subfigure}
     \hfill
     \begin{subfigure}[b]{0.24\textwidth}
         \centering
         \includegraphics[trim={0 0 0 0},clip, width=\textwidth]{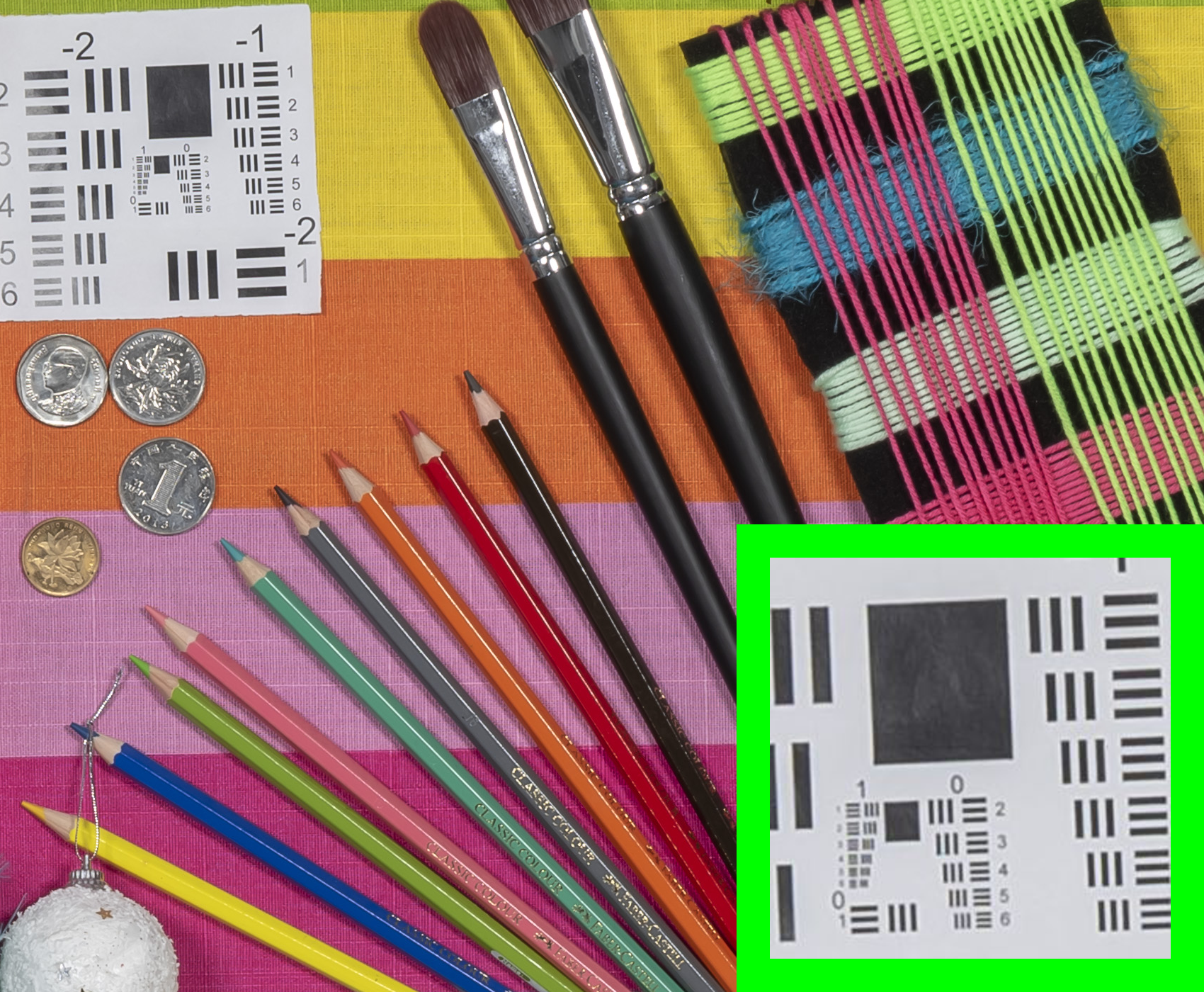}
         \caption*{GT~(HR)}
     \end{subfigure}
     \\
        \caption{
        \textbf{Qualitative comparisons of the generated images (LLR and SR) by ICF-SRSR for scale $\times4$.} 
        }
        \label{fig:supp_scales4}
\end{figure*}     

%% file: sections/figures/rebuttal_LRHR.tex
\renewcommand{\thefigure}{S7}
\begin{figure*}[t]
     %\vspace{-2mm}
     \centering
     %%%%%%%%%%%%%%%%%%%%%%%%%%% Airplane1 %%%%%%%%%%%%%%%%%%%%%%%%%%
     \begin{subfigure}[b]{0.15\textwidth}
         \centering
         \includegraphics[page=1, width=\textwidth]{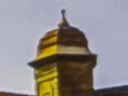}
     \end{subfigure}
     \hspace{0.01cm}
     \begin{subfigure}[b]{0.15\textwidth}
         \centering
         \includegraphics[page=1, width=\textwidth]{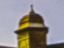}
     \end{subfigure}
     \hspace{0.01cm}
     \begin{subfigure}[b]{0.15\textwidth}
         \centering
         \includegraphics[page=1, width=\textwidth]{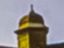}
     \end{subfigure} 
     \hspace{0.05cm}
     \begin{subfigure}[b]{0.15\textwidth}
         \centering
         \includegraphics[page=1,width=\textwidth]{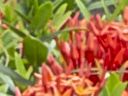}
     \end{subfigure}
          \hspace{0.01cm}
     \begin{subfigure}[b]{0.15\textwidth}
         \centering
         \includegraphics[page=1, width=\textwidth]{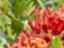}
     \end{subfigure} 
     \hspace{0.01cm}
     \begin{subfigure}[b]{0.15\textwidth}
         \centering
         \includegraphics[page=1,width=\textwidth]{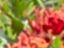}
     \end{subfigure}
     %\vspace{0.5mm}
    \\
    \vspace{0.5mm}
    %%%%%%%%%%%%%%%%%%%%%%%%%%% Cabinet %%%%%%%%%%%%%%%%%%%%%%%%%%
     \begin{subfigure}[b]{0.15\textwidth}
         \centering
         \includegraphics[page=1, width=\textwidth]{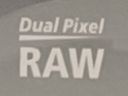}
         \caption*{HR}
     \end{subfigure}  
     \hspace{0.01cm}
     \begin{subfigure}[b]{0.15\textwidth}
         \centering
         \includegraphics[page=1, width=\textwidth]{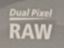}
         \caption*{Our LR}
     \end{subfigure}
     \hspace{0.01cm}   
     \begin{subfigure}[b]{0.15\textwidth}
         \centering
         \includegraphics[page=1, width=\textwidth]{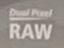}
         \caption*{Real LR}
     \end{subfigure}
     \hspace{0.05cm}
     \begin{subfigure}[b]{0.15\textwidth}
         \centering
         \includegraphics[page=1, width=\textwidth]{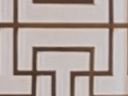}
         \caption*{HR}
     \end{subfigure}
          \hspace{0.01cm}   
     \begin{subfigure}[b]{0.15\textwidth}
         \centering
         \includegraphics[page=1, width=\textwidth]{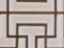}
         \caption*{Our LR}
     \end{subfigure}
     \hspace{0.01cm}
     \begin{subfigure}[b]{0.15\textwidth}
         \centering
         \includegraphics[page=1, width=\textwidth]{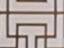}
         \caption*{Real LR}
     \end{subfigure}
     %\vspace{-5mm}
    \caption{\textbf{Qualitative comparisons of the real LR images and our generated LR images given HR images.}}
        \label{fig:reb_LRHR}
\end{figure*}     

%% file: sections/figures/supp/single_image.tex
\renewcommand{\thefigure}{S8}
\begin{figure*}[h]
    \centering
	\captionsetup[subfloat]{labelformat=empty,aboveskip=1pt}
	%\begin{center}
		\newcommand{\rowArg}{1.95cm}
		\newcommand{\fullSize}{4.92cm}
		\newcommand{\patchSize}{3.0cm}
		% 			\begin{adjustbox}{width=\linewidth, center=\linewidth}
		\setlength\tabcolsep{0.0cm}
		\begin{tabular}[b]{c c c}
			\multicolumn{2}{c}{\multirow{2}{*}[\rowArg]{
					\subfloat[Input~(LR)]
					% 			{\includegraphics[width = \fullSize, height = \fullSize]
					{\includegraphics[trim={0 0 375pt 0},clip, height=\fullSize]
						{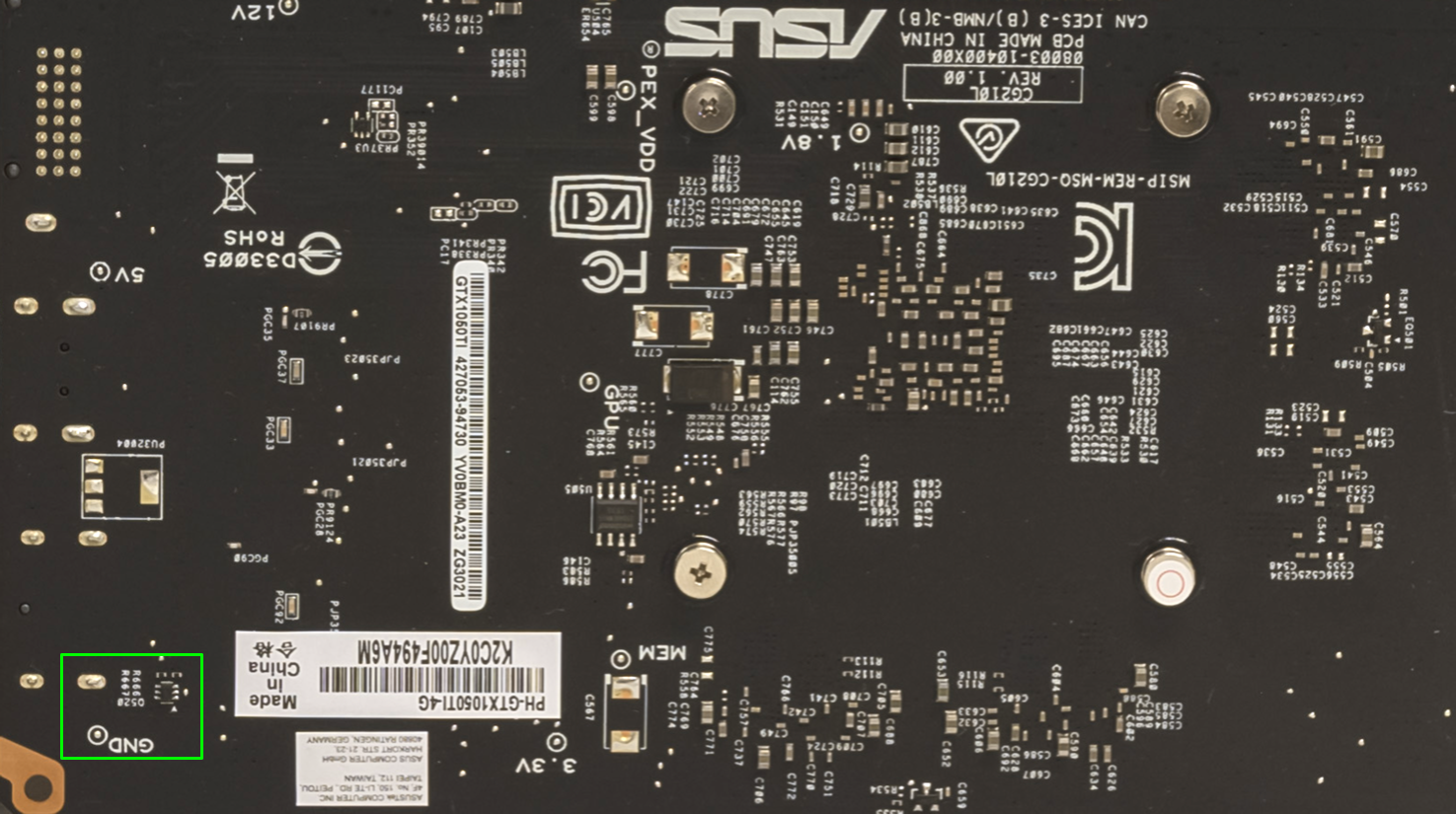}}}} & 
			\hspace{0.5cm}
			\subfloat[\centering  Input~(LR) ]{
				\includegraphics[width = \patchSize]
				{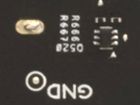}} \hspace{0.25cm}
			\subfloat[\centering  GT~(HR)  ]{
			    \includegraphics[width = \patchSize]
				{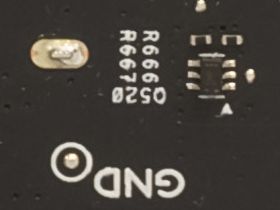}} \\  & &
												
            \hspace{0.5cm}
			\subfloat[\centering   Single~(29.82dB)]{
				\includegraphics[width = \patchSize]
				{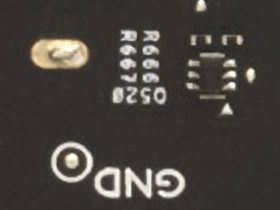}} \hspace{0.25cm}
				%\vspace{2.25mm}
			\subfloat[\centering  Multi~(\textbf{29.91dB})]{
				\includegraphics[width = \patchSize]
				{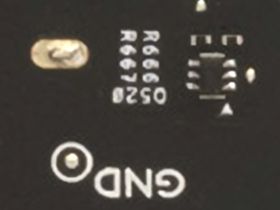}}
	\end{tabular}
	\\
	%%%%%%%%%%%%%%%%%%%%%%%%%%%%%%%%%
	\begin{tabular}[b]{c c c}
			\multicolumn{2}{c}{\multirow{2}{*}[\rowArg]{
					\subfloat[Input~(LR)]
					% 			{\includegraphics[width = \fullSize, height = \fullSize]
					{\includegraphics[trim={0 375pt 0 0},clip, height=\fullSize]
						{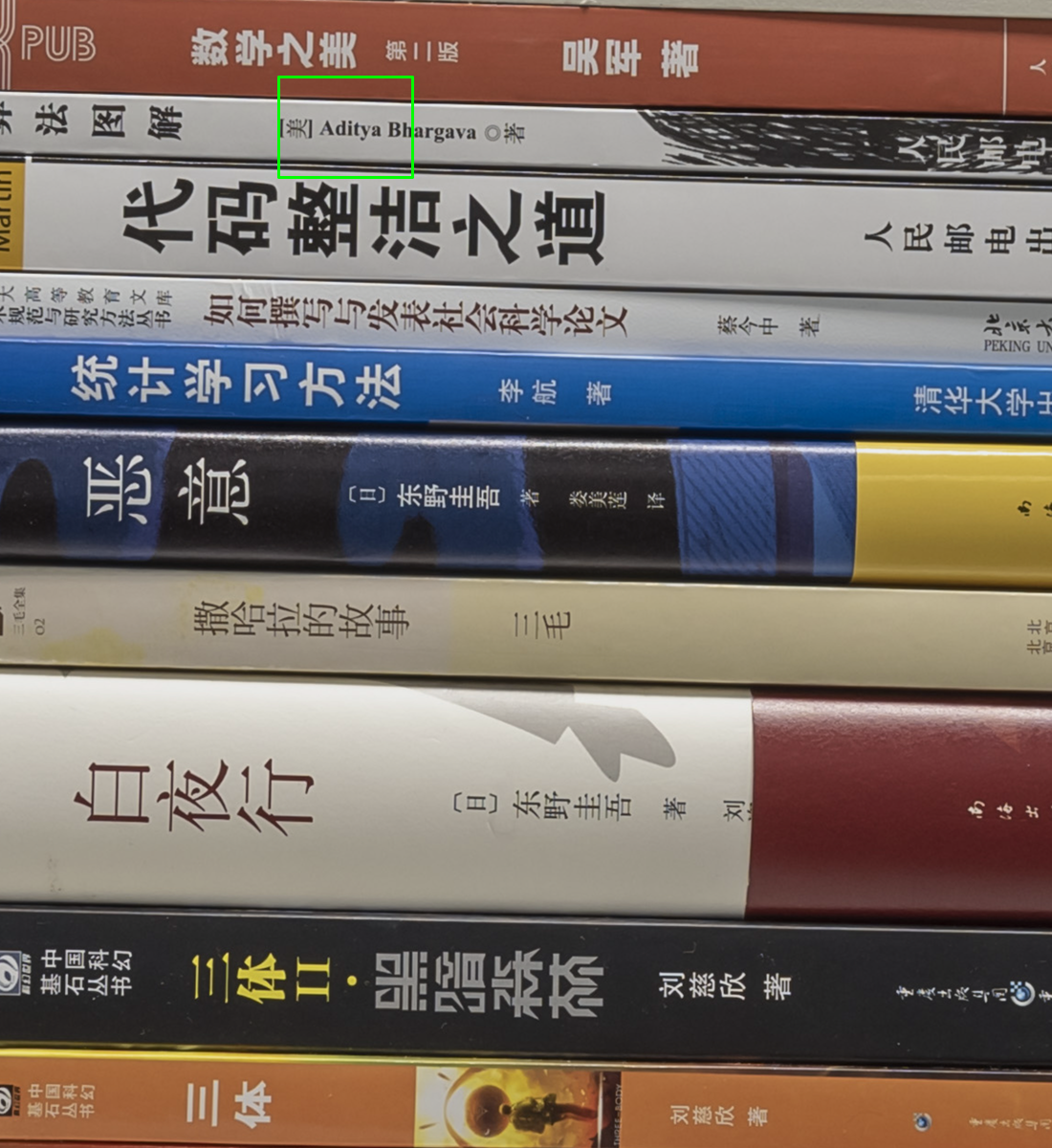}}}} &
			\hspace{0.5cm}
			\subfloat[\centering  Input~(LR) ]{
				\includegraphics[width = \patchSize]
				{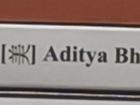}} \hspace{0.25cm}
			\subfloat[\centering  GT~(HR)]{
				\includegraphics[width = \patchSize]
				{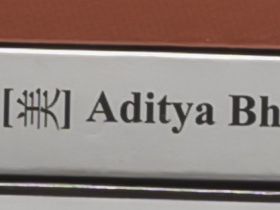}} \\  & &

			\hspace{0.5cm}
			\subfloat[\centering   Single~(\textbf{32.03dB})]{
				\includegraphics[width = \patchSize]
				{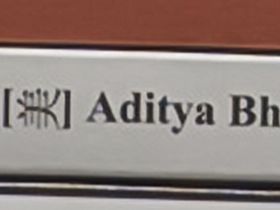}} \hspace{0.25cm}
				%\vspace{2.25mm}
			\subfloat[\centering  Multi~(31.89dB)]{
				\includegraphics[width = \patchSize]
				{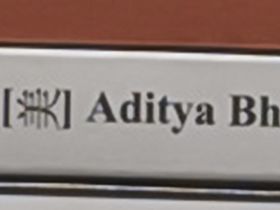}}
	\end{tabular}
	\\
	\begin{tabular}[b]{c c c}
			\multicolumn{2}{c}{\multirow{2}{*}[\rowArg]{
					\subfloat[Input~(LR)]
					% 			{\includegraphics[width = \fullSize, height = \fullSize]
					{\includegraphics[trim={0 475pt 0 0},clip, height=\fullSize]
						{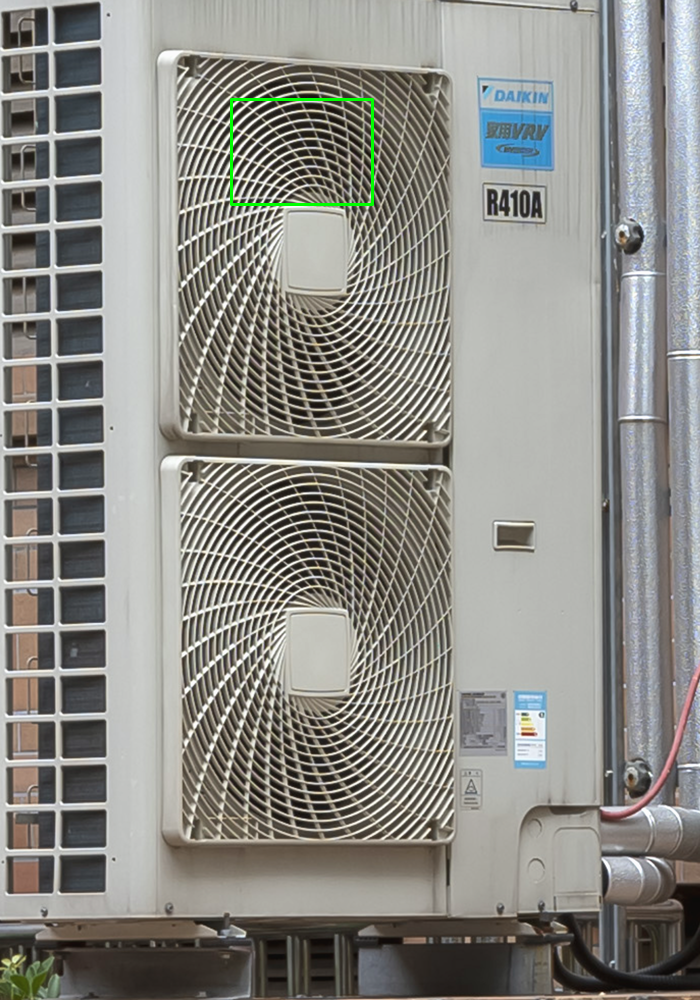}}}} &
			\hspace{0.5cm}
			\subfloat[\centering  Input~(LR) ]{
				\includegraphics[width = \patchSize]
				{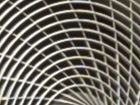}} \hspace{0.25cm}
			\subfloat[\centering  GT~(HR)  ]{
				\includegraphics[width = \patchSize]
				{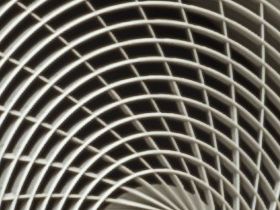}} \\  & &

			\hspace{0.5cm}
			\subfloat[\centering   Single~(\textbf{28.19dB})]{
				\includegraphics[width = \patchSize]
				{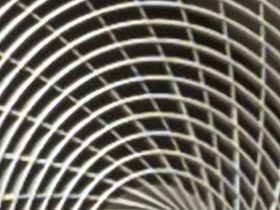}} \hspace{0.25cm}
				%\vspace{2.25mm}
			\subfloat[\centering Multi~(28.15dB)]{
				\includegraphics[width = \patchSize]
				{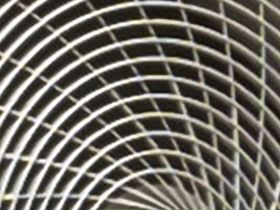}}
	\end{tabular}
	\\
	%%%%%%%%%%%%%%%%%%%%%%%%%%%%%%%%%
	\begin{tabular}[b]{c c c}
			\multicolumn{2}{c}{\multirow{2}{*}[\rowArg]{
					\subfloat[Input~(LR)]
					% 			{\includegraphics[width = \fullSize, height = \fullSize]
					{\includegraphics[height=\fullSize]
						{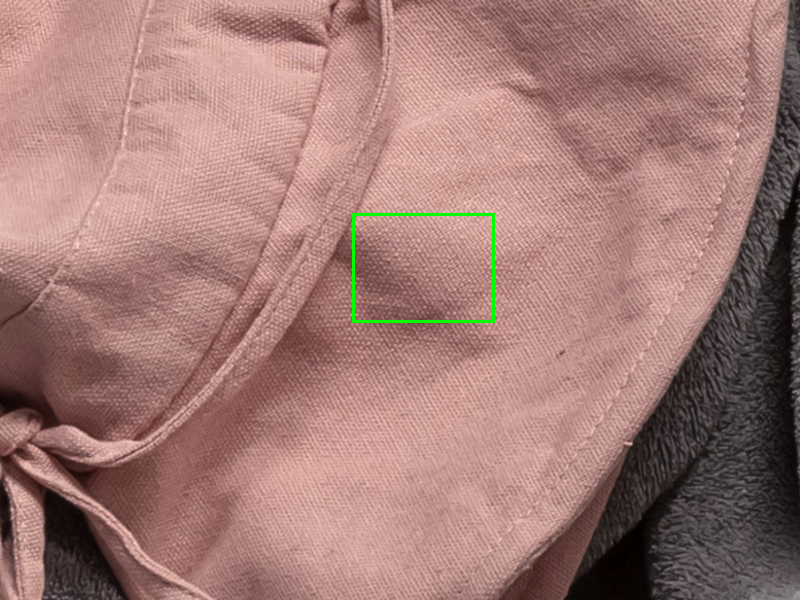}}}} &
			\hspace{0.5cm}
			\subfloat[\centering  Input~(LR) ]{
				\includegraphics[width = \patchSize]
				{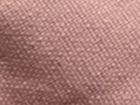}} \hspace{0.25cm}
			\subfloat[\centering  GT~(HR)]{
				\includegraphics[width = \patchSize]
				{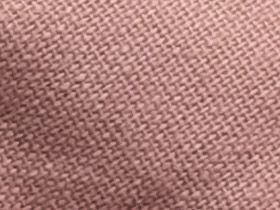}} \\  & &
			\hspace{0.5cm}
			\subfloat[\centering   Single~(\textbf{27.15dB})]{
				\includegraphics[width = \patchSize]
				{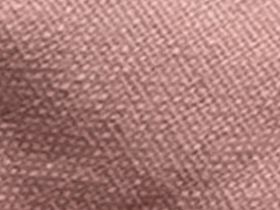}} \hspace{0.25cm}
				%\vspace{2.25mm}
			\subfloat[\centering Multi~(26.81dB)]{
				\includegraphics[width = \patchSize]
				{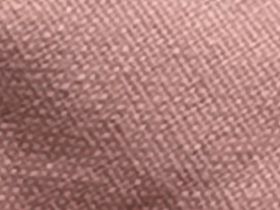}}
	\end{tabular}
	%\setlength{\abovecaptionskip}{0cm}
	%\vspace{-2mm}
	%\captionsetup{justification=raggedright,singlelinecheck=false}
	
	\caption{\textbf{Qualitative SR comparisons on single and multiple training images for scale $\times2$.}
	}
	\label{fig:supp_single}
	%\vspace{2mm}
\end{figure*}

%% file: sections/figures/supp/real_single.tex
\renewcommand{\thefigure}{S9}
\begin{figure*}[h]
    \centering
	\captionsetup[subfloat]{labelformat=empty,aboveskip=1pt}
	%\begin{center}
		\newcommand{\rowArg}{1.95cm}
		\newcommand{\fullSize}{4.92cm}
		\newcommand{\patchSize}{3.0cm}
		% 			\begin{adjustbox}{width=\linewidth, center=\linewidth}
		\setlength\tabcolsep{0.0cm}
		\begin{tabular}[b]{c c c}
			\multicolumn{2}{c}{\multirow{2}{*}[\rowArg]{
					\subfloat[Input~(LR)]
					% 			{\includegraphics[width = \fullSize, height = \fullSize]
					{\includegraphics[height=\fullSize]
						{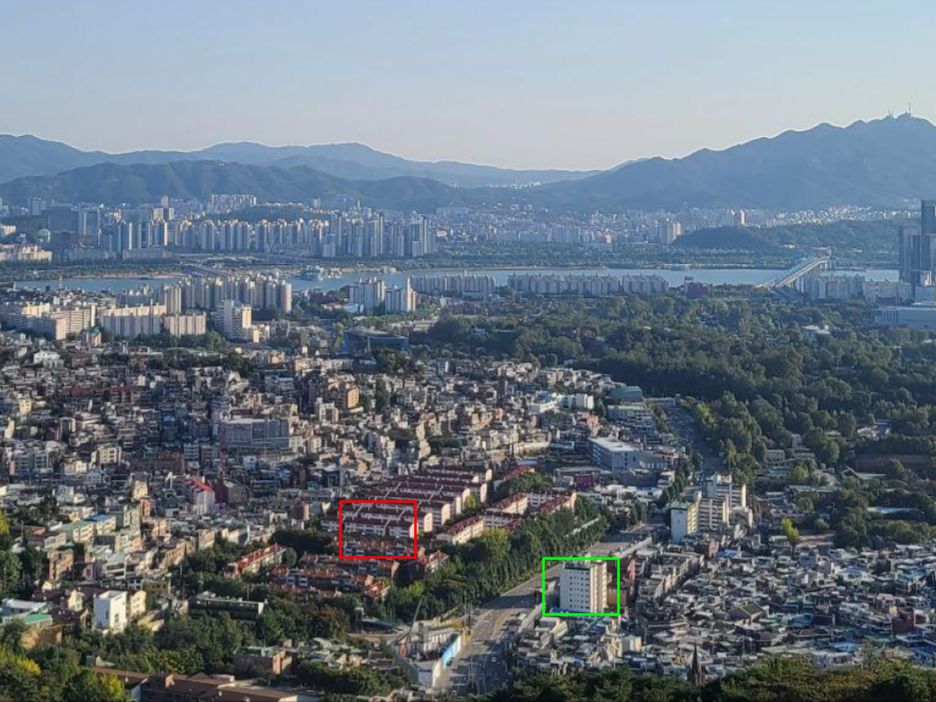}}}} & 
			\hspace{0.5cm}
			\subfloat[\centering  Input~(LR) ]{
				\includegraphics[width = \patchSize]
				{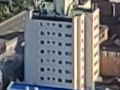}} \hspace{0.25cm}
			\subfloat[\centering  SR~(Ours)  ]{
			    \includegraphics[width = \patchSize]
				{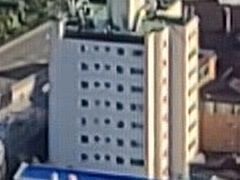}} \\  & &
												
            \hspace{0.5cm}
			\subfloat[\centering   Input~(LR)]{
				\includegraphics[width = \patchSize]
				{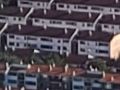}} \hspace{0.25cm}
				%\vspace{2.25mm}
			\subfloat[\centering  SR~(Ours)]{
				\includegraphics[width = \patchSize]
				{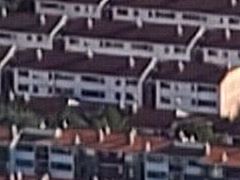}}
	\end{tabular}
	\\
	%%%%%%%%%%%%%%%%%%%%%%%%%%%%%%%%%
	\begin{tabular}[b]{c c c}
			\multicolumn{2}{c}{\multirow{2}{*}[\rowArg]{
					\subfloat[Input~(LR)]
					% 			{\includegraphics[width = \fullSize, height = \fullSize]
					{\includegraphics[ height=\fullSize]
						{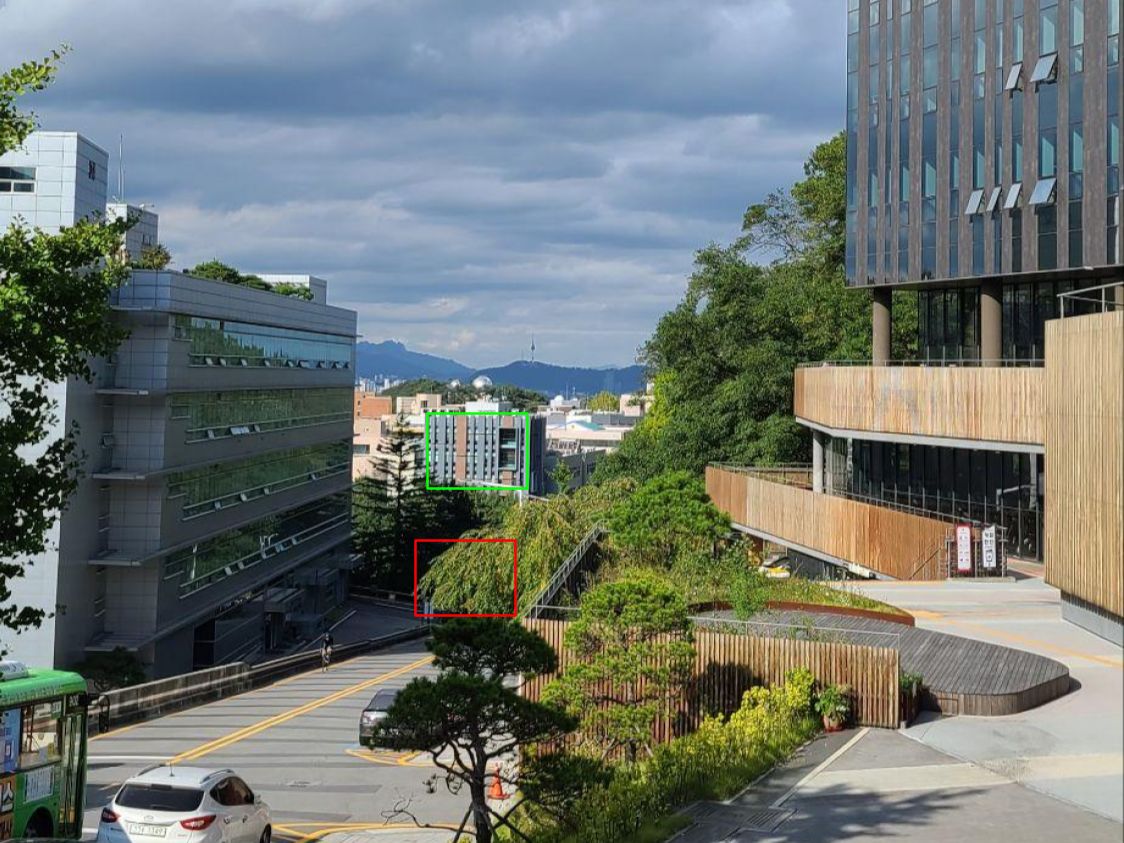}}}} &
			\hspace{0.5cm}
			\subfloat[\centering  Input~(LR) ]{
				\includegraphics[width = \patchSize]
				{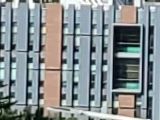}} \hspace{0.25cm}
			\subfloat[\centering  SR~(Ours)]{
				\includegraphics[width = \patchSize]
				{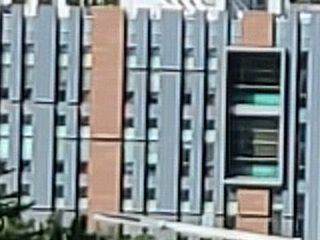}} \\  & &

			\hspace{0.5cm}
			\subfloat[\centering   Input~(LR)]{
				\includegraphics[width = \patchSize]
				{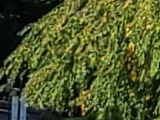}} \hspace{0.25cm}
				%\vspace{2.25mm}
			\subfloat[\centering  SR~(Ours)]{
				\includegraphics[width = \patchSize]
				{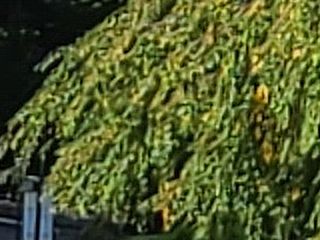}}
	\end{tabular}
	\\
	\begin{tabular}[b]{c c c}
			\multicolumn{2}{c}{\multirow{2}{*}[\rowArg]{
					\subfloat[Input~(LR)]
					% 			{\includegraphics[width = \fullSize, height = \fullSize]
					{\includegraphics[height=\fullSize]
						{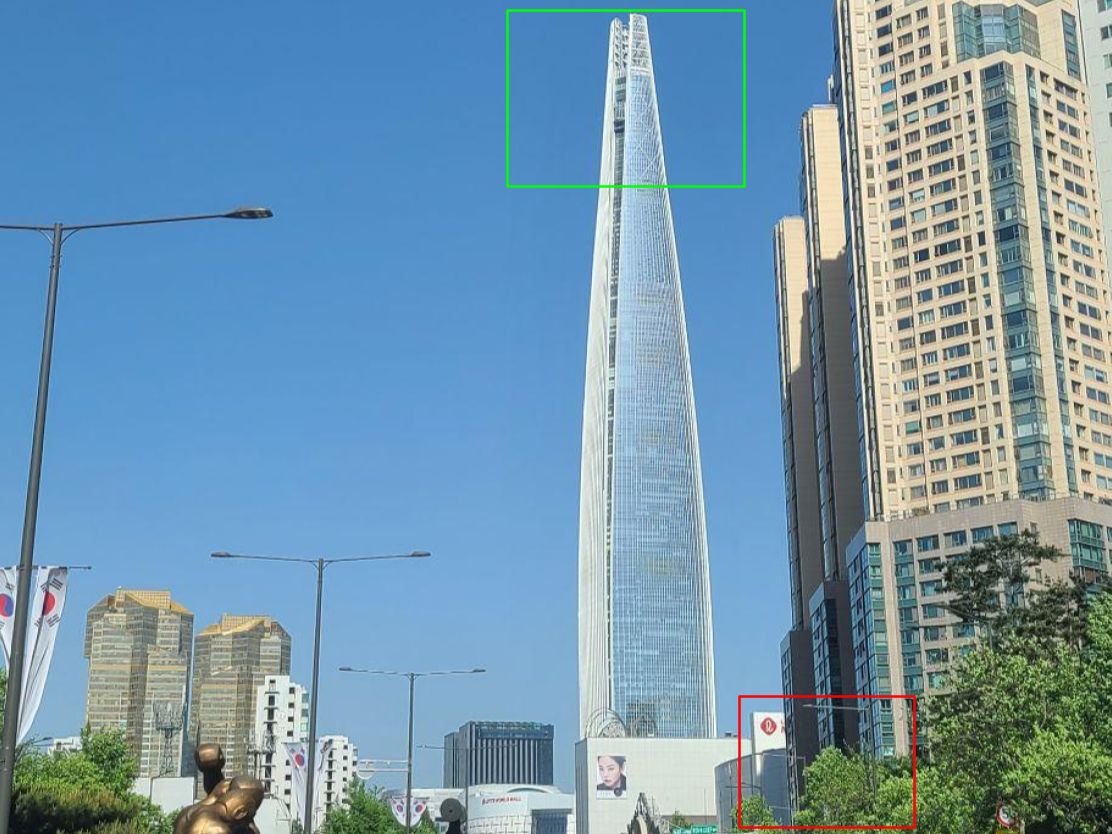}}}} &
			\hspace{0.5cm}
			\subfloat[\centering  Input~(LR) ]{
				\includegraphics[width = \patchSize]
				{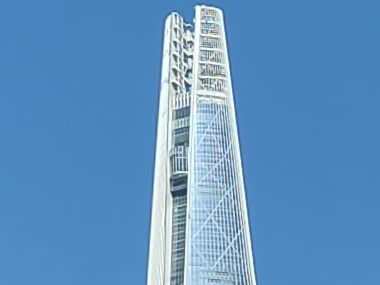}} \hspace{0.25cm}
			\subfloat[\centering  SR~(Ours)  ]{
				\includegraphics[width = \patchSize]
				{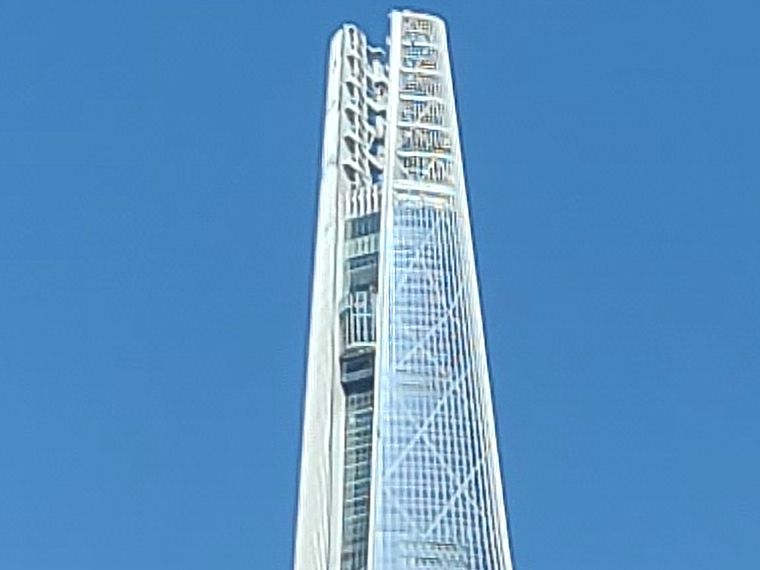}} \\  & &

			\hspace{0.5cm}
			\subfloat[\centering   Input~(LR)]{
				\includegraphics[width = \patchSize]
				{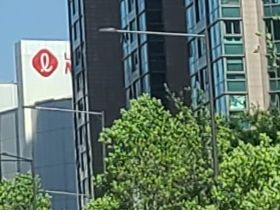}} \hspace{0.25cm}
				%\vspace{2.25mm}
			\subfloat[\centering SR~(Ours)]{
				\includegraphics[width = \patchSize]
				{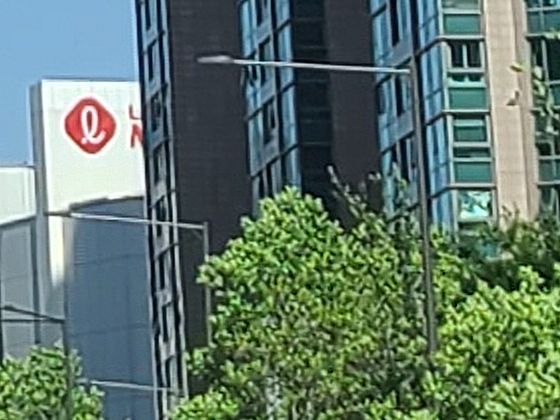}}
	\end{tabular}
	\\
	%%%%%%%%%%%%%%%%%%%%%%%%%%%%%%%%%
	\begin{tabular}[b]{c c c}
			\multicolumn{2}{c}{\multirow{2}{*}[\rowArg]{
					\subfloat[Input~(LR)]
					% 			{\includegraphics[width = \fullSize, height = \fullSize]
					{\includegraphics[height=\fullSize]
						{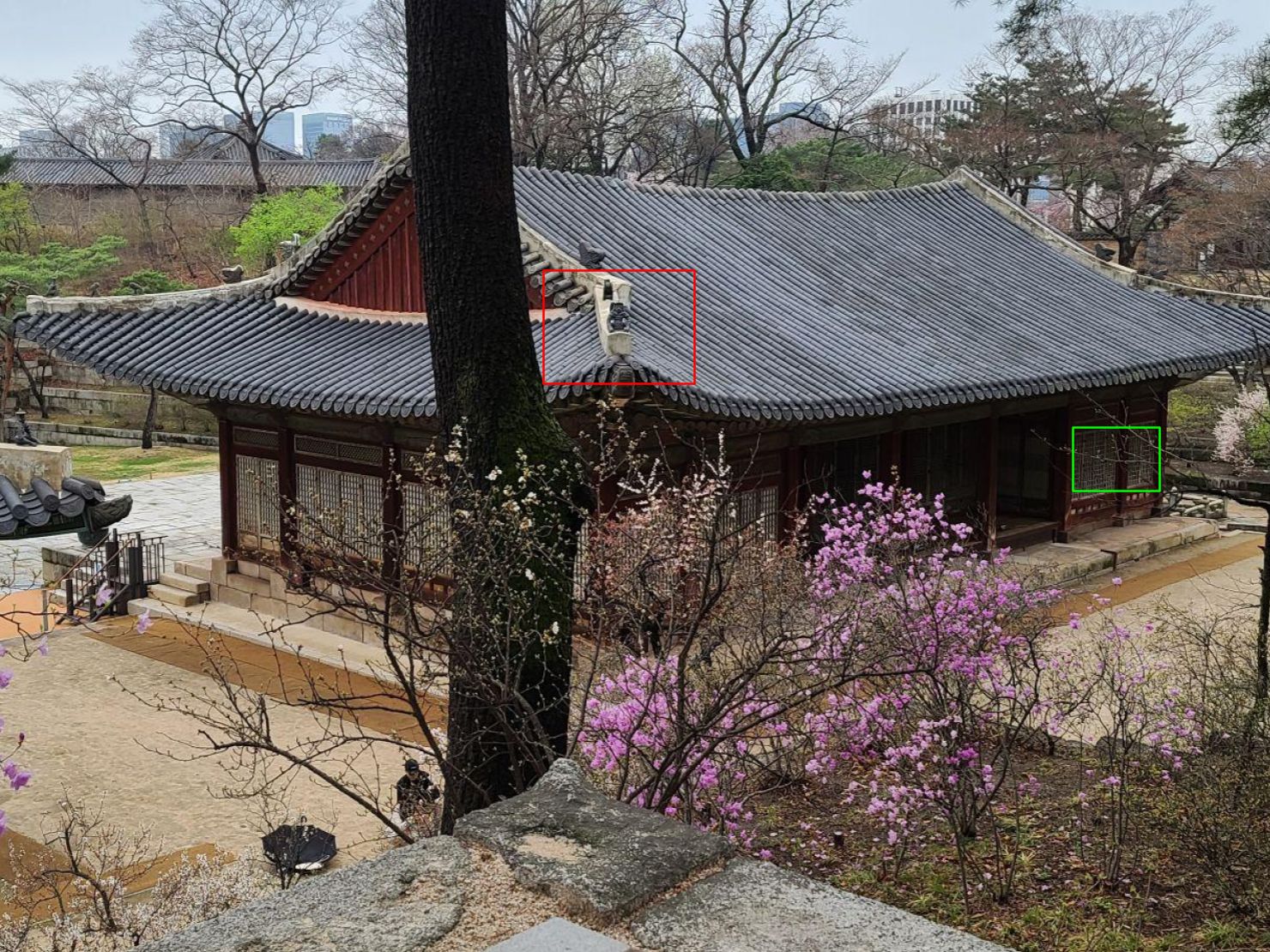}}}} &
			\hspace{0.5cm}
			\subfloat[\centering  Input~(LR) ]{
				\includegraphics[width = \patchSize]
				{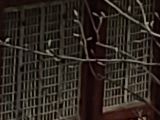}} \hspace{0.25cm}
			\subfloat[\centering  SR~(Ours)]{
				\includegraphics[width = \patchSize]
				{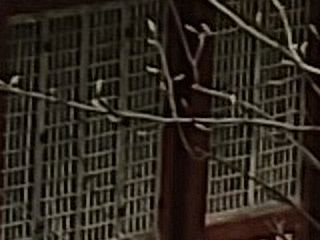}} \\  & &

			\hspace{0.5cm}
			\subfloat[\centering   Input~(LR)]{
				\includegraphics[width = \patchSize]
				{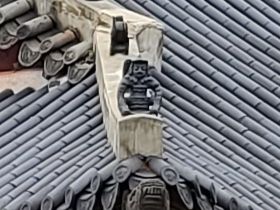}} \hspace{0.25cm}
				%\vspace{2.25mm}
			\subfloat[\centering SR~(Ours)]{
				\includegraphics[width = \patchSize]
				{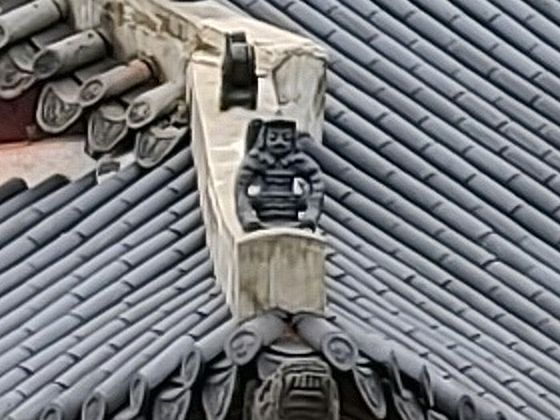}}
	\end{tabular}
	%\setlength{\abovecaptionskip}{0cm}
	%\vspace{-2mm}
	%\captionsetup{justification=raggedright,singlelinecheck=false}
	
	\caption{\textbf{SR results on single training images from our captured images with scale $\times2$.}
 	}
	\label{fig:supp_real_single}
	%\vspace{2mm}
\end{figure*}